\documentclass[11pt]{article}
\usepackage[utf8]{inputenc}
\usepackage[T1]{fontenc}
\usepackage{amsthm, amsmath}
\usepackage{amsfonts}
\usepackage[margin=1in]{geometry}
\usepackage[algo2e,ruled,noend,resetcount,linesnumbered]{algorithm2e}
\usepackage{graphicx}
\usepackage{comment}
\usepackage{enumitem}
\usepackage{thmtools}
\usepackage{thm-restate}
\usepackage{algorithm, algorithmicx, algpseudocode} 
\usepackage{array}
\usepackage{stmaryrd}
\usepackage{xcolor}

\usepackage[draft,margin=false,inline=true]{fixme}
\fxsetup{mode=multiuser,theme=color, layout=inline}
\FXRegisterAuthor{bs}{abs}{\color{blue} {\bf Barna}}

\setlist[enumerate]{nosep, topsep=1ex}
\setlist[itemize]{nosep, topsep=1ex}
\setlist[description]{nosep}
\allowdisplaybreaks
\usepackage[colorlinks,citecolor=blue,linkcolor=blue,urlcolor=red,pagebackref]{hyperref}
\usepackage[capitalise]{cleveref}
\usepackage{xspace}

\usepackage{caption}
\usepackage{subcaption}

\newcommand{\innerAlg}{\mathcal{A}}

\newcommand{\problem}{\mathcal{P}}

\newcommand{\bigO}[1]{O \left( #1 \right)}

\newcommand{\tO}[1]{\widetilde{O} (#1)}
\newcommand{\bigtO}[1]{\widetilde{O} \left( #1 \right)}
\newcommand{\OO}{\widetilde{O}}
\newcommand{\bigTh}[1]{\Theta \left( #1 \right)}

\newtheorem{theorem}{Theorem}[section]
\newtheorem{hypothesis}[theorem]{Hypothesis}
\newtheorem{corollary}[theorem]{Corollary}
\newtheorem{lemma}[theorem]{Lemma}
\newtheorem{proposition}[theorem]{Proposition}
\newtheorem{claim}[theorem]{Claim}

\newtheorem{open}{Open Question}
\theoremstyle{definition}
\newtheorem{definition}[theorem]{Definition}
\newtheorem{remark}[theorem]{Remark}

\newcommand{\true}{\mathrm{TRUE}}
\newcommand{\false}{\mathrm{FALSE}}

\newcommand{\eps}{\varepsilon}
\newcommand{\polylog}{\mathrm{polylog}}

\newcommand{\given}{\textrm{\xspace s.t. \xspace}}
\newcommand{\andT}{\textrm{\xspace and \xspace}}
\newcommand{\otherwise}{\textrm{\xspace o/w \xspace}}
\newcommand{\ceil}[1]{\left\lceil #1 \right\rceil}
\newcommand{\floor}[1]{\left\lfloor #1 \right\rfloor}
\newcommand{\set}[1]{\{#1\}}

\newcommand{\wt}{\mathsf{wt}}
\newcommand{\neighborhood}{N}

\newcommand{\maxmin}{(\max, \min)\xspace}
\newcommand{\minleq}{(\min, \leq)\xspace}
\newcommand{\maxleq}{(\max, \leq)\xspace}
\newcommand{\dominance}{\leq_{\#}}

\newcommand{\rowbal}{rb}
\newcommand{\colbal}{cb}

\newcommand{\arrival}{\mathbf{a}}
\newcommand{\arrivalPath}{\mathbf{ap}}

\newcommand{\distance}{\mathbf{d}}

\newcommand{\bottleneck}{\mathbf{b}}
\newcommand{\bottleneckPath}{\mathbf{bp}}

\newcommand{\stBP}{{\rm stBP}}
\newcommand{\SSBP}{{\rm SSBP}}
\newcommand{\APBP}{{\rm APBP}}

\newcommand{\stEA}{{\rm stEA}}
\newcommand{\SSEA}{{\rm SSEA}}
\newcommand{\APEA}{{\rm APEA}}

\newcommand{\stSP}{{\rm stSP}}
\newcommand{\SSSP}{{\rm SSSP}}
\newcommand{\APSP}{{\rm APSP}}

\newcommand{\nwstSP}{{\rm nw-stSP}}
\newcommand{\nwSSSP}{{\rm nw-SSSP}}
\newcommand{\nwAPSP}{{\rm nw-APSP}}

\newcommand{\SSR}{{\rm SSR}}

\newcommand{\OMv}{{\rm OMv}}
\newcommand{\OMvThree}{{\rm OMv3}}

\newcommand{\BMM}{{\rm BMM}}

\title{Fine-Grained Optimality of Partially Dynamic Shortest Paths and More}

\author{Barna Saha\thanks{University of California San Diego. \texttt{bsaha@ucsd.edu}. Supported by NSF grants 1652303, 1909046, 2112533, and HDR TRIPODS Phase II grant 2217058. This work was done in part while the author was visiting the Simons Institute for the Theory of Computing.}, Virginia Vassilevska Williams\thanks{Massachusetts Institute of Technology. \texttt{virgi@mit.edu}. Supported by NSF Grant CCF-2330048, BSF Grant 2020356 and a Simons Investigator Award. This work was done in part while the author was visiting the Simons Institute for the Theory of Computing.}, Yinzhan Xu\thanks{Massachusetts Institute of Technology. \texttt{xyzhan@mit.edu}. Partially supported by NSF Grant CCF-2330048, BSF Grant 2020356 and a Simons Investigator Award. This work was done in part while the author was visiting the Simons Institute for the Theory of Computing, and the Institute for Emerging CORE Methods in Data Science (EnCORE).}, Christopher Ye\thanks{University of California San Diego. \texttt{czye@ucsd.edu}. Partially supported by NSF grants 1652303, 1909046, 2112533, and HDR TRIPODS Phase II grant 2217058. This work was done in part while the author was visiting the Simons Institute for the Theory of Computing.}}
\date{\today}

\begin{document}

\maketitle

\setcounter{page}{0}
\thispagestyle{empty}

\begin{abstract}
    Single Source Shortest Paths ($\textrm{SSSP}$) is among the most well-studied problems in computer science. In the incremental (resp. decremental) setting, the goal is to maintain distances from a fixed source in a graph undergoing edge insertions (resp. deletions). A long line of research culminated in a near-optimal deterministic $(1 + \varepsilon)$-approximate data structure with $m^{1 + o(1)}$ total update time over all $m$ updates by Bernstein, Probst Gutenberg and Saranurak [FOCS 2021]. However, there has been remarkably little progress on the exact $\textrm{SSSP}$ problem beyond Even and Shiloach's algorithm [J. ACM 1981] for unweighted graphs. For weighted graphs, there are no exact algorithms beyond recomputing $\textrm{SSSP}$ from scratch in $\widetilde{O}(m^2)$ total update time, even for the simpler Single-Source Single-Target Shortest Path problem ($\textrm{stSP}$). Despite this lack of progress, known (conditional) lower bounds only rule out algorithms with amortized update time better than $m^{1/2 - o(1)}$ in dense graphs. 

In this paper, we give a tight (conditional) lower bound: any partially dynamic exact $\textrm{stSP}$ algorithm requires $m^{2 - o(1)}$ total update time for any sparsity $m$. We thus resolve the complexity of partially dynamic shortest paths, and separate the hardness of exact and approximate shortest paths, giving evidence as to why no non-trivial exact algorithms have been obtained while fast approximation algorithms are known.

Moreover, we give tight bounds on the complexity of combinatorial algorithms for several path problems that have been studied in the static setting since early sixties: Node-weighted shortest paths (studied alongside edge-weighted shortest paths), bottleneck paths (early work dates back to 1960), and earliest arrivals (early work dates back to 1958). These bounds rule out any nontrivial combinatorial algorithms for these problems in the partially dynamic setting. Interestingly, for all of the above path-variant problems, we obtain faster partially dynamic algorithms using fast matrix multiplication.

\end{abstract}

\newpage

\newpage
\setcounter{page}{1}

\section{Introduction}

Computing shortest paths is one of the most fundamental tasks in computer science. Given an $n$-node, $m$-edge {\em directed } graph\footnote{Throughout the paper, we will assume $m \ge n - 1$. } $G = (V, E)$ with nonnegative integer edge weights, the Single Source Shortest Paths (\SSSP{}) problem asks to compute the distance from a given node $s$ to all $v \in V$. Dijkstra's classical algorithm \cite{dijkstra1959, fredman1987fibonacci} solves this problem in $O(m + n \log n)$ time. \SSSP{} (and its variants for undirected graphs and for graphs with negative weights) has been extensively studied~\cite{shimbel1953structural, ford1956lr, bellman1958routing, moore1959shortest, DBLP:conf/focs/Gabow83, DBLP:journals/siamcomp/GabowT89,DBLP:journals/jcss/FredmanW93,  FREDMAN1994533, DBLP:journals/siamcomp/Goldberg95, DBLP:conf/esa/Raman96, DBLP:conf/soda/Thorup96, DBLP:journals/sigact/Raman97, DBLP:journals/jacm/Thorup99, Hagerup00, DBLP:journals/jal/Thorup00, DBLP:journals/siamcomp/PettieR05, DBLP:journals/jcss/Thorup04, DBLP:conf/focs/BrandLNPSS0W20, DBLP:conf/focs/AxiotisMV20, DBLP:conf/focs/BernsteinNW22, DBLP:conf/focs/ChenKLPGS22,negativeRealSSSP}. The current fastest \SSSP{} algorithm runs in $O(m + n \log \log n)$ time \cite{DBLP:journals/jcss/Thorup04}.  

For the all-pairs version of the problem, All-Pairs Shortest Paths (\APSP{}), one needs to compute the distance between every pair of nodes in the given graph. One could run Dijkstra's algorithm $n$ times to obtain an $O(mn + n^2 \log n)$ running time. For a dense graph where $m = \Theta(n^2)$, this running time becomes $O(n^3)$. An alternative classical algorithm, the Floyd–Warshall algorithm, also achieves an $O(n^3)$ running time. A long series of works (e.g. \cite{fredman1976new,Dob1990,Takaoka92,Takaoka98,Han04,pettie2004apsp,ZwickAPSP04,Chan05,Han06,DBLP:journals/siamcomp/Chan10,HanT12, orlin2022directed}) improved poly-logarithmic factors over the classic $O(n^3)$ running time, culminating in the current fastest algorithm by Williams \cite{Williams18} with $n^3 / 2^{\Theta(\sqrt{\log n})}$ running time, a super poly-logarithmic improvement over the classic $O(n^3)$ running time. Due to the lack of polynomial improvements, the \APSP{} hypothesis, stating that there is no $O(n^{3-\eps})$ time algorithm for any $\eps>0$ that computes \APSP{} for $n$-node graphs, is among the most popular hypotheses in Fine-Grained Complexity. 

Shortest paths problems have also been extensively studied in dynamic graphs. The main focus of our paper is the {\em partially dynamic } setting, where the dynamic graph only undergoes edge insertions (incremental) or only undergoes edge deletions (decremental). Partially dynamic problems are often easier than their fully-dynamic counterparts where both edge insertions and deletions are allowed. We are interested in the total running time of the algorithms over all the operations (starting from an empty graph in the incremental case, or deleting edges until the graph is empty in the decremental case). A seminal algorithm of Demetrescu and Italiano \cite{DemetrescuI04} shows how to maintain \APSP{} in amortized $\OO(n^2)$ time,\footnote{$\OO$ hides $\polylog(n)$ factors.} even for fully dynamic graphs. This amortized running time is essentially optimal if our goal is to minimize the total running time, as the output size after each operation is already $n^2$ for \APSP{}.\footnote{It is still an active line of research to improve the worst-case time guarantee for fully dynamic \APSP{} \cite{DBLP:conf/stoc/Thorup05, DBLP:conf/soda/AbrahamCK17, DBLP:conf/soda/GutenbergW20b, DBLP:conf/soda/ChechikZ23, Mao24}, and the current fastest algorithm is designed by Mao \cite{Mao24} which runs in $\OO(n^{2.5})$ time per operation. } Therefore, we will focus on partially dynamic $\SSSP$.

For unweighted graphs, Even and Shiloach \cite{EvenS81} gave a partially dynamic \SSSP{} algorithm with total time $O(m n)$, a bound that has stood for more than 40 years.
In the meantime, a long line of work has studied $(1 + \varepsilon)$-approximate $\SSSP$, where the algorithm must maintain approximate shortest paths after each update \cite{DBLP:conf/soda/BernsteinR11, DBLP:conf/stoc/HenzingerKN14, DBLP:conf/soda/HenzingerKN14, DBLP:conf/stoc/BernsteinC16, DBLP:journals/siamcomp/Bernstein16, DBLP:conf/soda/BernsteinC17, DBLP:conf/icalp/Bernstein17, DBLP:journals/jacm/HenzingerKN18, DBLP:conf/stoc/ChuzhoyK19, DBLP:conf/focs/BernsteinGS20, DBLP:conf/stoc/GutenbergWW20, DBLP:conf/soda/GutenbergW20a, DBLP:conf/soda/GutenbergW20, DBLP:conf/focs/BernsteinGS21}.

However, there has been little progress on maintaining {\em exact} \SSSP{} in partially dynamic weighted graphs. 
The naive algorithm for partially dynamic \SSSP{} is to rerun Dijkstra's algorithm after every update, which results in an $\OO(m^2)$ total running time, or $\OO(m)$ time per-update, even when amortized over all $m$ updates. 
For dense graphs, this running time is $\OO(n^4)$. This trivial re-computation is essentially the best known for partially dynamic \SSSP{}. Even for the simpler partially dynamic \stSP{} problem where we only need to maintain the distance between one source node $s$ and one sink node $t$, there are no better upper bounds. 

\begin{restatable}{open}{openQstSP}
    \label{open:n4-stSP}
    Can we solve partially dynamic \stSP{} in $O(m^{2-\eps})$ total time for some $\eps > 0$, for any sparsity $m$? In particular, can we solve partially dynamic \stSP{} in $O(n^{4-\eps})$ total time for some $\eps > 0$? 
\end{restatable}

There has been partial progress towards answering \cref{open:n4-stSP}. Roditty and Zwick~\cite{RodittyZ11} proved that under the \APSP{} hypothesis, any partially dynamic \SSSP{} algorithm handling $n$ updates requires total time $n^{3 - o(1)}$, which is a factor of $n$ away from the naive algorithm for dense graphs. This lower bound was later strengthened by Abboud and Vassilevska Williams \cite{abboud2014popular} to hold also for partially dynamic \stSP{}. While this lower bound rules out any algorithm improving upon re-computation from scratch when amortized over only $n$ updates, it only rules out algorithms with $m^{1/2 - \eps}$ amortized update time when amortizing over all $m$ updates. In particular, previous lower bounds do not rule out the possibility of an algorithm which has only $n$ costly updates with the rest of the updates being relatively fast, resulting in $m^{1/2 - o(1)}$ amortized time when amortizing over all $m$ updates.

More partial progress was made by Gutenberg, Vassilevska Williams and Wein \cite{DBLP:conf/stoc/GutenbergWW20}, who showed that in the special case of sparse graphs where $m = n^{1+o(1)}$, any algorithm solving partially dynamic \stSP{} requires total time $m^{2 - o(1)}$, under the $k$-Cycle hypothesis from Fine-Grained Complexity (see~\cite{DBLP:conf/soda/LincolnWW18, DBLP:conf/icalp/AnconaHRWW19}) which postulates that any algorithm that can detect whether a graph contains a $k$-cycle (for every constant $k\geq 3$) requires $m^{2-o(1)}$ time. In terms of $n$, detecting a $k$-cycle can be solved in $O(n^\omega)$ time\footnote{$\omega < 2.372$ \cite{DBLP:conf/focs/DuanWZ23, VXXZ24} is the square matrix multiplication exponent. We will also use $\omega(a, b, c)$ to denote the exponent for multiplying an $n^a \times n^b$ matrix and an $n^b \times n^c$ matrix. } for all constants $k$ (see e.g., \cite{DBLP:journals/siamcomp/ItaiR78}). When $m = \Omega(n^{1+c})$ for some $c > 0$, $O(n^\omega)$ can be faster than $m^2$, e.g. in the optimistic case where $\omega = 2$. Therefore, the $k$-Cycle hypothesis is most reasonable for graphs with $m = n^{1+o(1)}$. Thus, the lower bound in \cite{DBLP:conf/stoc/GutenbergWW20} does not help with \cref{open:n4-stSP} for graphs with $m = \Omega(n^{1 + c})$ for any $c > 0$. Furthermore, their result is not meaningful for partially dynamic SSSP with $m = n^{1+o(1)}$, because an $\Omega(mn) = m^{2-o(1)}$ lower bound is trivial, as $mn$ is the total output size ($n$ distances after each update). 

We remark that \emph{fully dynamic} \stSP{} was shown to require $n^{2 - o(1)}$ amortized update time over any (polynomial) number of updates for combinatorial\footnote{``Combinatorial'' algorithms are not well-defined, but the term has become synonomous with algorithms that do not use the heavy algebraic techniques used in fast matrix multiplication algorithms. Hardness under the \BMM{} hypothesis essentially implies that fast algorithms can only exist if fast matrix multiplication is used.} algorithms by Jin and Xu \cite{JinX22}. 
Indeed, any combinatorial algorithm requires $n^{2 - o(1)}$ amortized update time even for the simpler $(s, t)$-reachability problem in the fully dynamic setting. 

Roditty and Zwick \cite{RodittyZ11} considered partially dynamic \SSSP{} where instead of outputting the distances from $s$ to all nodes in $V$ after each update, the data structure only needs to support distance queries between $s$ and any queried node $v$. They explicitly stated the following open question:

\begin{open}
\label{open:RZ}
    Is partially dynamic $\SSSP$ with $m$ updates and $n^2$ queries in $\OO(mn)$ time?
\end{open}

In particular, if the answer to \cref{open:RZ} is affirmative, then one can solve partially dynamic $\stSP$ in $\OO(mn)$ time, as the total number of queries for partially dynamic $\stSP$ is $m \le n^2$. Therefore, if the answer to \cref{open:RZ} is affirmative, then the answer to \cref{open:n4-stSP} is also affirmative for $m = \Omega(n^{1 + c})$ for $c > 0$. Neither of the previously mentioned lower bounds~\cite{abboud2014popular, DBLP:conf/stoc/GutenbergWW20} can resolve \cref{open:RZ}. In the contrapositive, if one can resolve \cref{open:n4-stSP} in the negative, then it would also imply a resolution of \cref{open:RZ} in the negative. 

\paragraph{Strong Hardness for Exact Partially Dynamic \stSP{}.}
As the first main result of our paper, we fully resolve \cref{open:n4-stSP} in the negative based on a popular fine-grained hypothesis. Thus, the answer to \cref{open:RZ} is (conditionally) negative as well, based on the above discussion. 
In particular, we resolve the complexity of partially dynamic shortest paths, and separate the hardness of exact and approximate shortest paths, giving evidence as to why no non-trivial exact algorithms have been obtained while fast approximation algorithms are known.

Our lower bound is based on the Minimum-Weight 4-Clique hypothesis, which states that finding the $4$-clique with minimum total edge weights in an $n$-node graph requires $n^{4-o(1)}$ time. This hypothesis is a special case of the Minimum-Weight $k$-Clique hypothesis, which has been used in, e.g., \cite{DBLP:conf/icalp/AbboudWW14, DBLP:conf/icalp/BackursDT16, DBLP:conf/icml/BackursT17, DBLP:conf/soda/LincolnWW18, DBLP:conf/stoc/AbboudBDN18, DBLP:journals/talg/BringmannGMW20, DBLP:journals/corr/abs-2106-03131, BringmannFHKKR24}. In particular, the Minimum-Weight $3$-Clique hypothesis is equivalent to the \APSP{} hypothesis \cite{DBLP:journals/jacm/WilliamsW18}. Our result can be formally stated as follows:

\begin{restatable}{theorem}{stSPLB}
    \label{thm:s-t-sp-lb}
    Under the Minimum Weight 4-Clique hypothesis, any algorithm computing incremental/decremental \stSP{} on $n$-node undirected graphs requires $n^{4 - o(1)}$ total time.
\end{restatable}

Through a simple graph transformation, \cref{thm:s-t-sp-lb} implies that any partially dynamic \stSP{} algorithm requires $m^{2 - o(1)}$ total time over $m$ updates in an $n$-node $m$-edge undirected graph for {\em any sparsity} $m$. In our work, we mostly consider directed graphs, but notably, \cref{thm:s-t-sp-lb} even holds for undirected graphs. 

\subsection{Variants of Shortest Paths in the Dynamic Setting} In the static setting, there are several variants of shortest paths that have been studied extensively. 
Examples include shortest paths on structured graphs such as node-weighted graphs \cite{DBLP:journals/siamcomp/Chan10}, bottleneck paths \cite{pollack60bp, hu1961maximum} (the bottleneck path between two nodes is the path whose minimum weight edge is maximized), and earliest arrivals \cite{minty1958variant} (the earliest arrival between two nodes is the minimum weight of the last edge among all paths whose edges are increasing in weight; earliest arrivals is also known as nondecreasing paths). 
All these problems are extremely basic and well-motivated: 
Node-weighted graphs naturally arise in applications for many problems related to shortest paths \cite{DBLP:conf/stoc/VassilevskaW06, DBLP:conf/icalp/VassilevskaWY06, DBLP:journals/siamcomp/CzumajL09, DBLP:conf/soda/ShapiraYZ07, DuanP09, DBLP:journals/siamcomp/WilliamsW13}. 
Bottleneck paths \cite{gabow1988bottleneck, chechik2016bottleneck} have many applications, from 
 max-flow algorithms
(\cite{DBLP:journals/jacm/EdmondsK72}, Chapter 7.3 in \cite{networkflows}), to voting protocols~\cite{Schulze11}, to metabolic pathway analysis \cite{UllahLH09} and more.
Earliest arrivals has applications to public transportation scheduling such as train and planes (see e.g. \cite{minty1958variant,nondecpaths}).

Modifications of Dijkstra’s algorithm can solve the single-source version of all three variants in $\OO(m)$ time. Faster than Dijkstra's algorithms exist for all of them: Single Source Shortest Paths on node-weighted graphs (\nwSSSP{}) can be solved in $O(m + n \log \log n)$ time \cite{DBLP:journals/jcss/Thorup04} similar to $\SSSP{}$, Single Source Earliest Arrival (\SSEA{}) can be solved in $O(m\log\log n)$ time in the addition-comparison model and $O(m+n)$ time on the word-RAM \cite{nondecpaths}, and Single Source Bottleneck Paths (\SSBP{}) can be solved in $O(m \sqrt{\log n})$ time \cite{DuanLX18} which is faster than Dijkstra’s algorithm when $m = o(n \sqrt{\log n})$.

In contrast to \APSP{}, the all-pairs versions of these problems all have {\em truly subcubic} time algorithms as first shown by \cite{VassilevskaWY09,nondecpaths,DBLP:journals/siamcomp/Chan10}. The fastest algorithm to date for All-Pairs Shortest Paths on node-weighted graphs (\nwAPSP{}) runs in $\OO(n^{(9+\omega)/4}) = O(n^{2.843})$ time \cite{DBLP:journals/siamcomp/Chan10}. Using rectangular matrix multiplication, this running time can be improved to $O(n^{2.830})$~\cite{DBLP:conf/soda/Yuster09}.  The fastest algorithms for All-Pairs Bottleneck Paths (\APBP{})  \cite{DuanP09} and All-Pairs Earliest Arrivals (\APEA{}) \cite{DuanJW19} both run in $\OO(n^{(3+\omega)/2}) = O(n^{2.686})$ time. 

Notably, all of the aforementioned subcubic algorithms use fast matrix multiplication, and that is for a good reason. Under the Boolean Matrix Multiplication (\BMM{}) hypothesis, no combinatorial algorithm for these problems can achieve $O(n^{3-\eps})$ for $\eps > 0$ time. 

To the best of our knowledge, the dynamic versions of these natural variants of shortest paths have not been explicitly studied before, despite the extensive study of their static versions. In this paper, we initiate the study of the dynamic versions of node-weighted shortest paths, bottleneck paths, and earliest arrivals, obtaining algorithms and conditional lower bounds.

\paragraph{New Partially Dynamic Algorithms.} 
First, we give sub-quartic total time algorithms for the partially dynamic versions of all three variants, setting them apart from partially dynamic \SSSP{}. In particular, for partially dynamic \SSEA{}, our running time is near-linear. For node-weighted $s$-$t$ shortest paths (\nwstSP{}), we also give an algorithm that is faster than \nwSSSP{}. We give similar faster algorithms for Single Source Bottleneck Paths (\SSBP{}) and $s$-$t$ Bottleneck Paths (\stBP{}).

\begin{theorem}
\label{thm:intro:algos}
For $n$-node $m$-edge graphs, there are algorithms for 
    \begin{itemize}
        \item incremental \nwstSP{} in $\tO{n^{4 - \gamma/3}} = O(n^{3.887})$ total time;
        \item incremental \nwSSSP{} in $\tO{n^{4 - \gamma/4}} = O(n^{3.915})$ total time;
        \item incremental/decremental \stBP{} in $\tO{n^{2 + \varepsilon_1 + \varepsilon_2} + n^{2 + \omega(1, \varepsilon_1, \varepsilon_2) - \varepsilon_1} + n^{2 + \omega(1, 1, \varepsilon_2) - \varepsilon_2}} = O(n^{3.405})$ total time;
        \item incremental \SSBP{} in $\tO{n^{(5 + \omega)/2}} = O(n^{3.686})$ time; 
        \item incremental or decremental \SSEA{} in $\OO(m)$ total time, 
    \end{itemize}
    where $\gamma$ is the solution to the equation $\omega(1, 1, 1 + \gamma) = 3 - \gamma$ and $\varepsilon_1, \varepsilon_2$ are parameters to be optimized.
\end{theorem}

\paragraph{New Hardness Results.} All algorithms except the one for \SSEA{} in \cref{thm:intro:algos} use fast matrix multiplication, and it is natural to ask whether fast matrix multiplication is necessary in order to achieve $O(n^{4-\eps})$ running time for these problems. We show that this is indeed the case, under the Combinatorial $4$-Clique hypothesis, which states that any combinatorial algorithm detecting whether a graph contains a $4$-clique requires $n^{4-o(1)}$ time. This hypothesis is a natural generalization of the BMM hypothesis, as the latter  has been shown to be equivalent to the hypothesis that any combinatorial algorithm for Triangle Detection in $n$-node graphs requires $n^{3-o(1)}$ time \cite{DBLP:journals/jacm/WilliamsW18}. The Combinatorial $k$-Clique hypothesis has been widely used in prior works (e.g, \cite{DBLP:journals/comgeo/Chan10, DBLP:conf/focs/BringmannGL17, DBLP:conf/focs/Li19, DBLP:conf/icalp/AbboudGIKPTUW19, DBLP:conf/stoc/GutenbergWW20, JinX22, huang2023tight}). In fact, the reductions for this result are adapted from the reduction for \cref{thm:s-t-sp-lb}, showcasing the versatility of our technique. 

\begin{theorem}
\label{thm:intro:stBP-nwstSP-lb}
    Under the Combinatorial $4$-Clique hypothesis, any combinatorial algorithm computing incremental/decremental \stBP{} or \nwstSP{} requires $n^{4-o(1)}$ total time. 
\end{theorem}
Similar to before, \cref{thm:intro:stBP-nwstSP-lb} implies an $m^{2-o(1)}$ lower bound for any sparsity $m$. 

Even with fast matrix multiplication, the current best running time for detecting whether a graph contains a $4$-clique is $O(n^{\omega(1, 2, 1)})$ \cite{DBLP:journals/tcs/EisenbrandG04}.
Notably, the current best bound \cite{VXXZ24} on $\omega(1, 2, 1)$ is $3.251>3$; if $\omega=2$, then $\omega(1, 2, 1)=3$.

Our method implies that, unless we can improve this $4$-clique running time, any algorithm (even non-combinatorial ones) computing incremental/decremental \stBP{} or \nwstSP{} requires $n^{\omega(1, 2, 1)-o(1)}$ total time. We can further improve the exponent for the lower bound, under the hardness of the \OMvThree{} problem, proposed by~\cite{DBLP:conf/stoc/GutenbergWW20} as a generalization of the \OMv{} problem \cite{DBLP:conf/stoc/HenzingerKNS15}, and the Minimum-Witness $3$-Product problem, which is a generalization of the Minimum-Witness Product problem proposed in \cite{DBLP:journals/algorithms/KowalukL22}.

In the \OMvThree{} problem, we need to preprocess an $n \times n$ Boolean matrix $A$, so that we can support the following types of queries: Given three $n$-dimensional Boolean vectors $u, v, w$, determine whether there exist $i, j, k \in [n]$, so that $u_i \wedge v_j \wedge w_k \wedge A_{ij} \wedge A_{jk} \wedge A_{ki}$. A naive algorithm solves each query in $O(n^\omega)$ time, and it is hypothesized in \cite{DBLP:conf/stoc/GutenbergWW20} that this is essentially the best algorithm among all algorithms with polynomial preprocessing time. 

\begin{restatable}[\OMvThree{} hypothesis \cite{DBLP:conf/stoc/GutenbergWW20}]{hypothesis}{OMvThreeHypothesis}
    \label{conj:omv-3}%
    There is no algorithm with polynomial preprocessing time and total update and query time $O(n^{\omega + 1 - \varepsilon})$  solving the \OMvThree{} problem, for any $\varepsilon > 0$.
\end{restatable}

We show the following results under the \OMvThree{} hypothesis:
\begin{theorem}
    Under the \OMvThree{} hypothesis, any algorithm computing incremental/decremental \stBP{} or \nwstSP{} requires $n^{\omega+1-o(1)}$ total time. 
\end{theorem}

For \nwSSSP{}, we are able to show a higher lower bound, based on the hardness of a variant of the Minimum-Witness Product problem. 

In the Minimum-Witness Product problem, we are given two Boolean matrices $A$ and $B$, and we need to compute $\min \{k \in [n]: A_{i,k} \wedge B_{k, j}\}$ for every $(i, j) \in [n] \times [n]$. The current best algorithm for this problem runs in $O(n^{2+\lambda})$ time \cite{CzumajKL07}, where $\omega(1, \lambda, 1) = 1+2\lambda$. If $\omega = 2$, this running time is essentially $O(n^{2.5})$; the hypothesis that there is no $O(n^{2.5-\eps})$ time algorithm for Minimum-Witness Product has been considered before \cite{DBLP:conf/innovations/Lincoln0W20}. 

We consider the following natural generalization of Minimum-Witness Product to three matrices, first proposed by \cite{DBLP:journals/algorithms/KowalukL22}, which we call Minimum-Witness $3$-Product. In this problem, we are given three $n \times n$ Boolean matrices $A, B, C$, and for every $i_1, i_2, i_3 \in [n]$, we need to find the minimum value of $j \in [n]$ such that $A_{i_1, j} \wedge B_{i_2, j} \wedge C_{i_3, j}$. 

\cite{DBLP:journals/algorithms/KowalukL22} gave an $O(n^{3.5})$ time algorithm (when $\omega = 2$) for Minimum-Witness $3$-Product, which is an adaptation of the existing algorithm for Min-Witness Product \cite{CzumajKL07}. 
As $O(n^{3.5})$ is the best running time even when $\omega = 2$, the following hypothesis is plausible. See \cref{sec:min-witness-lb} for more discussion on why it is plausible.

\begin{restatable}[Minimum-Witness $3$-Product hypothesis]{hypothesis}{MinWitnessThreeProductHypothesis}
    \label{conj:min-witness-3-product}
    There is no $O(n^{3.5 - \varepsilon})$ time algorithm for the Minimum-Witness $3$-Product problem for $\eps > 0$. 
\end{restatable}

We obtain a conditional lower bound based on the Minimum-Witness $3$-Product hypothesis.

\begin{theorem}
    Under the Minimum-Witness $3$-Product hypothesis, any algorithm computing incremental/decremental \nwSSSP{} requires $n^{3.5-o(1)}$ total time. 
\end{theorem}

\begin{table}
    \footnotesize
    \begin{center}
        {\setlength{\extrarowheight}{5pt}%
        \begin{tabular}{>{\raggedright}p{0.17\textwidth}|>{\raggedright}p{0.13\textwidth}|>{\raggedright}p{0.18\textwidth}|>{\raggedright}p{0.17\textwidth}|>{\raggedright}p{0.17\textwidth}}
             \hline
             Problem & Upper Bound & Lower Bound & Combinatorial Lower Bound & Known Lower Bound\tabularnewline
             \hline
             \hline
             {\bf Shortest Paths} & & & & \tabularnewline
             $(s, t)$ & $n^4$ & $n^{4 - o(1)}$ (Thm. \ref{thm:s-t-sp-lb}) & - & $n^{3 - o(1)}$ \cite{abboud2014popular} \tabularnewline
             single source & $n^4$ & $n^{4 - o(1)}$ (Thm. \ref{thm:s-t-sp-lb}) & - & $n^{3 - o(1)}$ \cite{RodittyZ11} \tabularnewline
             {\bf Node-Weighted Shortest Paths} & & & & \tabularnewline
             {\bf inc.} $(s, t)$ & $n^{3.887}$ $[n^{3.834}]$ (Prop. \ref{prop:incremental-node-weighted-s-t-sp}) & $n^{\omega + 1 - o(1)}$ [$n^{3- o(1)}$] (Thm. \ref{thm:s-t-sp-node-weight-lb-omv-3}) & $n^{4 - o(1)}$ (Thm. \ref{thm:s-t-sp-node-weight-lb}) & $n^{2}$ \tabularnewline
             {\bf inc.} single source & $n^{3.915}$ $[n^{3.875}]$ (Prop. \ref{prop:incremental-node-weighted-sssp}) & $n^{3.5 - o(1)}$ (Thm. \ref{thm:sssp-node-weight-min-witness-lb}) & $n^{4 - o(1)}$ (Thm. \ref{thm:s-t-sp-node-weight-lb}) & $n^{2.75 - o(1)}$ \cite{RodittyZ11} \tabularnewline
             {\bf Bottleneck Paths} & & & & \tabularnewline
             $(s, t)$ & $n^{3.405}$ $[n^{3.25}]$ (Thm. \ref{prop:s-t-bp-reachability}) & $n^{\omega + 1 - o(1)}$ [$n^{3 - o(1)}$] (Thm. \ref{thm:s-t-bp-lb-omv-3}) & $n^{4 - o(1)}$ (Thm. \ref{thm:s-t-bp-lb}) & $n^{2}$ \tabularnewline
             {\bf inc.} single source & $n^{3.686}$ $[n^{3.5}]$ (Thm. \ref{prop:inc-ssbp-alg}) & $n^{\omega + 1 - o(1)}$ [$n^{3 - o(1)}$] (Thm. \ref{thm:s-t-bp-lb-omv-3})& $n^{4 - o(1)}$ (Thm. \ref{thm:s-t-bp-lb}) & $n^{2}$ \tabularnewline
            {\bf Earliest Arrivals} & & & & \tabularnewline
             single source & $n^2$ (Thm. \ref{thm:partially-dynamic-SSEA}) & $n^{2}$ & $n^{2}$ & $n^{2}$ \tabularnewline
             \hline
        \end{tabular}}
    \end{center}
    \caption{Summary of results for partially dynamic $(s, t)$ and single source problems with respect to number of nodes $n$. 
    The table shows total update and query time. 
    Unless otherwise stated, our bounds hold for both incremental and decremental versions of the problem. 
    Upper bounds are given ignoring polylogarithmic factors. 
    Bounds are given with respect to the current best upper bound of (rectangular) matrix multiplication \cite{VXXZ24} computed with \cite{Complexity}. 
    Bounds within brackets assume $\omega = 2$.}
    \label{tbl:partially-dynamic-results}
\end{table}

\begin{table}
    \footnotesize
    \begin{center}
        {\setlength{\extrarowheight}{5pt}%
        \begin{tabular}{>{\raggedright}p{0.17\textwidth}|>{\raggedright}p{0.22\textwidth}|>{\raggedright}p{0.17\textwidth}|>{\raggedright}p{0.29\textwidth}}
             \hline
             Problem & Lower Bound & Combinatorial Lower Bound & Known Lower Bound\tabularnewline
             \hline
             \hline
             {\bf Shortest Paths} & & & \tabularnewline
             $(s, t)$ & $m^{2 - o(1)}$ (Thm. \ref{thm:s-t-sp-lb}) & - & $m^{2 - o(1)}$ if $m = n^{1+o(1)}$ \cite{DBLP:conf/stoc/GutenbergWW20} \tabularnewline
             {\bf Node-Weighted Shortest Paths} & & & \tabularnewline
             $(s, t)$ & $m^{\frac{\omega + 1}{2} - o(1)}$ [$m^{1.5 - o(1)}$] (Thm. \ref{thm:s-t-sp-node-weight-lb-omv-3}) & $m^{2 - o(1)}$ (Thm. \ref{thm:s-t-sp-node-weight-lb}) & $m$ \tabularnewline
             single source & $n^{1.75 - o(1)}$ (Thm. \ref{thm:sssp-node-weight-min-witness-lb}) & $m^{2 - o(1)}$ (Thm. \ref{thm:s-t-sp-node-weight-lb}) & $m^{1.375 - o(1)}$ \cite{RodittyZ11} \tabularnewline
             {\bf Bottleneck Paths} & & & \tabularnewline
             $(s, t)$ & $m^{\frac{\omega + 1}{2} - o(1)}$ [$m^{1.5 - o(1)}$] (Thm. \ref{thm:s-t-bp-lb-omv-3}) & $m^{2 - o(1)}$ (Thm. \ref{thm:s-t-bp-lb}) & $m$ \tabularnewline
             \hline
        \end{tabular}}
    \end{center}
    \caption{Summary of lower bounds for partially dynamic $(s, t)$ and single source problems with respect to number of edges $m$. 
    The table shows total update and query time. 
    Our bounds hold for both incremental and decremental versions of the problem and all values of $m \leq n^2$ (Prop \ref{prop:graph-sparsification}).
    Bounds are given with respect to the current best upper bound $\omega$ \cite{VXXZ24}.
    Bounds within brackets assume $\omega = 2$.}
    \label{tbl:partially-dynamic-results-m}
\end{table}

\paragraph{Weight-Dynamic Algorithms}
Typically, updates to dynamic graphs are considered in the context of edge insertion and deletion.
\cite{DBLP:conf/networking/HenzingerP021} introduced \emph{weight-dynamic} graphs, on which updates can additionally modify the weight of an existing edge.
Note that this generalizes the insertion/deletion model, as edge insertions/deletions can be modeled by setting weights to $\infty$ or $- \infty$ depending on the setting of the problem.
In the partially dynamic versions of this model, one can either only increase the edges weights, or only decrease the edge weights. 
In this more general model, all our lower bounds hold, and we additionally show that combinatorial algorithms for $\stEA$ require $m^{2 - o(1)}$ time under the Combinatorial $4$-Clique hypothesis, whereas in the insertion/deletion model we obtain an $\OO(m)$ time algorithm for \SSEA{}, exhibiting a separation between the two dynamic models.

\paragraph{Fully Dynamic Algorithm}
The main focus of our paper is partially dynamic algorithms, but we also show a fully dynamic algorithm for \APBP{}. Namely, we show that fully dynamic \APBP{} can be solved in $\OO(n^2)$ amortized update time.

\subsection{Preliminaries}
\label{sec:prelim-slim}

Let $[n] = \set{1, 2, \dotsc, n}$.
Let $\vec{v}[i]$ denote the $i$-th entry of vector $\vec{v}$ and $A[i, j]$ denote the $[i, j]$-th entry of $A$.
Let $N(v)$ denote the neighborhood of a vertex $v$.
If $P_1, P_2$ are paths in $G$ with $P_1 = (a, \dotsc, b), P_2 = (b \dotsc, c)$, let $P_1 \circ P_2$ denote the path concatenation $(a, \dotsc, b, \dotsc, c)$.

In this work, we consider a dynamic edge update model.
Concretely, an incremental algorithm must handle edge insertions, a decremental must handle deletions, while a fully dynamic algorithm must handle both insertions and deletions.
Unless otherwise specified, our upper bounds obtain algorithms that answer queries in $O(1)$ time.
When specified, we also consider the weight update model.
In this case, an incremental algorithm must handle weight increases, a decremental algorithm must handle weight decreases, while a fully dynamic algorithm must handle arbitrary weight changes.

\subsection{Organization} The remainder of the paper is organized as follows. Section~\ref{sec:shortest-paths} is dedicated to showing a tight lower bound for partially dynamic shortest paths. Section~\ref{sec:node-weight}, Section~\ref{sec:bottleneck}, and Section~\ref{sec:earliest-arrivals} delve into node-weighted shortest paths, bottleneck paths, and the earliest-arrival problems respectively, giving both lower bounds and faster algorithms in the partially dynamic setting.

\section{A Tight Lower Bound for Partially Dynamic Shortest Paths}
\label{sec:shortest-paths}

We now present our main result: a tight conditional lower bound for partially dynamic \stSP{}.
As a warm-up, we prove a lower bound for graphs allowing parallel edges. 
Prior reductions have used APSP, triangle detection as well as $4$-Clique instances for dynamic shortest path lower bounds e.g. as in \cite{RodittyZ11,abboud2014popular,JinX22}. 
We extend these constructions with the addition of parallel edges with carefully chosen weights and insertion sequence that allow for a partially dynamic lower bound. 
This extension from previous works with parallel edges in the constructed graph is relatively simple and it was also observed by \cite{AdamPersonal}. 
Later we explain how to remove the parallel edges, which is the crux of our contribution.

\begin{proposition}
    \label{prop:multi-s-t-sp-lb-min-weight}
    Under the Minimum-Weight $4$-Clique hypothesis, any combinatorial algorithm computing incremental/decremental $\stSP$ on undirected (not necessarily simple) graphs with $n$ vertices and $m$ edges requires $n^{4 - o(1)}$ total time.
\end{proposition}

Suppose we are given a Minimum-Weight $4$-Clique instance. 
As is typical with $k$-clique problems,\footnote{For any $k$-clique instance $G=(V,E)$ we can create a $k$-partite $G'$ with vertex set containing $k$ copies of $V$, $V_1,\ldots,V_k$, each an independent set, and for every two copies $V_i$ and $V_j$ we add an edge $(u_i,v_j)$ for every edge $(u,v)$ in $G$, where $u_i\in V_i,v_j\in V_j$ are the copies of $u$ and $v$ in $V_i$ and $V_j$ respectively. 
If $G$ had weights, the vertices and edges of $G'$ inherit the weights of the vertices and edges in $G$ that they represent.} we may assume that the given graph is $k$-partite. 
Thus, let the Minimum-Weight $4$-Clique instance be a $4$-partite graph $G$ with $V(G) = A \cup B \cup C \cup D$, each part consisting of $n$ nodes.
We identify the vertices of $A, B, C, D$ with the integer set $[n]$.
We can also assume that $G$ is a complete $4$-partite graph, as the non-edges can be replaced by edges of large enough weight (for example $6 W + 1$ where $W$ is the maximum weight of any edge).
For any tuple $(a, b, c, d)\in A\times B\times C\times D$, let
\begin{equation*}
    \wt(a, b, c, d) = \wt(a, b) + \wt(a, c) + \wt(a, d) + \wt(b, c) + \wt(b, d) + \wt(c, d)
\end{equation*}
denote the weight of the $4$-clique $(a, b, c, d)$.
The Minimum-Weight $4$-Clique problem asks to compute
\begin{equation*}
    \min \set{\wt(a, b, c, d) \given a \in A, b \in B, c \in C, d \in D}.
\end{equation*}
A simple algorithm enumerates all cliques in $O(n^{4})$ time.
The Minimum-Weight $4$-Clique hypothesis states that this is essentially optimal.

\begin{hypothesis}[Minimum-Weight $k$-Clique hypothesis]
    \label{conj:min-weight-k-clique}
    There is no $O(n^{k - \varepsilon})$ algorithm for Minimum-Weight $k$-Clique on $n$-node graphs with non-negative weights, for any $\varepsilon > 0$.
\end{hypothesis}

\begin{proof}[Proof of \Cref{prop:multi-s-t-sp-lb-min-weight}]
    For simplicity, we consider the incremental case on directed graphs.
    Let $W$ be larger than the weight of any $4$-clique in the graph (say larger than $6$ times the maximum weight in the graph).
    We construct an \stSP{} instance as illustrated in Figures \ref{fig:4-clique-reduction} and \ref{fig:4-clique-multi-edge-gadget}.
    
    \begin{figure}[ht]
        \centering
        \includegraphics[width=0.6\columnwidth]{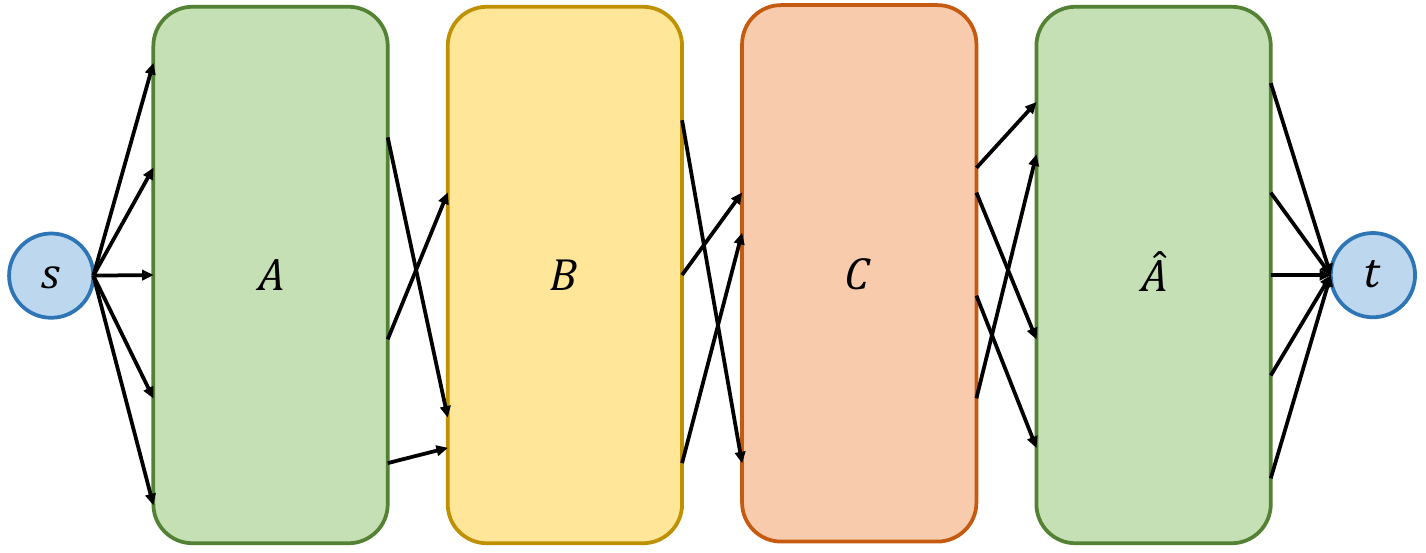}
        \caption{Overall structure of reduction of dynamic path problems.
        For each $d \in D$, the $A, B, C$ and $\hat{A}$ gadgets encode neighbors of $d$ in $A, B$, and $C$.
        Edges between gadgets encode edges between $A, B$, and $C$ in the Minimum-Weight $4$-Clique instance.}
        \label{fig:4-clique-reduction}
    \end{figure}
    
    We build gadgets of $O(n)$ vertices, with $A, \hat{A}$ encoding the adjacency information between $A$ and $D$, $B$ encoding adjacency between $B$ and $D$; and $C$ encoding adjacency between $C$ and $D$.
    The gadget construction is described in Figure \ref{fig:4-clique-multi-edge-gadget}.
    We describe only the $B$ gadget, noting that all gadgets are constructed analogously.
    The gadget consists of 2 layers $B_1, B_2$, each a copy of part $B \subset V(G)$ in the Minimum-Weight $4$-Clique instance.
    For each $b\in B$, we label the copy of $b$ in $B_{i}$ as $b^{(i)}$, $i=1,2$.
    Note that copies of $a$ in $\hat{A}_{i}$ are denoted $\hat{a}^{(i)}$.
    We have thus constructed a layered graph with vertex sets
    \begin{equation*}
        \set{s} \cup A_1 \cup A_2 \cup B_1 \cup B_2 \cup C_1 \cup C_2 \cup \hat{A}_1 \cup \hat{A}_2 \cup \set{t}.
    \end{equation*}
    
    We use edges between gadgets to encode the adjacency information between $A, B$, and $C$.
    Formally, for every edge between $A, B$ (resp. $B, C$ and $A, C$) we insert an edge with weight $\wt(a, b)$ between $(a^{(2)}, b^{(1)})$ (resp. $\wt(b, c)$ between $(b^{(2)}, c^{(1)})$ and $\wt(a, c)$ between $(c^{(2)}, \hat{a}^{(1)})$). 
    Finally, the source $s$ (resp. sink $t$) are connected to all nodes in the first layer $A_1$ (resp. last layer $\hat{A}_2$) with weight $0$ edges. %
    For the previous insertions, $s$ cannot reach $t$ and we do not need the \stSP{} information in the reduction. 
    
    \begin{figure}
        \centering
        \begin{subfigure}{0.45\columnwidth}
          \includegraphics[width=0.9\columnwidth]{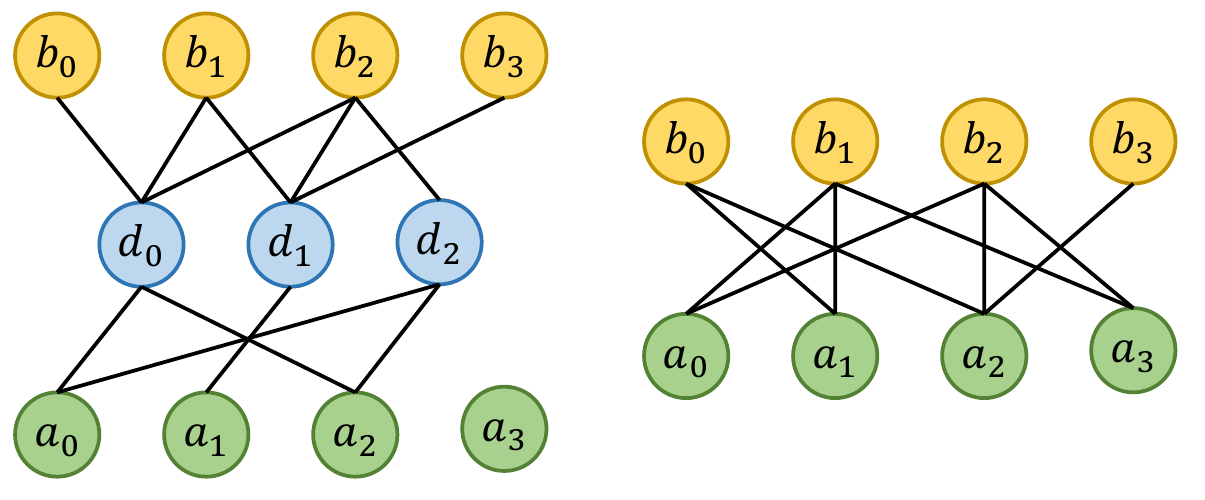}
          \caption{An example Minimum-Weight $4$-Clique instance.
          In the figure, we show only edges between $A, B$ and $D$.}
        \end{subfigure}
        \begin{subfigure}{0.45\columnwidth}
          \includegraphics[width=0.9\columnwidth]{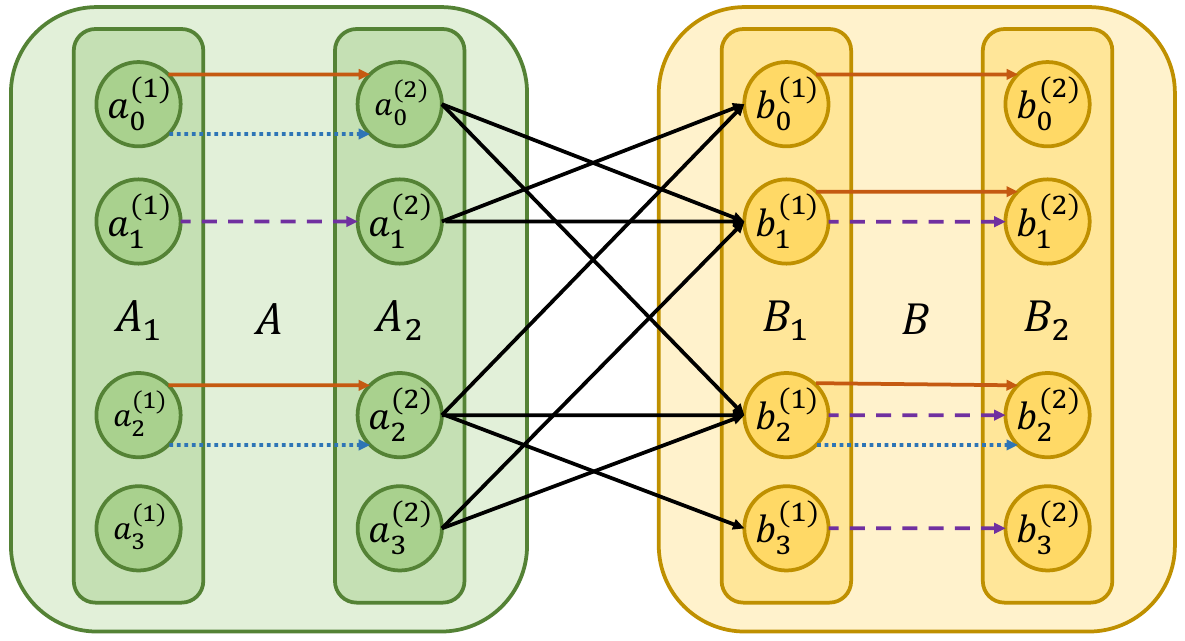}
          \caption{The $A, B$ gadgets with multi-edges.}
        \end{subfigure}
        \caption{An example of encoding a Minimum-Weight $4$-Clique instance into an incremental \stSP{} instance with parallel edges. 
        Solid black edges encode edges $A \times B$ with weight $\wt(a, b)$. 
        Edges within gadgets are inserted in decreasing order of weight. 
        Let $n = 4$.
        Solid red edges join neighbors of $d_2$ with weight $2 W + \wt(b_j, d_2)$ between $(b_j^{(1)}, b_j^{(2)})$ in $B$ and weight $2 n W + k W + \wt(a_k, d_2)$ between $(a_k^{(1)}, a_k^{(2)})$ in $A$.
        Dashed purple edges join neighbors of $d_1$ with weight $W + \wt(b_j, d_1)$ between $(b_j^{(1)}, b_j^{(2)})$ in $B$ and weight $n W + k W + \wt(a_k, d_1)$ between $(a_k^{(1)}, a_k^{(2)})$ in $A$.
        Dotted blue edges join neighbors of $d_0$ with weight $\wt(b_j, d_0)$ between $(b_j^{(1)}, b_j^{(2)})$ in $B$ and weight $k W + \wt(a_k, d_0)$ between $(a_k^{(1)}, a_k^{(2)})$ in $A$.
        The $(s, t)$-distance is maintained after each insertion.
        }
        \label{fig:4-clique-multi-edge-gadget}
    \end{figure}
    
    \begin{figure}[ht]
        \centering
        \includegraphics[width=0.9\columnwidth]{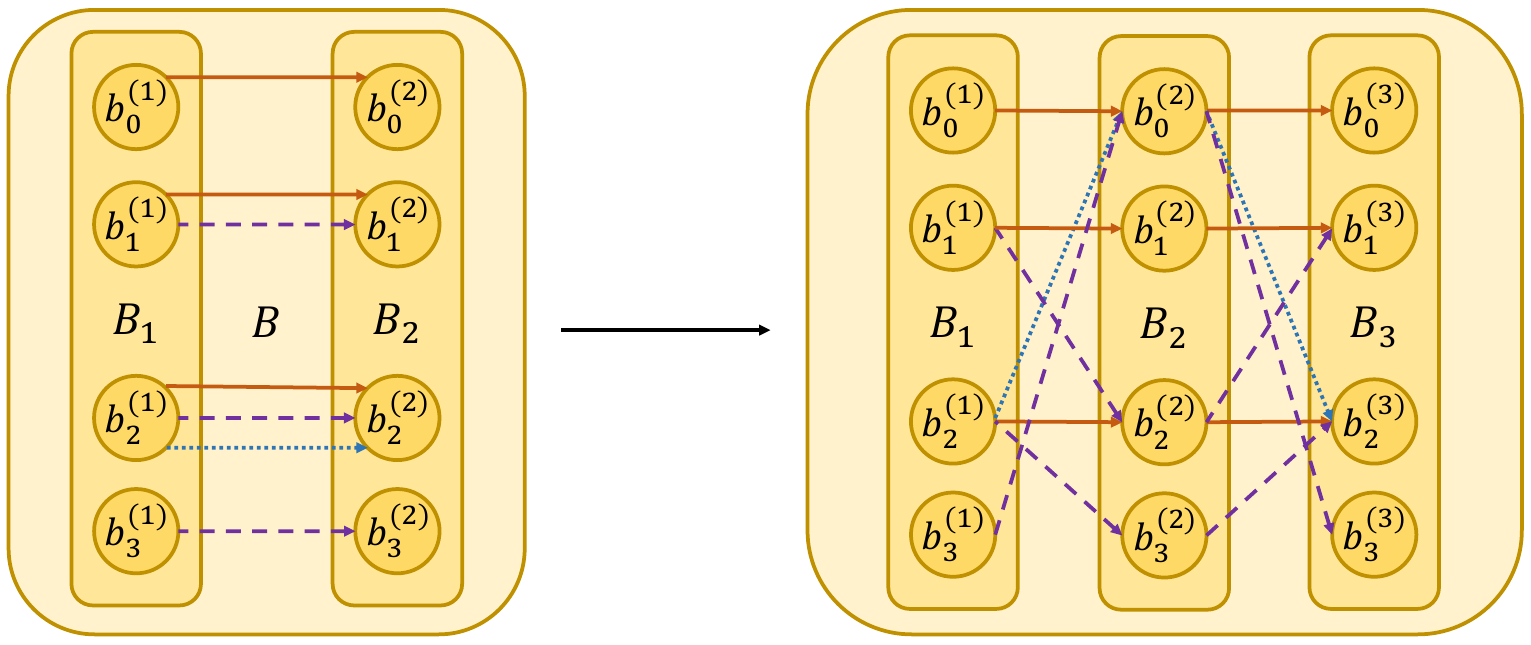}
        \caption{Removing parallel edges from the multi-edge gadget. 
        On the left, solid red edges join neighbors of $d_2$ with weight $2 W + \wt(b_j, d_2)$.
        Dashed purple edges join neighbors of $d_1$ with weight $W + \wt(b_j, d_1)$.
        Dotted blue edges join neighbors of $d_0$ with weight $\wt(b_j, d_0)$. On the right, 
        we insert an additional middle layer of $n = 4$ nodes.
        If $b_j$ is a neighbor of $d_i$, two edges of weight $i W + \frac{\wt(b_j, d_i)}{2}$ join $b_{j}^{(1)}, b_{j}^{(3)}$ via $b_{j + i \mod n}^{(2)}$.
        Note that each $b_{j}^{(2)}$ participates in at most one path of weight $< (2 i + 1) W$, where $d_i$ is the current smallest vertex of $d$.
        In particular, all paths with weight in this range are node disjoint, and must enter and leave the gadget from the copies of the same vertex $b \in B$.
        Furthermore, the simple graph only has $n$ additional nodes.}
        \label{fig:multi-edge-to-simple}
    \end{figure}
    
    Next, we will iterate over node pairs $(d, a)$ in decreasing lexicographic order, and use the dynamic \stSP{} instance to compute the minimum weight $4$-clique involving $(d, a)$.
    We begin with the iteration over a node $d$ in the outer loop.
    For every neighbor $b$ (resp. $c$) of $d$, we connect $(b^{(1)}, b^{(2)})$ (resp. $(c^{(1)}, c^{(2)})$) with an edge of weight $d W + \wt(b, d)$ (resp. weight $d W + \wt(c, d)$).
    Now, in the inner loop, we iterate over nodes $a \in A$ (also in decreasing order).
    We connect $(a^{(1)}, a^{(2)}), (\hat{a}^{(1)}, \hat{a}^{(2)})$ with edges of weight $d n W + a W + \frac{\wt(a, d)}{2}$.
    Note that each gadget is a matching (with parallel edges).
    Then, we claim that the construction of the graph ensures that if the minimum-weight $4$-clique involving $(d, a)$ has weight $w^*$, then the shortest path from $s$ to $t$ has length,
    \begin{equation*}
        \distance(s, t) = 2 d n W + (2 a + 2 d) W + w^*.
    \end{equation*}
    Consider the minimum-weight $4$-clique $(a, b, c, d)$ involving $(d, a)$ with weight $w^*$.
    Then by our construction there is a path
    \begin{equation*}
        (s, a^{(1)}, a^{(2)}, b^{(1)}, b^{(2)}, c^{(1)}, c^{(2)}, \hat{a}^{(1)}, \hat{a}^{(2)}, t)
    \end{equation*}
    with length
    \begin{equation*}
        2 d n W + (2 a + 2 d) W + \wt(a, b, c, d) = 2 d n W + (2 a + 2 d) W + w^*.
    \end{equation*}
    We claim that there is no shorter path.
    Let $P$ be the shortest path from $s$ to $t$.
    Since each gadget is a matching, the edges in the gadgets correspond to some node in the $4$-Clique instance and $P$ has the form
    \begin{equation*}
        P = (s, a_1^{(1)}, a_1^{(2)}, b^{(1)}, b^{(2)}, c^{(1)}, c^{(2)}, \hat{a}_2^{(1)}, \hat{a}_2^{(2)}, t).
    \end{equation*}
    Note that all paths from $s$ to $t$ in the graph must have length at least $2 d n W + (2 a + 2 d) W$.
    Since we iterate over $d$ in decreasing order, if we take an edge between $(b^{(1)}, b^{(2)})$ or $(c^{(1)}, c^{(2)})$ inserted by a previous iteration $d' > d$ over $D$, the path has length at least
    \begin{equation*}
        2 d n W + (2 a + d + d') W \geq 2 d n W + (2 a + 2 d + 1) W > 2 d n W + (2 a + 2 d) W + w^*,
    \end{equation*}
    so $P$ cannot be a shortest path.
    Similarly, if we take an edge between $(a^{(1)}, a^{(2)})$ or $(\hat{a}^{(1)}, \hat{a}^{(2)})$ inserted in a previous iteration $d' > d$, the path has length at least $(2 d + 1) n W + (2 a + 2 d) W$.
    Finally, since we iterate over $a$ in decreasing order, if $\max\set{a_1, a_2} > a$ then the path has length at least 
    \begin{equation*}
        2 d n W + (a_1 + a_2 + 2 d) W \geq 2 d n W + (2 a + 2 d + 1) W.
    \end{equation*}
    Thus, if $P$ is a shortest path, we may assume $a = a_1 = a_2$ and all edges in $P$ within gadgets are inserted in the current iteration $d \in D$.
    Thus, if $P$ is a shortest path, it must correspond to a $4$-clique $(a, b, c, d)$ involving $(d, a)$ and furthermore the length of this path is
    \begin{equation*}
        2 d n W + (2 a + 2 d) W + \wt(a, b, c, d).
    \end{equation*}
    Since we assumed that $w^*$ is the weight of the minimum-weight $4$-clique involving $(d, a)$, we conclude that $\distance(P) = 2 d n W + (2 a + 2 d) W + w^*$.
    
    Thus, for each $(d, a)$ we use the \stSP{} distance to compute,
    \begin{equation*}
        w^* = \distance(s, t) - 2 d n W - (2 a + 2 d) W.
    \end{equation*}
    The minimum-weight $4$-clique of $G$ is then computed by taking the minimum $w^*$ over all $(d, a)$.
    Thus, if Minimum-Weight $4$-Clique requires $n^{4 - o(1)}$ time, then incremental \stSP{} with $O(n^2)$ parallel edges requires $n^{4 - o(1)}$ time.
    To conclude, we observe the proof generalizes to the decremental case by running the reduction in reverse and to undirected graphs by adding a sufficiently large weight to each edge.
\end{proof}

\paragraph*{Removing Parallel Edges.} \Cref{prop:multi-s-t-sp-lb-min-weight} has shown that partially dynamic $\stSP$ on graphs with parallel edges requires total time $n^{4 - o(1)}$.
If we naively turn this into a simple graph (say by turning each parallel edge into a two-edge path) or modify parallel edges into a path as in \cite{DBLP:conf/stoc/GutenbergWW20}, this drastically increases the number of vertices to $n^2$.
The key idea behind our reduction is that we can in fact modify \Cref{prop:multi-s-t-sp-lb-min-weight} to a simple graph while only increasing the number of vertices by a constant factor. This idea also crucially underlies the lower bounds shown in later sections.

First, we give some intuition of how to avoid parallel edges. 
Note that parallel edges only occur within the gadgets $A, B, C$, and $\hat{A}$ of the reduction.
The parallel edges are inserted in the matching between the two layers of a single gadget to encode edge weights between $D$ and vertex sets $A, B$, and $C$ of $G$.
Specifically, we use parallel edges to ensure that any path must enter and leave a gadget from copies of the \emph{same} vertex.
We now describe how to construct these gadgets without the use of parallel edges.
Again, we describe only the $B$ gadget, noting that the others are constructed analogously.
Consider a $3$-layer gadget $B_1 \cup B_2 \cup B_3$, with each copy of $b \in B$ denoted by $b^{(1)}, b^{(2)}, b^{(3)}$.
Fix an iteration over a vertex $d$ in the outer loop.
We join $b^{(1)}, b^{(3)}$ with a length two path through $((b + d) \bmod n)^{(2)}$ in the central layer with edges of weight $d W + \frac{\wt(b, d)}{2}$.
Thus, there is a path from $b^{(1)}$ to $b^{(3)}$ through the gadget of the form
\begin{equation}
    \label{eq:b-1-b-3-path}
    (b^{(1)}, ((b + d) \bmod n)^{(2)}, b^{(3)})
\end{equation}
with weight $2 d W + \wt(b, d)$.
We claim any other path through the $B$ gadget has weight at least $(2 d + 1) W$.
Consider a path,
\begin{equation*}
    (b_1^{(1)}, b_2^{(2)}, b_3^{(3)}).
\end{equation*}
Since we iterate over $d$ in decreasing order, $b_2 \geq b_1 + d$ (and similarly $b_2 \geq b_3 + d$).
If $b_2 > b_1 + d$ or $b_2 > b_3 + d$, then
\begin{equation*}
    \wt(b_1^{(1)}, b_2^{(2)}) + \wt(b_2^{(2)}, b_3^{(3)}) \geq (2 b_2 - b_1 - b_3) W \geq (2 d + 1) W,
\end{equation*}
since $\wt(b_1^{(1)}, b_2^{(2)}) \geq (b_2 - b_1) W$ and $\wt(b_2^{(2)}, b_3^{(3)}) \geq (b_2 - b_3) W$.
Thus, we conclude $b_1 = b_3 = b_2 - d$.
Indeed, any shortest path through the $B$ gadget must entering and leave from copies of the \emph{same} $b \in B$ and this path has length $2 d W + \wt(b, d)$, just as in the parallel edge case.

Thus, we can construct a $B$ gadget without parallel edges and the rest of the reduction follows as above.
\Cref{fig:multi-edge-to-simple} illustrates the gadget transformation.
We give a full formal proof below.

\begin{proof}
    We describe the reduction for incremental $\stSP$.
    For the decremental setting, we can execute the reduction in reverse.
    
    Suppose for contradiction there is an algorithm $\innerAlg$ for incremental \stSP{} with $O(n^{4 - c})$ total time. 
    Consider a Minimum Weight 4-Clique instance with vertex sets $A, B, C, D$ of size $n$ and integer weights in $[W - 1]$.
    Throughout the reduction, assume that the vertex sets $A, B, C, D$ are indexed from $0$ to $n - 1$.
    We design an algorithm for Minimum-Weight $4$-Clique with total time $O(n^{4 - c})$.

    We construct a graph with vertices
    \begin{equation*}
        \set{s} \cup (A_1 \cup A_2 \cup A_3) \cup (B_1 \cup B_2 \cup B_3) \cup (C_1 \cup C_2 \cup C_3) \cup (\hat{A}_1 \cup \hat{A}_2 \cup \hat{A}_3) \cup \set{t}.
    \end{equation*}
    Each vertex set $A_1, A_2, A_3, B_1, B_2, B_3, C_1, C_2, C_3, \hat{A}_1, \hat{A}_2$ and $\hat{A}_3$ has $n$ vertices.
    For each $a \in A$, we create a copy $a^{(i)} \in A_i$ and a copy $\hat{a}^{(i)} \in \hat{A}_i$ for $i \in \set{1, 2, 3}$.
    Similarly, for each $b \in B$ and $c \in C$, we create a copy $b^{(i)} \in B_i$ and $c^{(i)} \in C_i$ for $i \in \set{1, 2, 3}$.
    Let $W_0 =100 n^2 W$.
    Initially, insert edges $(s, a^{(1)})$ and $(\hat{a}^{(3)}, t)$ with  weight $W_0$ for all $a \in A$.
    Furthermore, insert edges with weight $W_0 + \wt(a, b)$  between $(a^{(3)}, b^{(1)})$ (resp. weight $W_0 + \wt(b, c)$ between $(b^{(3)}, c^{(1)})$ and weight $W_0 + \wt(c, a)$ between $(c^{(3)}, \hat{a}^{(1)})$) if and only if $(a, b) \in E$ (resp. if $(b, c) \in E$ and $(c, a) \in E$).

    This creates a graph with $O(n)$ vertices. 
    We now proceed to more edge insertions of the reduction.
    Initialize a variable $M \gets \infty$.
    In the outer loop we iterate over $d_i \in D$ in decreasing order.
    \begin{enumerate}
        \item For each $b_j \in \neighborhood(d_i)$, 
        insert edges $\left(b_j^{(1)}, b_{j + i \mod n}^{(2)}\right)$ and $\left(b_{j + i \mod n}^{(2)}, b_j^{(3)} \right)$ with weight $W_0 + i \cdot W + \frac{\wt(b_j, d_i)}{2}$.
        
        \item For each $c_j \in \neighborhood(d_i)$, insert edges $\left(c_j^{(1)}, c_{j + i \mod n}^{(2)}\right)$ and $\left(c_{j + i \mod n}^{(2)}, c_j^{(3)} \right)$ with weight $W_0 + i \cdot W + \frac{\wt(c_j, d_i)}{2}$.
        
        \item In the inner loop, we iterate over $a_{k} \in A$ in decreasing order.
        \begin{enumerate}
            \item If $a_k \in \neighborhood(d_i)$, insert the following edges of weight $W_0 + i \cdot n W + k \cdot W + \frac{\wt(a_k, d_i)}{4}$: 
            \begin{equation*}
            	\left(a_{k}^{(1)}, a_{k + i \mod n}^{(2)}\right), \left(a_{k + i \mod n}^{(2)}, a_{k}^{(3)}\right), \left(\hat{a}_{k}^{(1)}, \hat{a}_{k + i \mod n}^{(2)}\right), \left(\hat{a}_{k + i \mod n}^{(2)}, \hat{a}_{k}^{(3)}\right).
            \end{equation*}
            
            \item Use $\innerAlg$ to compute $\distance(s, t)$. If $\distance(s, t) < 13 \cdot W_0 + 4 i \cdot n W + 4 (k + i + 1) W$ then update $M \gets \min\{M, \distance(s, t) - (13 \cdot W_0 + 4 i \cdot n W + 4 (k + i) W)\}$.
        \end{enumerate}
    \end{enumerate}
    Finally, return $M$ after all iterations.
    Clearly, the bottleneck of this algorithm is the total time of algorithm $\innerAlg$, so the running time of this algorithm is $O(n^{4-c})$.

    To show the correctness of this algorithm, we first show the following lemma.

\begin{lemma}
    \label{lemma:s-t-sp-min-weight-4-clique-equiv}
    Consider the graph  after adding all edges when iterating over $d_i$ and $a_k$.
    There is a one-to-one correspondence between:
    \begin{enumerate}
        \item paths $P$ with length $\distance(P) < 13 \cdot W_0 + 4 i \cdot n W + 4(k + i + 1) W$

        \item $4$-cliques $(a_k, b, c, d_i)$ containing $a_k, d_i$.
    \end{enumerate}

    Furthermore, each such path $P$ has length $\distance(P) = 13 \cdot W_0 + 4 i \cdot n W + 4(k + i) W + \wt(a_k, b, c, d_{i})$.
\end{lemma}

\begin{proof}
    Suppose $d_i, a_k$ are in a 4-clique with vertices $b_{j}, c_{\ell}$.
    Then, the following $(s, t)$-path exists in the graph and has weight $13 \cdot W_0 + 4 i \cdot n W + 4 (k + i) W + \wt(a_k, b_j, c_{\ell}, d_i)$:
    \begin{equation*}
        \left(s, a_{k}^{(1)}, a_{k + i \mod n}^{(2)},  a_{k}^{(3)}, b_{j}^{(1)}, b_{j + i \mod n}^{(2)}, b_{j}^{(3)}, c_{\ell}^{(1)}, c_{\ell + i \mod n}^{(2)}, c_{\ell}^{(3)}, \hat{a}_{k}^{(1)}, \hat{a}_{k + i \mod n}^{(2)},  \hat{a}_{k}^{(3)}, t\right). 
    \end{equation*}
    Note that by our choice of $W$, $\wt(a_k, b_j, c_{\ell}, d_i) \leq 3 W$ so this path has total weight less than the threshold $13 \cdot W_0 + 4 i \cdot n W + 4 (k + i + 1) W$.

    We verify that this path exists and has the required weight.
    The edge $(s, a_{k}^{(1)})$ has weight $W_0$.
    Each edge in the sub-path $\left(a_{k}^{(1)}, a_{k + i \mod n}^{(2)}, a_{k}^{(3)} \right)$ exists and has weight $W_0 + i \cdot n W + k W + \frac{\wt(a_k, d_i)}{4}$.
    Then, $(a_k, b_j) \in E$ implies that the edge $(a_{k}^{(3)}, b_{j}^{(1)})$ exists with weight $W_{0} + \wt(a, c)$.
    A similar argument shows that the remaining edges in the path exist.
    Summing over all 13 edges in the path, we conclude the length of the path is $13 \cdot W_0 + 4 i \cdot n W + 4(k + i) W + \wt(a_k, b_j, c_{\ell}, d_i) < 4 i \cdot n W + 4(k + i + 1)W$.
    
    Conversely, suppose there is some path of length $<13 \cdot W_0 + 4 i \cdot n W + 4 (k + i + 1) W$.
    Recall that all weights in the $4$-Clique instance are non-negative.
    Since each edge in the graph has weight at least $W_0$ and we have chosen $W_0$ so that
    \begin{equation*}
    	4 i \cdot n W + 4 (k + i + 1) W < W_0,
    \end{equation*}
	any $(s, t)$ path with the required length has at most 13 edges.
	Since we have a 14-layered graph, with edges only between adjacent layers, any $(s, t)$-path with the required length must have the following structure:
    \begin{equation*}
        P = \left(s, a_{i_1}^{(1)}, a_{i_2}^{(2)},  a_{i_3}^{(3)}, b_{i_4}^{(1)}, b_{i_5}^{(2)}, b_{i_6}^{(3)}, c_{i_7}^{(1)}, c_{i_8}^{(2)}, c_{i_9}^{(3)}, \hat{a}_{i_{10}}^{(1)}, \hat{a}_{i_{11}}^{(2)},  \hat{a}_{i_{12}}^{(3)}, t\right).
    \end{equation*}
    
    At the time of the query in the $i$-th iteration over $D$ and $k$-th iteration over $A$, the only edges in $(A_1 \times A_2) \cup (A_2 \times A_3) \cup (\hat{A}_1 \times \hat{A}_2) \cup (\hat{A}_2 \times \hat{A}_3)$ with weight less than $W_0 + i \cdot n W + (k + 1) W$ are $\left(a_{k}^{(1)}, a_{k + i \mod n}^{(2)}\right)$, $\left(a_{k + i \mod n}^{(2)}, a_{k}^{(3)}\right)$, $\left(\hat{a}_{k}^{(1)}, \hat{a}_{k + i \mod n}^{(2)}\right)$, and $\left(\hat{a}_{k + i \mod n}^{(2)}, \hat{a}_{k}^{(3)}\right)$ for $a_{k} \in N(d_{i})$.
    If $P$ uses these edges, then $k = i_1 = i_3 = i_{10} = i_{12}$.
    Similarly, the only edges in $B_1 \times B_2$ and $B_2 \times B_3$ with weight less than $W_0 + (i + 1) \cdot W$ are $\left(b_{j}^{(1)}, b_{j + i \mod n}^{(2)}\right)$, $\left(b_{j + i \mod n}^{(2)}, b_{j}^{(3)}\right)$ for $b_{j} \in N(d_{i})$.
    If $P$ uses these edges, then $i_4 = i_6 = j$ for some $b_j \in N(d_i)$, since all paths of this form are vertex disjoint.
    Similarly, the only edges in $C_1 \times C_2$ and $C_2 \times C_3$ with weight less than $W_0 + (i + 1) \cdot W$ are $\left(c_{\ell}^{(1)}, c_{\ell + i \mod n}^{(2)}\right)$, $\left(c_{\ell + i \mod n}^{(2)}, c_{\ell}^{(3)}\right)$ for $c_{\ell} \in N(d_{i})$.
    By a similar argument, if $P$ uses these edges then $i_7 = i_9 = \ell$.
    If any of the above edges do not exist in $P$, the $(s, t)$-path in question has length at least $13 \cdot W_0 + 4 i \cdot n W + 4 (k + i + 1) W$, contradicting the assumption.
    
    Finally, it remains to show $a_k, b_j, c_{\ell}$ form a triangle.
    Since the edge $(a_{k}^{(3)}, b_{j}^{(1)})$ exists, $(a_{k}, b_{j}) \in E$.
    A similar argument shows that $(a_{k}, c_{\ell}), (b_j, c_{\ell}) \in E$.

    Thus, we have shown paths of length less than $13 \cdot W_0 + 4 i \cdot n W + 4(k + i + 1)W$ correspond to $4$-cliques containing $d_i, a_k$.
    Furthermore, any such path has weight $4 i \cdot n W + 4 (k + i) W + \wt(a_k, b, c, d_i)$ for $b, c$ in the clique.
\end{proof}
    
    From Lemma \ref{lemma:s-t-sp-min-weight-4-clique-equiv}, at the time of computing $\distance(s, t)$, each path $P$ with length $\distance(P) < 13 \cdot W_0 + 4 i \cdot n W + 4(k + i + 1) W$ corresponds to a $4$-clique $(a_k, b, c, d_i)$ in the input graph containing $d_i, a_k$ and furthermore
    \begin{equation*}
        \distance(P) = 13 \cdot W_0 + 4 i \cdot n W + 4(k + i) W + \wt(a_k, b, c, d_{i}),
    \end{equation*}
    where $\wt(a_k, b, c, d_{i})$ denotes the weight of the $4$-clique.
    Thus, if $\distance(s, t) < 13 \cdot W_0 + 4 i \cdot n W + 4(k + i + 1) W$ (which must happen if $a_k$ and $d_i$ are in any $4$-clique), then $\distance(s, t) - (13 \cdot W_0 + 4 i \cdot n W + 4(k + i) W)$ is the minimum weight $4$-clique containing $a_k, d_i$.
    Since we iterate over all $a_k, d_i$, the above procedure solves the Minimum Weight $4$-Clique problem, contradicting the Minimum-Weight $4$-Clique hypothesis.
\end{proof}

\section{Node-Weighted Shortest Paths}
\label{sec:node-weight}

In this section, we give our algorithms and lower bounds for partially dynamic shortest paths on \emph{node-weighted} graphs.
We start the section with some simple algorithms that use fast matrix multiplication (FMM) to beat the naive algorithm recomputing shortest paths with each update. Next, we show that using FMM is in fact necessary by establishing a conditional lower bound that rules out any nontrivial combinatorial partially dynamic algorithms for node-weighted shortest paths. Finally, we show some nontrivial lower bounds that hold even for algebraic algorithms.

\subsection{Faster Algebraic Algorithms for Partially Dynamic \texorpdfstring{\nwSSSP{}}{nwSSSP}}

We prove the following result that utilizes fast matrix multiplication.

\begin{proposition} 
    \label{prop:incremental-node-weighted-sssp}
    There is an algorithm solving incremental \nwSSSP{} in $\bigtO{n^{2 - \gamma/4}}$ amortized time where $\gamma$ is the solution to the equation $\omega(1, 1, 1 + \gamma) = 3 - \gamma$.
\end{proposition}

To prove \cref{prop:incremental-node-weighted-sssp}, we require the static node-weighted APSP algorithm of Yuster~\cite{DBLP:conf/soda/Yuster09}.

\begin{lemma}[Theorem 1.1 of \cite{DBLP:conf/soda/Yuster09}]
    \label{lemma:node-weighted-apsp}
    There is an algorithm computing \nwAPSP{} in $\bigO{n^{3 - \gamma/2}}$-time, where $\gamma$ is the solution to the equation $\omega(1, 1, 1 + \gamma) = 3 - \gamma$.
\end{lemma}

With this in hand, we prove Proposition \ref{prop:incremental-node-weighted-sssp}.
At a high level, we handle the updates to the graph in batches.
At the start of a batch, we compute \nwAPSP{} on the current graph.
Then, with each edge insertion, we construct a layered graph with $O(n^{1+t})$ edges, by using either the inserted edges or an edge encoding the shortest path computed by the \nwAPSP{} instance, so that computing \SSSP{} on the layered graph finds the shortest paths in the updated graph (since any paths not using edges inserted in this batch will use edges computed by the \nwAPSP{} instance).

\begin{proof}[Proof of \cref{prop:incremental-node-weighted-sssp}]
    Let $t$ be a parameter to be fixed later. 
    We will process updates in batches of size $n^{t}$.
    At the start of each batch, compute \nwAPSP{} on the current graph using Lemma \ref{lemma:node-weighted-apsp}.
    Let $\distance_0(u, v)$ denote the distances computed at the start of each batch.
    Then, after $k$ updates $\set{(u_i, v_i)}_{i = 1}^{k}$, construct the following 4-layer graph with vertices
    \begin{equation*}
        \set{s} \cup \set{u_i}_{i = 1}^{k} \cup \set{v_i}_{i = 1}^{k} \cup V,
    \end{equation*}
    with edge weights $\wt(s, u_i) = \distance_0(s, u_i)$ for all $i \in [k]$, $\wt(u_i, v_i) = 0$ (since the distances $\distance_0$ include the weights on endpoints) for all $i \in [k]$, $\wt(v_i, u_j) = \distance_0(v_i, u_j)$ for all $i, j \in [k]$ and $\wt(v_i, v) = \distance_0(v_i, v)$ for all $i \in [k]$ and $v \in V$.
    We also insert edge weights $\wt(s, v) = \distance_0(s, v)$ for all $v \in V$.
    On this graph with $O(n^{1 + t})$ edges, we compute $\SSSP$ in $\tO{n^{1 + t}}$-time.
    The amortized running time is 
    \[
    \OO\left(n^{3-\gamma / 2 - t} + n^{1+t} \right),
    \]
    where $\gamma$ is the solution to the equation $\omega(1, 1, 1 + \gamma) = 3 - \gamma$.
    Optimizing for parameter $t = 1 - \frac{\gamma}{4}$ gives an amortized running time of
    \begin{equation*}
        \bigtO{n^{2 - \frac{\gamma}{4}}}.
    \end{equation*}

    We claim this computes incremental \nwSSSP{}.
    First, since every distance in the constructed graph is realized by a path in the input graph, it suffices to show that the shortest path is encoded into the constructed graph.
    After $i$ updates in the batch, let $(s = w_0, w_1, \dotsc, w_d = v)$ be a shortest path from $s$ to $v$.
    If none of these edges are inserted during the batch, then they must have been present during the \nwAPSP{} computation, and therefore the constructed graph has a path of length $\distance_0(s, v)$.
    Otherwise, let $\set{(w_{i_j}, w_{i_j + 1})}_{j}$ denote the edges inserted in this batch.
    Then, these edges are present in our graph, while the remaining portions of the path are encoded in the graph using the \nwAPSP{} solution $\distance_0$, so a path with length at most the distance exists in the constructed graph.
\end{proof}

Following a similar technique, we obtain an incremental algorithm for \nwstSP{}.

\begin{proposition} 
    \label{prop:incremental-node-weighted-s-t-sp}
    There is an algorithm solving incremental \nwstSP{} in $\bigtO{n^{2 - \gamma/3}}$ amortized time where $\gamma$ is the solution to the equation $\omega(1, 1, 1 + \gamma) = 3 - \gamma$.
\end{proposition}

\begin{proof}
    Let $t$ be a parameter to be fixed later. 
    We will process updates in batches of size $n^{t}$.
    At the start of each batch, compute \nwAPSP{} on the current graph using Lemma \ref{lemma:node-weighted-apsp}.
    Let $\distance_0(u, v)$ denote the distances computed at the start of each batch.
    Then, after $k$ updates $\set{(u_i, v_i)}_{i = 1}^{k}$, construct the following 4-layer graph with vertices
    \begin{equation*}
        \set{s} \cup \set{u_i}_{i = 1}^{k} \cup \set{v_i}_{i = 1}^{k} \cup \set{t},
    \end{equation*}
    with edge weights $\wt(s, u_i) = \distance_0(s, u_i)$ for all $i \in [k]$, $\wt(u_i, v_i) = 0$ for all $i \in [k]$ (again the distance $\distance_0$ include endpoints), $\wt(v_i, u_j) = \distance_0(v_i, u_j)$ for all $i, j \in [k]$ and $\wt(v_i, t) = \distance_0(v_i, t)$ for all $i \in [k]$.
    We additionally insert edge weight $\wt(s, t) = \distance_0(s, t)$.
    On this graph with $O(n^{2t})$ edges, we compute $\stSP$ in $\tO{n^{2t}}$-time.
    The amortized running time is 
    \[
        \bigtO{n^{3-\gamma / 2 - t} + n^{2t}},
    \]
    where $\gamma$ is the solution to the equation $\omega(1, 1, 1 + \gamma) = 3 - \gamma$.
    Optimizing for parameter $t = 1 - \frac{\gamma}{6}$ gives an amortized running time of
    \begin{equation*}
        \bigtO{n^{2 - \frac{\gamma}{3}}}.
    \end{equation*}

    We claim this computes incremental \nwstSP{}.
    First, since every distance in the constructed graph is realized by a path in the input graph, it suffices to show that the shortest path is encoded into the constructed graph.
    After $i$ updates in the batch, let $(s = w_0, w_1, \dotsc, w_d = t)$ be a shortest path from $s$ to $t$.
    If none of these edges are inserted during the batch, then they must have been present during the \nwAPSP{} computation, and therefore the constructed graph has a path of length $\distance_0(s, t)$.
    Otherwise, let $\set{(w_{i_j}, w_{i_j + 1})}_{j}$ denote the edges inserted in this batch.
    Then, these edges are present in our graph, while the remaining portions of the path are encoded in the graph using the \nwAPSP{} solution $\distance_0$, so a path with length at most the distance exists in the constructed graph.
\end{proof}

\subsection{Conditional Lower Bounds for Partially Dynamic \texorpdfstring{\nwSSSP{}}{nwSSSP}}

We have exploited FMM to design algorithms faster than trivial recomputation.
We show that using FMM is in fact necessary: any combinatorial algorithm that can solve partially dynamic \nwstSP{} requires total time $n^{4 - o(1)}$.
However, our previous graph simplification technique (\Cref{fig:multi-edge-to-simple}) no longer suffices to produce vertex-disjoint paths.
Instead, we construct a new gadget for each multi-edge vertex set that ensures all short paths are vertex disjoint. 
The resulting simple graph is node-weighted, and the number of vertices increases only by a constant factor.

Before stating the theorem, we state the relevant hardness conjecture.

\begin{hypothesis}[Combinatorial $k$-Clique hypothesis]
    \label{conj:k-clique}
    There is no $O(n^{k - \varepsilon})$ combinatorial algorithm for $k$-Clique Detection on $n$-node graphs, for any $\varepsilon > 0$.
\end{hypothesis}

\begin{theorem}
    \label{thm:s-t-sp-node-weight-lb}
    Under the Combinatorial $4$-Clique hypothesis, any combinatorial algorithm computing incremental/decremental \nwstSP{} on undirected graphs requires $n^{4 - o(1)}$ total time.
\end{theorem}

\begin{proof}
    For simplicity, we describe only the reduction for incremental \nwstSP{}.
    For the decremental setting, we can execute the reduction in reverse.
    
    Suppose for contradiction there is a combinatorial algorithm $\innerAlg$ for partially dynamic \nwstSP{} with total time $O(n^{4 - c})$ for some $c > 0$.
    Consider a 4-Clique instance with vertex sets $A, B, C, D$ of size $n$.
    Throughout the reduction, assume that the vertex sets $A, B, C, D$ are indexed from $0$ to $n - 1$.
    We design an algorithm with total time $O(n^{4 - c})$, thus contradicting the Combinatorial 4-Clique hypothesis.
    In the following, let $W = 100 n$.

    \paragraph{Vertex Set Gadget}
    Before describing the construction of the full graph, we describe the key gadget in the reduction.
    First, consider vertex set $B$, and we construct the gadget $G_R(B)$.
    The gadget construction is shown in Figure \ref{fig:multi-edge-to-simple-node}.

    \begin{figure}[ht]
        \centering
        \includegraphics[width=0.9\columnwidth]{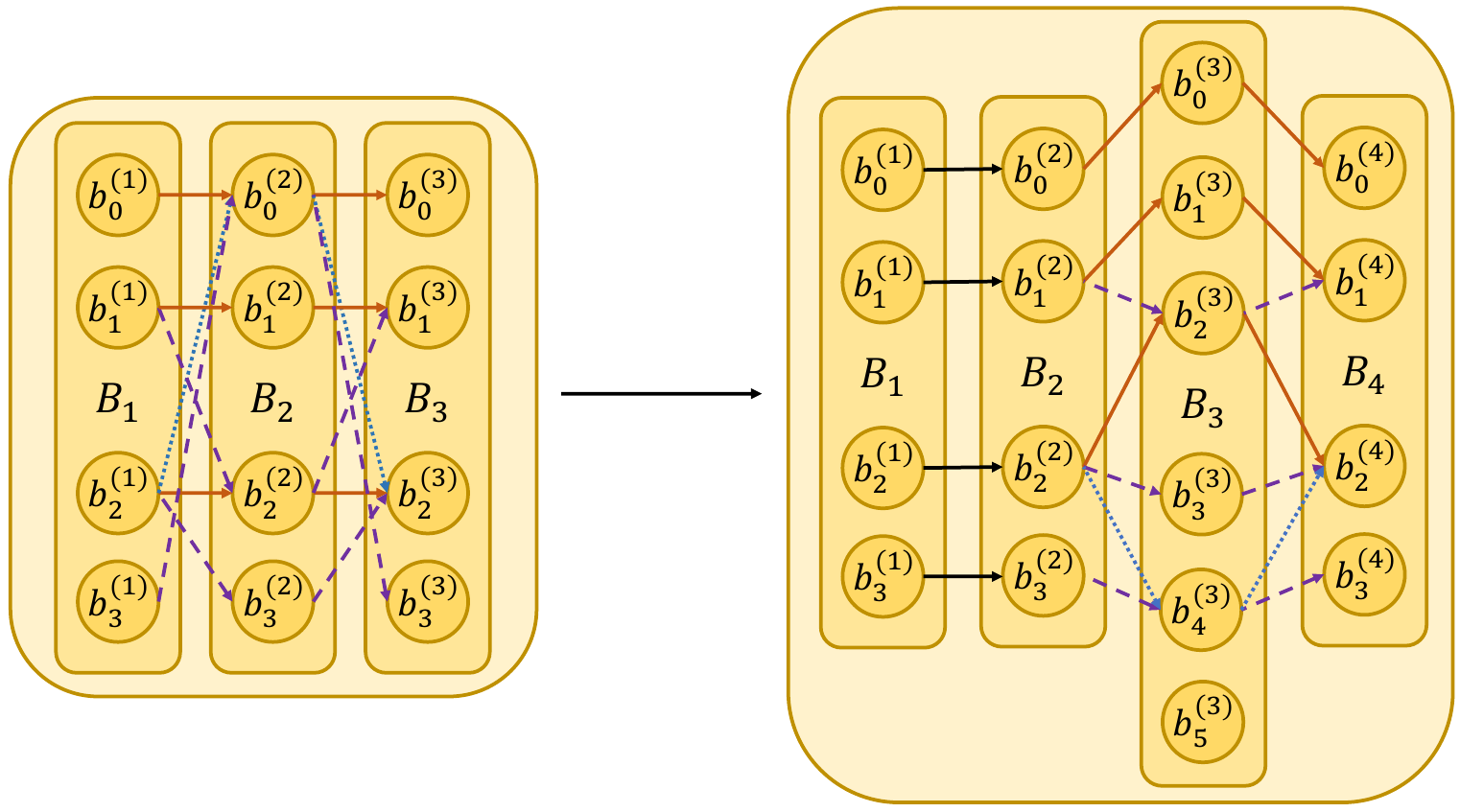}
        \caption{Comparison of node-weighted gadget with an edge-weighted gadget.
        The gadget on the left is the edge-weighted gadget from \Cref{thm:s-t-sp-lb}.
        We now create a $4$-layer node-weighted graph where the weight of $b_{j}^{(i)}$ vertices in $B_1, B_2, B_4$ are $j \cdot W$ for $i \in \set{1}$ and $j \cdot W^{3}$ for $i \in \set{2, 4}$ and the weight of $b_{j}^{(3)}$ vertices are $(4 n - 2 j) \cdot W^{3}$ for some large constant $W$.
        While the edges are no longer weighted, we format the edges as before to illustrate corresponding paths.
        Note that each $b_{j}^{(3)}$ participates in at most one path of weight $< (4n - 2i + 1) \cdot W^{3}$ where $d_i$ is the current smallest vertex of $d$.
        As in \Cref{fig:multi-edge-to-simple}, all paths with weight in this range are vertex disjoint, and must enter and leave the gadget from copies of the same vertex $b \in B$.
        Furthermore, the simple graph only has $4n$ additional vertices.}
        \label{fig:multi-edge-to-simple-node}
    \end{figure}

    In our previous lower bounds, we used a 3-layer graph to ensure that at each stage, there are \emph{vertex disjoint} paths between $b_j^{(1)}, b_j^{(3)}$ for all $b_j \in N(d_i)$.
    This in turn ensured that  the \emph{same} vertex $b_j$ was adjacent to both $a_{k}, c_{\ell}$ (and $d_i$) in the $4$-Clique instance.
    Crucially, we used edge weights to ensure that through each middle vertex, there was only one path with edges of weight between $[W_0 + i \cdot W, W_0 + (i+1) \cdot W)$.
    In a node-weighted graph, we describe the following modification.

    $G_R(B)$ consists of $4$ layers of vertices
    \begin{equation*}
        B_1 \cup B_2 \cup B_3 \cup B_4,
    \end{equation*}
    where $|B_1| = |B_2| = |B_4| = n$ and $|B_3| = 2n$.\footnote{For the current reduction, it suffices to construct a $3$-layer gadget ($B_2 \cup B_3 \cup B_4$), but we require the $4$-layer gadget for a later reduction (\Cref{thm:sssp-node-weight-min-witness-lb}).}
    For each $b \in B$, we create a copy $b^{(i)} \in B_i$ for $i \in \set{1, 2, 4}$.
    We index $B_3 = \set{b_0^{(3)}, \dotsc, b_{2n - 1}^{(3)}}$ and assign weights as follows,
    \begin{align*}
        \wt\left(b_{j}^{(2)}\right) &= \wt\left(b_{j}^{(4)}\right) = W^{4} + j \cdot W^{3} \\
        \wt\left(b_{j}^{(3)}\right) &= W^{4} + (4n - 2j) \cdot W^{3} \\
        \wt\left(b_{j}^{(1)}\right) &= W^{4} + j \cdot W
    \end{align*}
    We similarly define the vertex sets $G_R(C) = \bigcup_{i = 1}^{4} C_i$, $G_R(A) = \bigcup_{i = 1}^{4} A_i$ and $G_R(\hat{A}) = \bigcup_{i = 1}^{4} \hat{A}_i$ with the same vertex set sizes.
    The weights are defined as follows,
    \begin{align*}
        \wt(s) &= \wt(t) = W^{4}, \\
        \wt\left(c_{\ell}^{(1)}\right) &= W^4 + \ell, \\
        \wt\left(a_{k}^{(1)}\right) &= \wt\left(\hat{a}_{k}^{(1)}\right) = W^4 + k \cdot W^{2}, \\
        \wt\left(c_{i}^{(2)}\right) &= \wt\left(a_{i}^{(2)}\right) = \wt\left(\hat{a}_{i}^{(2)}\right) = W^4 + i \cdot W^{3}, \\
        \wt\left(c_{i}^{(3)}\right) &= \wt\left(a_{i}^{(3)}\right) = \wt\left(\hat{a}_{i}^{(3)}\right) = W^4 + (4n - 2i) \cdot W^{3}, \\
        \wt\left(c_{i}^{(4)}\right) &= \wt\left(a_{i}^{(4)}\right) = \wt\left(\hat{a}_{i}^{(4)}\right) = W^4 + i \cdot W^{3}.
    \end{align*}
    Note that all vertex weights are nonnegative.

    \paragraph{Lower Bound Construction}
    We construct a graph with the following vertices:
    \begin{equation*}
        \set{s} \cup G_R(A) \cup G_R(B) \cup G_R(C) \cup G_R(\hat{A}) \cup \set{t},
    \end{equation*}
    with vertex weights described above.
    
    Initially, insert edges $(s, a^{(1)})$ and $(\hat{a}^{(4)}, t)$ for all $a \in A$.
    Furthermore, insert edges between $(a^{(4)}, b^{(1)})$ (resp. between $(b^{(4)}, c^{(1)})$, between $(c^{(4)}, \hat{a}^{(1)})$) if and only if $(a, b) \in E$ (resp. if $(b, c) \in E$, $(c, a) \in E$).
    Insert edges $(a^{(1)}, a^{(2)}), (b^{(1)}, b^{(2)}), (c^{(1)}, c^{(2)}), (\hat{a}^{(1)}, \hat{a}^{(2)})$ for all $a, b, c$.

    We now proceed to more edge insertions of the reduction.
    In the outer loop we iterate over $d_i \in D$ in increasing order.
    \begin{enumerate}
        \item For each $b_j \in \neighborhood(d_i)$, insert edges, $\left(b_j^{(2)}, b_{j + i}^{(3)}\right)$ and $\left(b_{j + i}^{(3)}, b_{j}^{(4)} \right)$.
        
        \item For each $c_j \in \neighborhood(d_i)$, insert edges $\left(c_j^{(2)}, c_{j + i}^{(3)}\right)$ and $\left(c_{j + i}^{(3)}, c_{j}^{(4)} \right)$.
        
        \item In the inner loop, we iterate over $a_{k} \in A$ in decreasing order.
        \begin{enumerate}
            \item If $a_k \in \neighborhood(d_i)$, insert the following edges:
            \begin{align*}
                \left(a_k^{(2)}, a_{k + i}^{(3)}\right), \left(a_{k + i}^{(3)}, a_{k}^{(4)} \right), \left(\hat{a}_{k}^{(2)}, \hat{a}_{k + i}^{(3)}\right), \left(\hat{a}_{k + i}^{(3)}, \hat{a}_{k}^{(4)} \right).
            \end{align*}
            
            \item Query $\distance(s, t)$ from $\innerAlg$.
            If $\distance(s, t) < 18 \cdot W^4 + (16n - 8i) \cdot W^{3} + (2 k + 1) \cdot W^{2}$ return $\true$.
        \end{enumerate}
    \end{enumerate}
    Finally, after all iterations, return $\false$ if no query satisfied the required distance constraint.

    Clearly, the bottleneck of this algorithm is running $\innerAlg$, so the total running time is $O(n^{4-c})$. 
    Here, emphasize that $D$ is iterated over in increasing order while $A$ is iterated over in decreasing order.

    To show correctness, we first prove the following lemma.

\begin{lemma}
    \label{lemma:s-t-sp-node-weight-4-clique-equiv}
    Consider the query when iterating over $d_i$ and $a_k$.
    Then, there is a one-to-one correspondence between:
    \begin{enumerate}
        \item paths $P$ with $\distance(P) < 18 \cdot W^{4} + (16n - 8i) \cdot W^{3} + (2k + 1) \cdot W^{2}$
        \item $4$-cliques $(a_k, b, c, d_i)$ containing $a_k, d_i$.
    \end{enumerate}

    Furthermore, each such path $P$ has length
    \begin{equation*}
        \distance(P) = 18 \cdot W^{4} + (16 n - 8 i) \cdot W^{3} + 2 k \cdot W^{2} + b \cdot W + c.
    \end{equation*}
\end{lemma}

\begin{proof}    
    Suppose $d_i, a_k$ are in a 4-clique with vertices $b_{j}, c_{\ell}$.
    Then, the following $(s, t)$-path exists: 
    \begin{align*}
        \left(s, a_{k}^{(1)}, a_{k}^{(2)}, a_{k + i}^{(3)}, a_{k}^{(4)}, b_{j}^{(1)}, b_{j}^{(2)}, b_{j + i}^{(3)}, b_{j}^{(4)}, c_{\ell}^{(1)}, c_{\ell}^{(2)}, c_{\ell + i}^{(3)}, c_{\ell}^{(4)}, \hat{a}_{k}^{(1)}, \hat{a}_{k}^{(2)}, \hat{a}_{k + i}^{(3)}, \hat{a}_{k}^{(4)}, t\right).
    \end{align*}
    Furthermore, this path has length
    \begin{align*}
        18 \cdot W^{4} + (16n - 8i) \cdot W^{3} + 2 k \cdot W^{2} + j \cdot W + \ell.
    \end{align*}
    
    Note that by our choice of $W$, $j \cdot W + \ell < W^{2}$, so this path satisfies the desired constraint.
    We verify that this path exists and has the required length.
    Recall that the edge $(s, a_{k}^{(1)})$ exists.
    Since $a_{k} \in N(d_i)$, each edge in the sub-path $\left(a_{k}^{(1)}, a_{k}^{(2)}, a_{k + i}^{(3)}, a_{k}^{(4)} \right)$ exists and the vertices have combined weight
    \begin{align*}
        4 \cdot W^{4} + (4n - 2(k + i) + 2k) \cdot W^{3} + k \cdot W^{2} = 4 \cdot W^{4} + (4n - 2i) \cdot W^{3} + k \cdot W^{2}.
    \end{align*}
    Next, $(a_k, b_j) \in E$ implies that the edge $(a_{k}^{(4)}, b_{j}^{(1)})$ exists.
    Since $b_{j} \in N(d_i)$, each edge in the sub-path $\left(b_{j}^{(1)}, b_{j}^{(2)}, b_{j + i}^{(3)}, b_{j}^{(4)} \right)$ exists and the vertices have combined weight
    \begin{align*}
        4 \cdot W^{4} + (4n - 2(j + i) + 2 j) \cdot W^{3} + j \cdot W = 4 \cdot W^{4} + (4n - 2i) \cdot W^{3} + j \cdot W.
    \end{align*}
    Then, $(b_j, c_{\ell}) \in E$ implies that the edge $(b_{j}^{(4)}, c_{\ell}^{(1)})$ exists.
    Since $c_{\ell} \in N(d_i)$, each edge in the sub-path $\left(c_{\ell}^{(1)}, c_{\ell}^{(2)}, c_{\ell + i}^{(3)}, c_{\ell}^{(4)} \right)$ exists and the vertices have combined weight
    \begin{align*}
        4 \cdot W^{4} + (4n - 2(\ell + i) + 2 \ell) \cdot W^{3} + \ell = 4 \cdot W^{4} + (4n - 2i) \cdot W^{3} + \ell.
    \end{align*}
    Then, $(c_{\ell}, a) \in E$ implies that the edge $(c_{\ell}^{(4)}, \hat{a}_{k}^{(1)})$ exists.
    Finally, as $a_{k} \in N(d_i)$, each edge in the sub-path $\left(\hat{a}_{k}^{(1)}, \hat{a}_{k}^{(2)}, \hat{a}_{k + i}^{(3)}, \hat{a}_{k}^{(4)} \right)$ exists and the vertices have combined weight
    \begin{align*}
        4 \cdot W^{4} + (4n - 2(k + i) + 2k) \cdot W^{3} + k \cdot W^{2} = 4 \cdot W^{4} + (4n - 2i) \cdot W^{3} + k \cdot W^{2}.
    \end{align*}
    We conclude by observing that edge $(\hat{a}_{k}^{(4)}, t)$ exists.
    Summing over all vertices in the path, we conclude the length of the path is
    \begin{align*}
        18 \cdot W^{4} + (16n - 8i) \cdot W^{3} + 2 k \cdot W^{2} + j \cdot W + \ell.
    \end{align*}

    Conversely, suppose there is some path $P$ of length 
    \begin{equation*}
        \distance(P) < 18 \cdot W^{4} + (16n - 8i) \cdot W^{3} + (2k + 1) \cdot W^{2}.
    \end{equation*}
    Since all vertices have weight at least $W^{4}$ and 
    \begin{equation*}
        (16n - 8i) \cdot W^{3} + (2k + 1) \cdot W^{2} < W^{4},
    \end{equation*}
    any $(s, t)$-path $P$ satisfying the length constraint has at most 18 vertices.
    Since we have a layered graph and  edges only exist between adjacent layers, any such path $P$ must have form
    \begin{align*}
        P = \left(s, a_{i_{1}}^{(1)}, a_{i_{2}}^{(2)}, a_{i_{3}}^{(3)}, a_{i_{4}}^{(4)}, b_{i_{5}}^{(1)}, b_{i_{6}}^{(2)}, b_{i_{7}}^{(3)}, b_{i_{8}}^{(4)}, c_{i_{9}}^{(1)}, c_{i_{10}}^{(2)}, c_{i_{11}}^{(3)}, c_{i_{12}}^{(4)}, \hat{a}_{i_{13}}^{(1)}, \hat{a}_{i_{14}}^{(2)}, \hat{a}_{i_{15}}^{(3)}, \hat{a}_{i_{16}}^{(4)}, t\right).
    \end{align*}
    
    At the time of the query in the $i$-th iteration over $D$ and $k$-th iteration over $A$ in the inner-loop, the minimum weight of any path from $A_2$ to $A_4$ is $3 \cdot W^{4} + (4 n - 2 i) \cdot W^{3}$, and this is attained only by paths of the form
    \begin{equation}
    \label{eq:a-2-a-4-path-node-weight}
        \left( a_{k'}^{(2)}, a_{k' + i}^{(3)}, a_{k'}^{(4)} \right)
    \end{equation}
    for $k' \geq k$ since we iterate over $A$ in decreasing order.
    In fact, any other path from $A_2, A_4$ has weight at least $3\cdot W^4 + (4 n - 2 i + 1) \cdot W^{3}$.
    This is because $a_{i_3}^{(3)}$ is adjacent only to vertices $a_{i_2}^{(2)}, a_{i_4}^{(4)}$ with $i_2, i_4 \geq i_3 - i$.
    Thus, if either $i_2$ or $i_4$ is  $> i_3 - i$, the three vertices have weight at least
    \begin{equation*}
        3\cdot W^4 + (4n - 2i_3 + 2i_3 - 2i + 1) \cdot W^{3} = 3\cdot W^4+ (4n - 2i + 1) \cdot W^{3}.
    \end{equation*}
    
    By a similar argument, the minimum weight of any path from $B_2$ to $B_4$, $C_2$ to $C_4$, $\hat{A}_2$ to $\hat{A}_4$ is $3 \cdot W^{4} + (4 n - 2 i) \cdot W^{3}$ and any path not satisfying the form of \cref{eq:a-2-a-4-path-node-weight} has weight at least $3 \cdot W^{4} + (4 n - 2 i + 1) \cdot W^{3}$.
    Since 
    \begin{equation*}
        \distance(P) < 18 \cdot W^{4} + (16 n - 8 i) \cdot W^{3} + (2 k + 1) \cdot W^{2} < 18 \cdot W^{4} + (16 n - 7 i) \cdot W^{3}, 
    \end{equation*}
    we conclude
    \begin{align*}
        0 &= i_{2} - i_{4} = i_{6} - i_{8} = i_{10} - i_{12} = i_{14} - i_{16}, \\
        i &= i_{3} - i_{2} = i_{7} - i_{6} = i_{11} - i_{10} = i_{15} - i_{14}.
    \end{align*}
    
    Next, observe that for all $x \in \set{a, b, c, \hat{a}}$ and $X \in \set{A, B, C, \hat{A}}$, $x_{j}^{(2)}$ has a unique neighbor in $X_1$, namely $x_{j}^{(1)}$.
    In particular, there are indices $k', j', \ell', \hat{k}'$ such that
    \begin{align*}
        P = \left(s, a_{k'}^{(1)}, a_{k'}^{(2)}, a_{k' + i}^{(3)}, a_{k'}^{(4)}, b_{j'}^{(1)}, b_{j'}^{(2)}, b_{j' + i}^{(3)}, b_{j'}^{(4)}, c_{\ell'}^{(1)}, c_{\ell'}^{(2)}, c_{\ell' + i}^{(3)}, c_{\ell'}^{(4)}, c_{\ell'}^{(5)}, \hat{a}_{\hat{k}'}^{(1)}, \hat{a}_{\hat{k}'}^{(2)}, \hat{a}_{\hat{k}' + i}^{(3)}, \hat{a}_{\hat{k}'}^{(4)}, t\right).
    \end{align*}

    Next, we claim $k = k' = \hat{k}'$.
    Because we enumerate $A$ in decreasing order, we know that $k', \hat{k}' \geq k$.
    If $k' > k$, then
    \begin{equation*}
        \wt\left(a_{k'}^{(1)} \right) > k \cdot W^{2}.
    \end{equation*}
    Similarly, if $\hat{k}' > k$, then
    \begin{equation*}
        \wt\left(\hat{a}_{\hat{k}'}^{(1)} \right) > k \cdot W^{2}.
    \end{equation*}
    If either condition holds, then $\distance(P) \geq 18 \cdot W^{4} + (16 n - 8 i) \cdot W^{3} + (2 k + 1) \cdot W^{2}$, a contradiction.
    Thus, $k = k' = \hat{k}'$.
    
    In particular, we have $a_k, b_{j'}, c_{\ell'} \in N(d_i)$ and furthermore $a_k, b_{j'}, c_{\ell'}$ form a triangle as edges $\left( a_{k}^{(4)}, b_{j'}^{(1)} \right)$, $\left( b_{j'}^{(4)}, c_{\ell'}^{(1)} \right)$, and $\left( c_{\ell'}^{(4)}, \hat{a}_{k}^{(1)} \right)$ exist, so that $a_k, d_i$ form a $4$-clique with $b_{j'}, c_{\ell'}$.
    Note that this path $P$ has length
    \begin{equation*}
        \distance(P) = 18 \cdot W^{4} + (16n - 8i) \cdot W^{3} + 2 k \cdot W^{2} + j' \cdot W + \ell'.
    \end{equation*}
    as desired.
\end{proof}

    From Lemma \ref{lemma:s-t-sp-node-weight-4-clique-equiv}, at the time of the query,  path $P$ with length $\distance(P) < 18 \cdot W^{4} + (16n - 8i) \cdot W^{3} + (2k + 1) \cdot W^{2}$ corresponds to $4$-clique $(a_k, b, c, d_i)$ in the input graph containing $d_i, a_k$.
    Since we iterate over all $a_k, d_i$, the above procedure solves the 4-Clique Detection instance in total time $O(n^{4 - c})$, contradicting the $4$-Clique hypothesis.
\end{proof}

\subsubsection{Non-Combinatorial Lower Bounds}

Using an essentially identical reduction, we show that under the $\OMvThree$ hypothesis, any algorithm computing partially dynamic \nwstSP{} with polynomial preprocessing requires $n^{\omega + 1 - o(1)}$ total update and query time.
The $\OMvThree$ hypothesis, introduced by \cite{DBLP:conf/stoc/GutenbergWW20} is a generalization of the $\OMv$ hypothesis \cite{DBLP:conf/stoc/HenzingerKNS15}.

\paragraph{$\OMvThree$ hypothesis.}
Beyond its inability to deal with non-combinatorial algorithms, the Combinatorial $k$-Clique hypothesis (Hypothesis \ref{conj:k-clique}) has one further weakness: It does not rule out algorithms with polynomially large preprocessing time. 
Conceptually, this may not be a large issue in the incremental setting (since in the preprocessing phase the algorithm only has access to an empty graph), it is not unreasonable to expect a decremental algorithm, given sufficient preprocessing time, to handle updates and queries efficiently. 
In the $k = 3$ case, the $\OMv$ hypothesis \cite{DBLP:conf/stoc/HenzingerKNS15} strengthens conditional lower bounds based on the combinatorial BMM hypothesis \cite{abboud2014popular}, ruling out arbitrary algorithms with polynomial preprocessing.
The $\OMvThree$ hypothesis was introduced as a natural generalization of the $\OMv$ hypothesis, ruling out arbitrary algorithms with polynomial preprocessing time \cite{DBLP:conf/stoc/GutenbergWW20}.

\begin{definition}[$\OMvThree$ Problem \cite{DBLP:conf/stoc/GutenbergWW20}]
    \label{def:omv-3}
    During preprocessing, the algorithm is given an $n \times n$ boolean matrix $A$.
    During the query phase, the algorithm receives $n$ queries.
    Each query consist of three $n$-dimensional boolean vectors $\vec{u}, \vec{v}, \vec{w}$ and must output
    \begin{equation*}
        \bigvee_{i, j, k} \left( \vec{u}_i \wedge \vec{v}_j \wedge \vec{w}_k \wedge A_{ij} \wedge A_{jk} \wedge A_{ki} \right)
    \end{equation*}
    before the next query is received.
\end{definition}

Naturally, each $\OMvThree$ query can be answered in $O(n^{\omega})$ time using fast matrix multiplication.
The $\OMvThree$ hypothesis states that this is essentially optimal.

\OMvThreeHypothesis*

\begin{theorem}
    \label{thm:s-t-sp-node-weight-lb-omv-3}
    Under the $\OMvThree$ hypothesis, any algorithm computing incremental/decremental \nwstSP{} on undirected graphs with polynomial preprocessing time requires $n^{\omega + 1 - o(1)}$ total update and query time.
\end{theorem}

\begin{proof}
    Suppose for contradiction there is an incremental $\stSP$ algorithm $\innerAlg$ with polynomial preprocessing time and total update and query time $O(n^{\omega + 1 - c})$ for some $c > 0$.
    We will describe appropriate modifications for the decremental case where necessary.
    We design an efficient algorithm for $\OMvThree$.

    In the preprocessing phase we receive a Boolean matrix $A$.
    We construct a graph with the same vertex set as in Theorem \ref{thm:s-t-sp-node-weight-lb}:
    \begin{equation*}
        \set{s} \cup G_R(A) \cup G_R(B) \cup G_R(C) \cup \hat{G}_R(A) \cup \set{t}.
    \end{equation*}
    As before, insert edges $(s, a^{(1)})$ and $(\hat{a}^{(4)}, t)$ for all $a \in A$.
    Furthermore, insert edges between $(a^{(4)}, b^{(1)})$ (resp. $(b^{(4)}, c^{(1)})$, $(c^{(4)}, \hat{a}^{(1)})$) if and only if $A[a, b] = 1$ (resp. $A[b, c] = 1$,  $A[a, c] = 1$).
    Insert edges $(a^{(1)}, a^{(2)}), (b^{(1)}, b^{(2)}), (c^{(1)}, c^{(2)}), (\hat{a}^{(1)}, \hat{a}^{(2)})$ for all $a, b, c$.

    In the decremental case, we begin with a graph that additionally contains the following edges.
    \begin{enumerate}
        \item For all $i, j$, insert edges$\left(b_j^{(2)}, b_{j + i}^{(3)}\right)$ and $\left(b_{j + i}^{(3)}, b_{j}^{(4)} \right)$.
        \item For all $i, j$, insert edges $\left(c_j^{(2)}, c_{j + i}^{(3)}\right)$ and $\left(c_{j + i}^{(3)}, c_{j}^{(4)} \right)$.
        \item For all $i, k$, insert the following edges:
        \begin{align*}
            \left(a_k^{(2)}, a_{k + i}^{(3)}\right), \left(a_{k + i}^{(3)}, a_{k}^{(4)} \right), \left(\hat{a}_{k}^{(2)}, \hat{a}_{k + i}^{(3)}\right), \left(\hat{a}_{k + i}^{(3)}, \hat{a}_{k}^{(4)} \right).
        \end{align*}
    \end{enumerate}
    We then run the preprocessing step of $\innerAlg$ on this $O(n)$-vertex graph.

    We now proceed to the dynamic phase of the reduction.
    Throughout, we assume that both queries and coordinates are indexed starting at $0$ (just as vertices are indexed from $0$ in Theorem \ref{thm:s-t-sp-node-weight-lb}).
    Recall that in the vertex gadgets $G_R(A), G_R(B)$ the third layer $A_3, B_3$ has $2n - 1$ vertices.
    Suppose we have received the $i$-th query, $\vec{u}_i, \vec{v}_i, \vec{w}_i$.
    \begin{enumerate}
        \item If $\vec{v}_i[j] = 1$, insert edges $\left(b_j^{(2)}, b_{j + i}^{(3)}\right)$ and $\left(b_{j + i}^{(3)}, b_{j}^{(4)} \right)$.
        In the decremental case, remove edges $(b_j^{(2)}, b_{j + n - i}^{(3)})$ and $(b_{j + n - i}^{(3)}, b_{j}^{(4)})$ if $\vec{v}_i[j] = 0$.
        \item If $\vec{w}_i[j] = 1$, insert edges $\left(c_j^{(2)}, c_{j + i}^{(3)}\right)$ and $\left(c_{j + i}^{(3)}, c_{j}^{(4)} \right)$.
        In the decremental case, remove edges $(c_j^{(2)}, c_{j + n - i}^{(3)})$ and $(c_{j + n - i}^{(3)}, c_{j}^{(4)})$ if $\vec{w}_i[j] = 0$.
        \item Now, we iterate over the coordinates $k$ of vector $\vec{v}_i$ in decreasing order.
        In the decremental case, we instead iterate over the coordinates $k$ in increasing order.
        \begin{enumerate}
            \item If $\vec{u}_i[k] = 1$, insert the edges
            \begin{align*}
                \left(a_k^{(2)}, a_{k + i}^{(3)}\right), \left(a_{k + i}^{(3)}, a_{k}^{(4)} \right), \left(\hat{a}_{k}^{(2)}, \hat{a}_{k + i}^{(3)}\right), \left(\hat{a}_{k + i}^{(3)}, \hat{a}_{k}^{(4)} \right).
            \end{align*}
            In the decremental case, remove the edges
            \begin{equation*}
                \left(a_k^{(2)}, a_{k + n - i}^{(3)}\right), \left(a_{k + n - i}^{(3)}, a_{k}^{(4)} \right), \left(\hat{a}_{k}^{(2)}, \hat{a}_{k + n - i}^{(3)}\right), \left(\hat{a}_{k + n - i}^{(3)}, \hat{a}_{k}^{(4)} \right),
            \end{equation*}
            whenever $\vec{u}_i[k] = 0$.
            
            \item Query $\distance(s, t)$ from $\innerAlg$.
            If $\distance(s, t) < 18 \cdot W^4 + (16n - 8i) \cdot W^{3} + (2 k + 1) \cdot W^{2}$ return $\true$.

            \item In the decremental case, we additionally remove all remaining edges 
            \begin{equation*}
                \left(a_k^{(2)}, a_{k + n - i}^{(3)}\right), \left(a_{k + n - i}^{(3)}, a_{k}^{(4)} \right), \left(\hat{a}_{k}^{(2)}, \hat{a}_{k + n - i}^{(3)}\right), \left(\hat{a}_{k + n - i}^{(3)}, \hat{a}_{k}^{(4)} \right).
            \end{equation*}
        \end{enumerate}
        Otherwise, return $\false$ if none of the queries satisfied the required distance constraint.

        \item In the decremental case, we additionally remove all remaining edges 
        \begin{equation*}
            \left(b_j^{(2)}, b_{j + n - i}^{(3)}\right), \left(b_{j + n - i}^{(3)}, b_{j}^{(4)}\right), \left(c_j^{(2)}, c_{j + n - i}^{(3)}\right), \left(c_{j + n - i}^{(3)}, c_{j}^{(4)}\right).
        \end{equation*}
    \end{enumerate}

    Clearly, the bottleneck of this algorithm is running $\innerAlg$, so the total time over all $n$ queries is $O(n^{\omega+1 - c})$.
    
    Note that at the query after the $i$-th query and the $k$-th iteration of the inner loop, we have constructed the same graph as in Theorem \ref{thm:s-t-sp-node-weight-lb}. In particular, we have the following lemma, whose proof is deferred to Appendix \ref{app:omv-3-lb} as it is identical to the proof of  Lemma \ref{lemma:s-t-sp-node-weight-4-clique-equiv}.

    The following lemma shows that $\stSP$ queries compute $\OMvThree$ correctly.
    
    \begin{restatable}{lemma}{stSPNodeOMvThreeEquiv}
        \label{lemma:s-t-sp-node-omv-3-equiv}
        Consider the query $(\vec{u}_i, \vec{v}_i, \vec{w}_i)$ and the query after updating the $k$-th coordinate of $\vec{u}_i$.
        The following are equivalent:
        \begin{enumerate}
            \item $d(s, t) < 18 \cdot W^{4} + (16 n - 8 i) \cdot W^{3} + (2 k + 1) \cdot W^{2}$
            \label{item:s-t-s-p-node-weight-omv-3:distance};
            
            \item $\bigvee_{j, \ell} \left( \vec{u}_i[k] \wedge \vec{v}_i[j] \wedge \vec{w}_i[\ell] \wedge A[k, j] \wedge A[j, \ell] \wedge A[\ell, k] \right) = \true$.
            \label{item:s-t-s-p-node-weight-omv-3:omv-3}
        \end{enumerate}
    \end{restatable}
    
    In particular, Lemma \ref{lemma:s-t-sp-node-omv-3-equiv} shows that this query returns true if and only if
    \begin{equation*}
        \bigvee_{j, \ell} \left( \vec{u}_i[k] \wedge \vec{v}_i[j] \wedge \vec{w}_i[\ell] \wedge A[k, j] \wedge A[j, \ell] \wedge A[\ell, k] \right).
    \end{equation*}
    
    Thus, if the above is satisfied for any $k$, we return $\true$.
    Otherwise, we return $\false$, answering the query correctly in either case.
    Therefore, using $\innerAlg$ we have obtained an algorithm for the $\OMvThree$ instance with polynomial preprocessing and total update and query time $O(n^{\omega + 1 - c})$, contradicting the $\OMvThree$ hypothesis.
\end{proof}

We prove that the \stSP{} instance computes \OMvThree{}.

\begin{proof}[Proof of Lemma \ref{lemma:s-t-sp-node-omv-3-equiv}]
    We prove the equivalence in the incremental case, leaving the modifications for the decremental case to the end of the proof.
    Suppose \ref{item:s-t-s-p-node-weight-omv-3:omv-3} is true and let $j, \ell$ index one clause that is $\true$.
    Then, the following $(s, t)$-path exists 
    \begin{align*}
        \left(s, a_{k}^{(1)}, a_{k}^{(2)}, a_{k + i}^{(3)}, a_{k}^{(4)}, b_{j}^{(1)}, b_{j}^{(2)}, b_{j + i}^{(3)}, b_{j}^{(4)}, c_{\ell}^{(1)}, c_{\ell}^{(2)}, c_{\ell + i}^{(3)}, c_{\ell}^{(4)}, \hat{a}_{k}^{(1)}, \hat{a}_{k}^{(2)}, \hat{a}_{k + i}^{(3)}, \hat{a}_{k}^{(4)}, t\right).
    \end{align*}
    and has length
    \begin{align*}
        18 \cdot W^{4} + (16n - 8i) \cdot W^{3} + 2 k \cdot W^{2} + j \cdot W + \ell.
    \end{align*}
    
    Note that by our choice of $W$, $j \cdot W + \ell < W^{2}$ so this path satisfies the desired constraint.
    We verify that this path exists and has the required length.
    Recall that the edge $(s, a_{k}^{(1)})$ exists.
    Since $\vec{u}_{i}[k] = 1$, each edge in the sub-path $\left(a_{k}^{(1)}, a_{k}^{(2)}, a_{k + i}^{(3)}, a_{k}^{(4)}\right)$ exists and the vertices have combined weight,
    \begin{align*}
        4 \cdot W^{4} + (4n - 2(k + i) + 2k) \cdot W^{3} + k \cdot W^{2} = 4 \cdot W^{4} + (4n - 2i) \cdot W^{3} + k \cdot W^{2}
    \end{align*}
    Next, $A[k, j] = 1$ implies that the edge $(a_{k}^{(4)}, b_{j}^{(1)})$ exists.
    Since $\vec{v}_{i}[k] = 1$, each edge in the sub-path $\left(b_{j}^{(1)}, b_{j}^{(2)}, b_{j + i}^{(3)}, b_{j}^{(4)}\right)$ exists and the vertices have combined weight,
    \begin{align*}
        4 \cdot W^{4} + (4n - 2(j + i) + 2 j) \cdot W^{3} + j \cdot W = 4 \cdot W^{4} + (4n - 2i) \cdot W^{3} + j \cdot W
    \end{align*}
    Then, $A[j, \ell] = 1$ implies that the edge $(b_{j}^{(4)}, c_{\ell}^{(1)})$ exists.
    Since $\vec{w}_{i}[\ell] = 1$, each edge in the sub-path $\left(c_{\ell}^{(1)}, c_{\ell}^{(2)}, c_{\ell + i}^{(3)}, c_{\ell}^{(4)}\right)$ exists and the vertices have combined weight,
    \begin{align*}
        4 \cdot W^{4} + (4n - 2(\ell + i) + 2 \ell) \cdot W^{3} + \ell = 4 \cdot W^{4} + (4n - 2i) \cdot W^{3} + \ell
    \end{align*}
    Then, $A[\ell, k] = 1$ implies that the edge $(c_{\ell}^{(4)}, \hat{a}_{k}^{(1)})$ exists.
    Finally, as $\vec{u}_{i}[k] = 1$, each edge in the sub-path $\left(\hat{a}_{k}^{(1)}, \hat{a}_{k}^{(2)}, \hat{a}_{k + i}^{(3)}, \hat{a}_{k}^{(4)} \right)$ exists and the vertices have combined weight,
    \begin{align*}
        4 \cdot W^{4} + (4n - 2(k + i) + 2k) \cdot W^{3} + k \cdot W^{2} = 4 \cdot W^{4} + (4n - 2i) \cdot W^{3} + k \cdot W^{2}
    \end{align*}
    We conclude by observing that edge $(\hat{a}_{k}^{(4)}, t)$ exists.
    Summing over all vertices in the path, we conclude the length of the path is,
    \begin{align*}
        18 \cdot W^{4} + (16n - 8i) \cdot W^{3} + 2 k \cdot W^{2} + j \cdot W + \ell
    \end{align*}

    Conversely, suppose there is some path $P$ of length 
    \begin{equation*}
        \distance(P) < 18 \cdot W^{4} + (16n - 8i) \cdot W^{3} + (2k + 1) \cdot W^{2}.
    \end{equation*}
    
    Since all vertices have weight at least $W^{4}$ and 
    \begin{equation*}
        (16n - 8i) \cdot W^{3} + (2k + 1) \cdot W^{2} < W^{4},
    \end{equation*}
    any $(s, t)$-path $P$ satisfying the length constraint has at most 18 vertices.
    Since we have a layered graph and only edges between adjacent layers, any such path $P$ must have the form,
    \begin{align*}
        P = \left(s, a_{i_{1}}^{(1)}, a_{i_{2}}^{(2)}, a_{i_{3}}^{(3)}, a_{i_{4}}^{(4)}, b_{i_{5}}^{(1)}, b_{i_{6}}^{(2)}, b_{i_{7}}^{(3)}, b_{i_{8}}^{(4)}, c_{i_{9}}^{(1)}, c_{i_{10}}^{(2)}, c_{i_{11}}^{(3)}, c_{i_{12}}^{(4)}, \hat{a}_{i_{13}}^{(1)}, \hat{a}_{i_{14}}^{(2)}, \hat{a}_{i_{15}}^{(3)}, \hat{a}_{i_{16}}^{(4)}, t\right)
    \end{align*}
    
    At the time of the query after updating the $k$-th coordinate of $\vec{u}$ in the $\OMvThree$ query $\vec{u}_i, \vec{v}_i, \vec{w}_i$, the minimum weight of any path from $A_2$ to $A_4$ is $3 \cdot W^{4} + (4 n - 2 i) \cdot W^{3}$, and as argued in Lemma \ref{lemma:s-t-sp-node-weight-4-clique-equiv} this is attained only by paths of the form of \Cref{eq:a-2-a-4-path-node-weight}.
    
    Similarly, the minimum weight of any path from $B_2$ to $B_4$, $C_2$ to $C_4$, $\hat{A}_2$ to $\hat{A}_4$ is $3 \cdot W^{4} + (4 n - 2 i) \cdot W^{3}$ and any path not satisfying the form of \Cref{eq:a-2-a-4-path-node-weight} has weight at least $3 \cdot W^{4} + (4 n - 2 i + 1) \cdot W^{3}$.
    Since 
    \begin{equation*}
        \distance(P) < 18 \cdot W^{4} + (16 n - 8 i) \cdot W^{3} + (2 k + 1) \cdot W^{2} < 18 \cdot W^{4} + (16 n - 7 i) \cdot W^{3},
    \end{equation*}
    we conclude
    \begin{align*}
        0 &= i_{2} - i_{4} = i_{6} - i_{8} = i_{10} - i_{12} = i_{14} - i_{16} \\
        i &= i_{3} - i_{2} = i_{7} - i_{6} = i_{11} - i_{10} = i_{15} - i_{14}.
    \end{align*}
    
    Next, observe that for all $x \in \set{a, b, c, \hat{a}}$ and $X \in \set{A, B, C, \hat{A}}$, $x_{j}^{(2)}$ has a unique neighbor in $X_1$, namely $x_{j}^{(1)}$.
    In particular, there are indices $k', j', \ell', \hat{k}'$ such that,
    \begin{align*}
        P = \left(s, a_{k'}^{(1)}, a_{k'}^{(2)}, a_{k' + i}^{(3)}, a_{k'}^{(4)}, b_{j'}^{(1)}, b_{j'}^{(2)}, b_{j' + i}^{(3)}, b_{j'}^{(4)}, c_{\ell'}^{(1)}, c_{\ell'}^{(2)}, c_{\ell' + i}^{(3)}, c_{\ell'}^{(4)}, \hat{a}_{\hat{k}'}^{(1)}, \hat{a}_{\hat{k}'}^{(2)}, \hat{a}_{\hat{k}' + i}^{(3)}, \hat{a}_{\hat{k}'}^{(4)}, t\right)
    \end{align*}

    Next, we claim $k = k' = \hat{k}'$.
    As before, we have $k', \hat{k}' \geq k$ from \Cref{eq:a-2-a-4-path} since we iterate over $k$ in decreasing order.
    If $k' > k$, then
    \begin{equation*}
        \wt\left(a_{k'}^{(1)} \right) > (k + 1) \cdot W^{3}.
    \end{equation*}
    Similarly, if $\hat{k}' > k$, then
    \begin{equation*}
        \wt\left(\hat{a}_{\hat{k}'}^{(1)} \right) > (k + 1) \cdot W^{2}.
    \end{equation*}
    If either condition fails, then $\distance(P) \geq 18 \cdot W^{4} + (16 n - 8 i) \cdot W^{3} + (2 k + 1) \cdot W^{2}$, a contradiction.
    Thus, $k = k' = \hat{k}'$.
    
    In particular, we have $\vec{u}_{i}[k] = \vec{v}_{i}[j'] = \vec{w}_{i}[\ell'] = 1$ and furthermore $A[k, j] = A[j, \ell] = A[\ell, k] = 1$ as edges $\left( a_{k}^{(4)}, b_{j'}^{(1)} \right)$, $\left( b_{j'}^{(4)}, c_{\ell'}^{(1)} \right)$, and $\left( c_{\ell'}^{(4)}, \hat{a}_{k}^{(1)} \right)$ exist, so that
    \begin{equation*}
        \vec{u}_i[k] = \vec{v}_i[j'] = \vec{w}_{i}[\ell'] = A[k, j] = A[j, \ell] = A[\ell, k] = \true.
    \end{equation*}

    \paragraph{Correctness of Decremental Reduction}
    We now describe the modifications for the decremental case.
    In the decremental case, we instead have the path
    \begin{align*}
        \left(s, a_{k}^{(1)}, a_{k}^{(2)}, a_{k + n - i}^{(3)}, a_{k}^{(4)}, b_{j}^{(1)}, b_{j}^{(2)}, b_{j + n - i}^{(3)}, b_{j}^{(4)}, c_{\ell}^{(1)}, c_{\ell}^{(2)}, c_{\ell + n - i}^{(3)}, c_{\ell}^{(4)}, \hat{a}_{k}^{(1)}, \hat{a}_{k}^{(2)}, \hat{a}_{k + n - i}^{(3)}, \hat{a}_{k}^{(4)}, t\right),
    \end{align*}
    which has length
    \begin{equation*}
        18 \cdot W^{4} + (8n + 8i) \cdot W^{3} + 2 k \cdot W^{2} + j \cdot W + \ell.
    \end{equation*}
    
    We verify that this path exists and has the required length.
    Since $\vec{u}_{i}[k] = 1$, each edge in the sub-path $\left(a_{k}^{(1)}, a_{k}^{(2)}, a_{k + n - i}^{(3)}, a_{k}^{(4)}\right)$ exists and the vertices have combined weight
    \begin{align*}
        4 \cdot W^{4} + (4n - 2(k + n - i) + 2k) \cdot W^{3} + k \cdot W^{2} = 4 \cdot W^{4} + (2n + 2i) \cdot W^{3} + k \cdot W^{2}.
    \end{align*}
    Following similar arguments as above the path exists with the required length.

    Conversely, suppose there is some path $P$ of length 
    \begin{equation*}
        \distance(P) < 18 \cdot W^{4} + (8n + 8i) \cdot W^{3} + (2k + 1) \cdot W^{2}.
    \end{equation*}
    
    As above, the path has at most $18$ vertices and therefore has the form
    \begin{align*}
        P = \left(s, a_{i_{1}}^{(1)}, a_{i_{2}}^{(2)}, a_{i_{3}}^{(3)}, a_{i_{4}}^{(4)}, b_{i_{5}}^{(1)}, b_{i_{6}}^{(2)}, b_{i_{7}}^{(3)}, b_{i_{8}}^{(4)}, c_{i_{9}}^{(1)}, c_{i_{10}}^{(2)}, c_{i_{11}}^{(3)}, c_{i_{12}}^{(4)}, \hat{a}_{i_{13}}^{(1)}, \hat{a}_{i_{14}}^{(2)}, \hat{a}_{i_{15}}^{(3)}, \hat{a}_{i_{16}}^{(4)}, t\right).
    \end{align*}
    
    At the time of the query after updating the $k$-th coordinate of $\vec{u}$ in the $\OMvThree$ query $\vec{u}_i, \vec{v}_i, \vec{w}_i$, the minimum weight of any path from $A_2$ to $A_4$ is $3 \cdot W^{4} + (2 n + 2 i) \cdot W^{3}$ and this is attained only by paths of the form of
    \begin{equation}
        \label{eq:a-2-a-4-path-node-weight-dec-omv-3}
        (a_{k'}^{(2)}, a_{k' + n - i}^{(3)}, a_{k'}^{(4)}),
    \end{equation}
    for $k' \geq k$ since we iterate over coordinates $k$ in increasing order.
    In fact, any other path from $A_2$ to $A_4$ has weight at least $3 \cdot W^4 + (2 n + 2 i + 1) \cdot W^{3}$.
    This is because $a_{i_3}^{(3)}$ is adjacent only to vertices $a_{i_2}^{(2)}, a_{i_4}^{(4)}$ with $i_3 \leq i_2 + n - i$ or $i_2 \geq i_3 - n + i$.
    Thus, if either $i_2$ or $i_4$ is  $> i_3 - n + i$, the three vertices have weight at least
    \begin{equation*}
        3\cdot W^4 + (4 n - 2 i_3 + 2 i_3 - 2 n + 2 i + 1) \cdot W^{3} = 3 \cdot W^4+ (2 n + 2 i + 1) \cdot W^{3}.
    \end{equation*}
    
    Similarly, the minimum weight of any path from $B_2$ to $B_4$, $C_2$ to $C_4$, $\hat{A}_2$ to $\hat{A}_4$ is $3 \cdot W^{4} + (2 n + 2 i) \cdot W^{3}$ and any path not satisfying the form of \Cref{eq:a-2-a-4-path-node-weight-dec-omv-3} has weight at least $3 \cdot W^{4} + (2 n + 2 i + 1) \cdot W^{3}$.
    Since 
    \begin{equation*}
        \distance(P) < 18 \cdot W^{4} + (8 n + 8 i) \cdot W^{3} + (2 k + 1) \cdot W^{2} < 18 \cdot W^{4} + (8 n + 9 i) \cdot W^{3},
    \end{equation*}
    we conclude
    \begin{align*}
        0 &= i_{2} - i_{4} = i_{6} - i_{8} = i_{10} - i_{12} = i_{14} - i_{16} \\
        i &= i_{3} - i_{2} = i_{7} - i_{6} = i_{11} - i_{10} = i_{15} - i_{14}.
    \end{align*}
    
    Following previous arguments, there are indices $k', j', \ell', \hat{k}'$ such that,
    \begin{align*}
        P = \left(s, a_{k'}^{(1)}, a_{k'}^{(2)}, a_{k' + i}^{(3)}, a_{k'}^{(4)}, b_{j'}^{(1)}, b_{j'}^{(2)}, b_{j' + i}^{(3)}, b_{j'}^{(4)}, c_{\ell'}^{(1)}, c_{\ell'}^{(2)}, c_{\ell' + i}^{(3)}, c_{\ell'}^{(4)}, \hat{a}_{\hat{k}'}^{(1)}, \hat{a}_{\hat{k}'}^{(2)}, \hat{a}_{\hat{k}' + i}^{(3)}, \hat{a}_{\hat{k}'}^{(4)}, t\right)
    \end{align*}

    Next, we claim $k = k' = \hat{k}'$.
    As before, we have $k', \hat{k}' \geq k$ from \Cref{eq:a-2-a-4-path} since we iterate over $k$ in increasing order.
    Thus, $k = k' = \hat{k}'$ as in the incremental case.
    Following an identical proof as in the incremental case, we conclude
    \begin{equation*}
        \vec{u}_i[k] = \vec{v}_i[j'] = \vec{w}_{i}[\ell'] = A[k, j] = A[j, \ell] = A[\ell, k] = \true.
    \end{equation*}
\end{proof}

\subsubsection{Lower Bounds under Minimum Witness Hypothesis}
\label{sec:min-witness-lb}

Under the Minimum-Witness $3$-Product hypothesis (Hypothesis \ref{conj:min-witness-3-product}), we show a stronger lower bound for the single source version of the problem.
Before proving the lower bound, we provide some further discussion on the Minimum-Witness hypothesis.
The Minimum-Witness Product problem is defined below.

\begin{definition}[Min-Witness Product] 
    \label{def:min-witness}
    Given two Boolean matrices $A, B$, their min-witness product $C$ is defined by $C_{ij} = \min \set{k \mid A_{ik} \wedge B_{kj}}$.
\end{definition}

\cite{CzumajKL07} computes the product in $O(n^{2+\lambda})$ time , where $\omega(1, \lambda, 1) = 1+2\lambda$. 
If $\omega = 2$, this running time is essentially $O(n^{2.5})$.
\cite{DBLP:conf/innovations/Lincoln0W20} conjectures that this is essentially optimal.

\begin{hypothesis}[Minimum-Witness Product hypothesis]
    \label{hypo:min-witness-product}
    There is no $O(n^{2.5 - \varepsilon})$ algorithm for computing Minimum-Witness Product between two $n \times n$ Boolean matrices, for any $\varepsilon > 0$.
\end{hypothesis}

Minimum-Witness Product can also be generalized to a problem between $k > 2$ input Boolean matrices \cite{DBLP:journals/algorithms/KowalukL22}.  In this work, we only need the following version where $k = 3$. 

\begin{definition}[Minimum-Witness $3$-Product]
    \label{def:min-witness-3-product}
    Given three $n \times n$ Boolean matrices $A, B, C$, for every $i_1, i_2, i_3 \in [n]$, find the minimum value of $j \in [n]$ such that $A_{i_1, j} \wedge B_{i_2, j} \wedge C_{i_3, j}$. 
\end{definition}

The current best algorithm for Minimum-Witness $3$-Product runs in $O(n^{3.5})$ time when $\omega = 2$~\cite{DBLP:journals/algorithms/KowalukL22}, and it seems that new ideas are required in order to break this $O(n^{3.5})$ running time. 
Below is some evidence, similar to \cite{JinX22}'s argument for why the so-called OuMv$_k$ hypothesis is plausible. First, we observe that we can use the Minimum-Witness $3$-Product problem to solve $4$-Clique Detection.
Given a $4$-partite graph $G = (V_1 \cup V_2 \cup V_3 \cup V_4, E)$, we can use the matrix $A$ to encode the edges between $V_1$ and $V_4$, matrix $B$ to encode the edges between $V_2$ and $V_4$ and matrix $C$ to encode the edges between $V_3$ and $V_4$. 
Thus, the result of Minimum-Witness 3-Product can tell us, for every triple $(v_1, v_2, v_3) \in V_1 \times V_2 \times V_3$, whether there is some $v_4 \in V_4$ that is connected to all of them. 
Then we can detect whether the graph has a $4$-clique in $O(n^3)$ time. 
Currently, the all truly sub-quartic time algorithms for $4$-Clique Detection essentially groups two node parts together, then solves a Triangle Detection instance on an unbalanced tripartite graph.
Therefore, assuming there is no better framework for $4$-Clique Detection, we also need to perform similar groupings for the Minimum-Witness $3$-Product problem, as it can be used to solve $4$-Clique Detection. If we group any two of the indices $i_1, i_2, i_3, j$ together, we obtain a Min-Witness Product instance between an $n \times n^2$ matrix and an $n^2 \times n$ matrix, or between an $n^2 \times n$ matrix and an $n \times n$ matrix. 
Either way, the current fastest algorithm for them has running time $O(n^{3.5})$ if $\omega = 2$, which essentially follows the same framework for the square inputs case \cite{CzumajKL07}. 
Therefore, in order to refute the following hypothesis, it seems that one would need new ideas for either $4$-Clique Detection  or Min-Witness Product. 

\MinWitnessThreeProductHypothesis*

\subsubsection{Lower Bound for \texorpdfstring{\nwSSSP{}}{nw-SSSP}}

We now give our lower bound for \nwSSSP{}.

\begin{theorem}
    \label{thm:sssp-node-weight-min-witness-lb}
    Under the Minimum-Witness $3$-Product hypothesis, any algorithm computing incremental/decremental \nwSSSP{} on undirected graphs requires $n^{3.5 - o(1)}$ total update and query time.
\end{theorem}

\begin{proof}
    As before, we describe only the reduction for incremental $\stSP$.
    For the decremental setting, we can execute the reduction in reverse.
    
    Suppose for contradiction there is an algorithm $\innerAlg$ for incremental \nwSSSP{} in $O(n^{3.5 - c})$  total time. 
    Consider a Minimum-Witness $3$-Product instance with matrices $A, C, D$ of dimension $n \times n$.
    Throughout the reduction, assume that the matrices $A, C, D$ are indexed from $0$ to $n - 1$.
    We design an algorithm computing min-witness $3$-product with total time $O(n^{3.5 - c})$, thus contradicting the Minimum-Witness $3$-Product hypothesis.
    
    We construct a graph consisting of the gadgets 
    \begin{equation*}
        \set{s} \cup G_R(A) \cup G_R(B) \cup C_1
    \end{equation*}
    of Theorem \ref{thm:s-t-sp-node-weight-lb} with the following modifications.
    Whereas $A$ was a vertex set of $n$ nodes, think of $A$ now as a collection of $n$ rows of the matrix $A$.
    Whereas $B$ was a vertex set of $n$ nodes, think of $B$ now as a collection of $n$ columns.
    Whereas $C_1 = \set{c^{(1)}: c \in \{0, \ldots, n - 1\}}$ was a copy of the vertex set $C$ in the $4$-Clique instance, think of $c^{(1)}$ now as representing the $c$-th row of matrix $C$.
    
    We describe the formal reduction below.
    Initially, insert edges $(s, a_{i}^{(1)})$ all $i \in \set{0, 1, \dotsc, n - 1}$. 
    Next, insert edges between $(a_{i}^{(4)}, b_j^{(1)})$ (respectively between $(b_{j}^{(4)}, c_{i}^{(1)})$) if and only if $A[i, j] = \true$ (respectively $C[i, j] = \true$).
    Insert edges $(a^{(1)}, a^{(2)}), (b^{(1)}, b^{(2)})$ for all $a, b$.

    We now proceed to insert more edges in the reduction.
    In the outer loop we iterate over rows $i \in \set{0, 1, \dotsc, n-1}$ of $D$ in increasing order. 
    \begin{enumerate}
        \item For each $D[i, j] = \true$, insert edges $\left(b_j^{(2)}, b_{j + i}^{(3)}\right)$, $\left(b_{j + i}^{(3)}, b_{j}^{(4)} \right)$.
        
        \item In the inner loop, we iterate over rows $k \in \set{0, 1, \dotsc, n - 1}$ of $A$ in decreasing order.
        \begin{enumerate}
            \item Insert edges between $\left(a_k^{(2)}, a_{k + i}^{(3)}\right)$ and $\left(a_{k + i}^{(3)}, a_{k}^{(4)} \right)$.
            
            \item For each $c_{\ell}^{(1)} \in C_1$, query $\distance(s, c_{\ell}^{(1)})$ from $\innerAlg$.

            If $\distance(s, c_{\ell}^{(1)}) < 10 \cdot W^4 + (8n - 4i) \cdot W^{3} + (k + 1) \cdot W^{2}$, let
            \begin{equation*}
                \distance(s, c_{\ell}^{(1)}) = 10 \cdot W^4 + (8n - 4i) \cdot W^{3} + k \cdot W^{2} + b' \cdot W + \ell.
            \end{equation*}
            for some integer $0 \le b' < W$. 
            Then, assign the minimum witness of $(k, \ell, i)$ to be $b'$.
            Otherwise, if the $\distance(s, c)$ does not satisfy the required constraint, assign the minimum witness to $\bot$, or in other words there is no witness $j$ such that $A[k, j] = C[\ell, j] = D[i, j] = \true$.
        \end{enumerate}
    \end{enumerate}

    Again, we emphasize that we iterate over $D$ in increasing order and $A$ in decreasing order.
    Clearly, the bottleneck is running $\innerAlg$, so the total time of this algorithm is $O(n^{3.5 - c})$. 

    To show correctness, we first prove the following lemma. 
    
    \begin{lemma}
        \label{lemma:sssp-node-weight-3-prod-equiv}
        Consider the query when iterating over the $i$-th row of $D$ and the $k$-th row of $A$.
        Let $\ell$ be a row of $C$.
        Then, there is a one-to-one correspondence between:
        \begin{enumerate}
            \item Paths $P$ between $s, c_{\ell}^{(1)}$ with $\distance(P) < 10 \cdot W^{4} + (8n - 4i) \cdot W^{3} + (k + 1) \cdot W^{2}$;
            \item Witnesses $j$ such that $A[k, j] \wedge C[\ell, j] \wedge D[i, j]$.
        \end{enumerate}
    
        Furthermore, each such path $P$ has length
        \begin{equation*}
            \distance(P) = 10 \cdot W^{4} + (8n - 4i) \cdot W^{3} + k \cdot W^{2} + j \cdot W + \ell.
        \end{equation*}
    \end{lemma}
    
    \begin{proof}    
        Suppose there is a witness $j$ such that $A[k, j] = C[\ell, j] = D[i, j] = 1$.
        Then, the following $(s, c_{\ell}^{(1)})$-path exists :
        \begin{align*}
            \left(s, a_{k}^{(1)}, a_{k}^{(2)}, a_{k + i}^{(3)}, a_{k}^{(4)}, b_{j}^{(1)}, b_{j}^{(2)}, b_{j + i}^{(3)}, b_{j}^{(4)}, c_{\ell}^{(1)}\right).
        \end{align*}
        Furthermore, this path has length
        \begin{align*}
            10 \cdot W^{4} + (8n - 4i) \cdot W^{3} + k \cdot W^{2} + j \cdot W + \ell.
        \end{align*}
        
        Note that by our choice of $W$, $j \cdot W + \ell < W^{2}$ so this path satisfies the desired constraints.
        We verify that this path exists and has the claimed length.
        Recall that the edge $(s, a_{k}^{(1)})$ exists.
        Also, each edge in the sub-path $\left(a_{k}^{(1)}, a_{k}^{(2)}, a_{k + i}^{(3)}, a_{k}^{(4)} \right)$ exists and the vertices have combined weight $4 \cdot W^{4} + (4n - 2i) \cdot W^{3} + k \cdot W^{2}$.
        Next, $A[k, j] = \true$ implies that the edge $(a_{k}^{(4)}, b_{j}^{(1)})$ exists.
        Since $D[i, j] = \true$, each edge in the sub-path $\left(b_{j}^{(1)}, b_{j}^{(2)}, b_{j + i}^{(3)}, b_{j}^{(4)} \right)$ exists and the vertices have combined weight $4 \cdot W^{4} + (4n - 2i) \cdot W^{3} + j \cdot W$.
        Then, $C[\ell, j] = \true$ implies that the edge $(b_{j}^{(4)}, c_{\ell}^{(1)})$ exists, thus verifying the existence of the path.
        Summing over all vertices in the path, we conclude the length of the path is
        \begin{align*}
            10 \cdot W^{4} + (8n - 4i) \cdot W^{3} + k \cdot W^{2} + j \cdot W + \ell.
        \end{align*}
    
        Conversely, suppose there is some path $P$ of length 
        \begin{equation*}
            \distance(P) < 10 \cdot W^{4} + (8n - 4i) \cdot W^{3} + (k + 1) \cdot W^{2}.
        \end{equation*}
        Since all vertices have weight at least $W^{4}$ and 
        \begin{equation*}
            (8n - 4i) \cdot W^{3} + (k + 1) \cdot W^{2} < W^{4},
        \end{equation*}
        any $(s, t)$-path $P$ satisfying the length constraint has at most 10 vertices.
        Since we have a layered graph and  edges only exist between adjacent layers, any such path $P$ must have the form
        \begin{align*}
            P = \left(s, a_{i_{1}}^{(1)}, a_{i_{2}}^{(2)}, a_{i_{3}}^{(3)}, a_{i_{4}}^{(4)}, b_{i_{5}}^{(1)}, b_{i_{6}}^{(2)}, b_{i_{7}}^{(3)}, b_{i_{8}}^{(4)}, c_{\ell}^{(1)}\right).
        \end{align*}
        At the time of the query in the $i$-th iteration over $D$ and $k$-th iteration over $A$ in the inner-loop, the minimum weight of any path from $A_2$ to $A_4$ is $3 W^{4} + (4 n - 2 i) \cdot W^{3}$, and this is attained only by paths of the form
        \begin{equation}
            \label{eq:a-2-a-4-path}
            \left( a_{k}^{(2)}, a_{k + i}^{(3)}, a_{k}^{(4)} \right).
        \end{equation}
        As before, any other path from $A_2, A_4$ has weight at least $3 \cdot W^{4} + (4 n - 2 i + 1) \cdot W^{3}$.
        By a similar argument, the minimum weight of any path from $B_2$ to $B_4$ is $3 \cdot W^{4} + (4 n - 2 i) \cdot W^{3}$ and any path not satisfying the form of  \cref{eq:a-2-a-4-path} has weight at least $3 \cdot W^{4} + (4 n - 2 i + 1) \cdot W^{3}$.
        Since 
        \begin{equation*}
            \distance(P) < 10 \cdot W^{4} + (8 n - 4 i) \cdot W^{3} + (k + 1) \cdot W^{2} < 12 \cdot W^{4} + (8 n - 3 i) \cdot W^{3},
        \end{equation*}
        we conclude
        \begin{align*}
            0 &= i_{2} - i_{4} = i_{6} - i_{8}, \\
            i &= i_{3} - i_{2} = i_{7} - i_{6}.
        \end{align*}
        
        Next, observe that for all $x \in \set{a, b}$ and $X \in \set{A, B}$, $x_{j}^{(2)}$ has a unique neighbor in $X_1$, namely $x_{j}^{(1)}$.
        In particular, there are indices $k', j'$ such that,
        \begin{align*}
            P = \left(s, a_{k'}^{(1)}, a_{k'}^{(2)}, a_{k' + i}^{(3)}, a_{k'}^{(4)}, b_{j'}^{(1)}, b_{j'}^{(2)}, b_{j' + i}^{(3)}, b_{j'}^{(4)}, c_{\ell}^{(1)} \right).
        \end{align*}
    
        Following a similar argument as Lemma \ref{lemma:s-t-sp-node-weight-4-clique-equiv}, we see that $k = k'$.
        In particular, we have $D[i, j'] = \true$ and $A[k, j'] = \true$ since $(a_{k}^{(4)}, b_{j'}^{(1)})$ is an edge, and $C[\ell, j'] = \true$ since $(b_{j'}^{(4)}, c_{\ell}^{(1)})$ is an edge.
        In particular, $j'$ is a witness  such that $A[k, j'] = C[\ell, j'] = D[i, j'] = \true$.
        Note that this path $P$ has length
        \begin{equation*}
            \distance(P) = 10 \cdot W^{4} + (8n - 4i) \cdot W^{3} + k \cdot W^{2} + j' \cdot W + \ell
        \end{equation*}
        as desired.
    \end{proof}

    From Lemma \ref{lemma:sssp-node-weight-3-prod-equiv}, at the time of the query, each path $P$ from $s$ to $c_{\ell}^{(1)}$ with length $\distance(P) < 10 \cdot W^{4} + (8n - 4i) \cdot W^{3} + (k + 1) \cdot W^{2}$ corresponds to a witness $j$ such that $A[k, j] = C[\ell, j] = D[i, j] = \true$.
    Furthermore, Lemma \ref{lemma:sssp-node-weight-3-prod-equiv} shows that
    \begin{equation*}
        \distance(P) = 10 \cdot W^{4} + (8n - 4i) \cdot W^{3} + k \cdot W^{2} + j \cdot W + \ell,
    \end{equation*}
    so that the shortest path returns the minimum $j$ such that $A[k, j] \wedge C[\ell, j] \wedge D[i, j]$. In particular it computes the minimum-witness $3$-product for this $(k, \ell, i)$ triplet.
    If there is no path, then there is no witness $j$ such that $A[k, j] \wedge C[\ell, j] \wedge D[i, j]$.
    Thus, we compute the Minimum-Witness $3$-Product instance in $O(n^{3.5 - c})$ time, contradicting the hypothesis.
\end{proof}

\section{Bottleneck Paths}
\label{sec:bottleneck}

In this section, we give our algorithms and lower bounds for partially dynamic  bottleneck paths.
The bottleneck path problem is formally defined below.

\begin{definition}[Bottleneck Paths]
    \label{def:bottleneck-paths}
    Let $G = (V, E, \wt)$ be a directed, weighted graph. 
    For any path $P$, let $\bottleneck(P) = \min_{e \in P} \wt(e)$ denote the {\bf bottleneck capacity} of path $P$.
    For any pair of nodes $u, v$, let $\bottleneckPath(u, v)$ denote the {\bf bottleneck path} (breaking ties arbitrarily), the path between $u, v$ maximizing $\bottleneck(P)$.
    Let $\bottleneck(u, v) = \bottleneck(\bottleneckPath(u, v))$.

    The {\bf $(s, t)$-Bottleneck Path} problem ($\stBP$) asks to compute $\bottleneck(s, t)$ for fixed nodes $s, t$.
    The {\bf Single Source Bottleneck Paths} problem ($\SSBP$) asks to compute $\bottleneck(s, v)$ for a single source $s$  and for all $v \in V$.
    The {\bf All Pairs Bottleneck Paths} problem ($\APBP$) asks to compute $\bottleneck(u, v)$ for all $u, v \in V$.
\end{definition}

\subsection{Incremental/Decremental Algorithm for \stBP{}}

We begin with partially dynamic algorithms for bottleneck paths.
First, we present a simple algorithm using a fully dynamic $(s, t)$-reachability data structure to compute partially dynamic $\stBP$, in which we need to maintain an $n$-vertex graph undergoing arbitrary edge insertions and deletions, and we need to support querying whether there is a path from $s$ to $t$. 

\begin{proposition}
    \label{prop:s-t-bp-reachability}
    The partially dynamic $\stBP$ can be solved in $O(m \cdot T(n))$ time, where $T(n) = \tO{n^{\varepsilon_1 + \varepsilon_2} + n^{\omega(1, \varepsilon_1, \varepsilon_2) - \varepsilon_1} + n^{\omega(1, 1, \varepsilon_2) - \varepsilon_2}} = O(n^{1.405})$ \cite{BrandNS19} is the update/query time for fully dynamic $(s, t)$-reachability in an $n$-node graph. 
\end{proposition}

As it seems that \cite{BrandNS19}'s algorithm works against oblivious, adaptive adversaries, our algorithm also does.

\begin{proof}
    We first consider the decremental setting.
    We initialize a fully dynamic $(s, t)$-reachability data structure. We keep adding edges to the data structure from large weight to small, as long as $s$ cannot reach $t$. 
    Once we remove an edge, if $s$ and $t$ are not connected anymore, we again add more edges, from large weight to small, until $s$ can reach $t$ again. 
    It is not difficult to see that the last edge we add to the data structure has the weight equal to $\bottleneck(s, t)$. 
    To bound the running time, observe that each edge is inserted and deleted at most one time each.

    The incremental setting can be solved similarly. 
\end{proof}

\subsection{Incremental Algorithm for \SSBP{}}

Next, we will give an incremental \SSBP{} algorithm. 
First, we formally define the relevant matrix products below, for which there are known subcubic algorithms using FMM.

\begin{definition}[Dominance and Equality Product]
    \label{def:dom-eq-product}
    Given two matrices $A, B$ over a totally ordered set, their dominance product $C = A \dominance B$ is defined by $C_{ij} = |\set{k \mid A_{ik} \leq B_{kj}}|$.
    
    Their equality product is defined by $C_{ij} = |\set{k \mid A_{ik} = B_{kj}}|$
\end{definition}

\begin{definition}[$\minleq$-Product and $\maxleq$-Product]
    \label{def:min-leq-product}
    Given two matrices $A, B$ over a totally ordered set, their $\minleq$ product $C$ is defined by $C_{ij} = \min_{k} \set{B_{kj} \mid A_{ik} \leq B_{kj}}$.
    The $\maxleq$ product $C$ is defined by $C_{ij} = \max_{k} \set{A_{ik} \mid A_{ik} \leq B_{kj}}$.
\end{definition}

\begin{definition}[$\maxmin$-Product]
    \label{def:max-min-product}
    Given two matrices $A, B$ over a totally ordered set, their $\maxmin$ product $C = A \varovee B$ is defined by $C_{ij} = \max_{k} \min\{A_{ik}, B_{kj}\}$.
\end{definition}

As observed in \cite{DBLP:conf/stoc/VassilevskaWY07}, the $\maxmin$ product can be computed by taking the entry-wise maximum of two $\maxleq$ products, $A, B$ and $B^T, A^T$.

Our \SSBP{} algorithm requires the following lemma, which modified an existing algorithm for computing the $(\max, \min)$-product \cite{DuanP09, GrandoniILPU21}.
Formally, we split the computation of the $(\max, \min)$-product into a preprocessing phase and a query phase.

\begin{lemma}
\label{lem:max-min-rect-datastructure}
    Given an $n \times n^b$ matrix $A$ and an $n^b \times n$ matrix $B$, and a parameter $0 \leq g \leq b$, we can preprocess in $\OO(n^{\omega(1, b, 1) + b - g})$ time, so that for every given $i$, we can output the $i$-th row of the $(\max, \min)$-product between $A$ and $B$ in $\OO(n^{1 + g})$ time.
    
    In particular, we can compute the $(\max, \min)$-product between $A$ and $B$ in $\OO(n^{(\omega(1, b, 1) + 2 + b) / 2})$ time, by setting $g = (\omega(1, b, 1) + b - 2) / 2$ and perform $n$ queries. 
\end{lemma}

We recall the definition of row (resp. column) balancing \cite{DuanP09}, which constructs  a matrix $A'$ given any matrix $A$ ensuring that no row (resp. column) of $A'$ has too many finite elements. 

Let $A$ be an $n \times p$ matrix with $m$ finite elements (the others are either all $\infty$ or $-\infty$). 
The {\bf row-balancing} of $A$, denoted $\rowbal(A)$ is a pair of matrices $(A', A'')$ each with at most $k = \ceil{m/n}$ finite elements in each row.
The {\bf column-balancing} of $A$, denoted $\colbal(A)$ is the pair of matrices $(A'^{T}, A''^{T})$ where $(A', A'') = \rowbal(A^T)$.

The row-balancing of $A$ is computed as follows.
Assume that all non-finite elements are $\infty$.
For each row $i$, sort the finite elements of row $i$ in increasing order, and partition this list into parts $T_{i}^1, \dotsc, T_{i}^{a_i}$ where the first $a_i - 1$ parts have exactly $k = \ceil{m/n}$ elements while the last part has at most $k$ elements. 
Now, define $A'[i, j] = A[i, j]$ if $A[i, j] \in T_i^{a_i}$ and $\infty$ otherwise to be the matrix containing all elements in the last part.
Now, since there are at most $\frac{m}{k} \leq n$ parts $T_i^{q}$, we choose an arbitrary injective mapping $\rho: (i, q) \mapsto [n]$ so that the part $T_i^{q}$ is placed into the $\rho(i, q)$-th row of $A''$.
In particular, $A''[i', j] = A[i, j]$ if $i' = \rho(i, q)$ and $A[i, j] \in T_i^{q}$.
Otherwise, $A''[i, j] = \infty$.

Given row and column balanced matrices, \cite{DuanP09} give an efficient algorithm for computing the dominance product (\Cref{def:dom-eq-product}) of sparse matrices.
We repeat the proof for completeness.

\begin{lemma}[Sparse Dominance Product \cite{DuanP09}]
    \label{lemma:sparse-dominance-product}
    Suppose $A$ is an $n^{a} \times n^{b}$ matrix with $m_1$ entries less than $\infty$ and $B$ is an $n^{b} \times n^{c}$ matrix with $m_2$ entries greater than $- \infty$.
    The dominance product $A \dominance B$ can be computed in time $\bigO{\frac{m_1 m_2}{n^b} + n^{\omega(a, b, c)}}$.
\end{lemma}

\begin{proof}
    Let $(A', A'') = \colbal(A)$ be the column-balancing of $A$. 
    Note that $A', A''$ have at most $\ceil{\frac{m_1}{n^b}}$ finite entries per column.
    We construct the following two matrices
    \begin{equation*}
        \hat{A}[i, k] = \begin{cases}
            1 & A''[i, k] < \infty \\
            0 & \otherwise 
        \end{cases},~
        \hat{B}[k, j] = \begin{cases}
            1 & B[k', j] \geq \max T_{k'}^{q'} \andT \rho(k', q') = k \\
            0 & \otherwise 
        \end{cases}.
    \end{equation*}
    We compute the matrix product $\hat{A} \hat{B}$.
    $\hat{A}[i, k] \hat{B}[k, j] = 1$ implies that $A''[i, k] < \infty$ and $B[k', j] \geq \max T_{k'}^{q'}$ where $\rho(k', q') = k$.
    In other words, $B[k', j]$ is larger than all elements in column $k$ of $A''$, which is a subset of column $k'$ of $A$ containing $A''[i, k] = A[i, k']$.
    Therefore, $(\hat{A}\hat{B})[i, j]$ counts the number of $k'$ where $A[i, k'] \leq B[k', j]$ and $A[i, k'] \in T_{k'}^{q'}$ where $q' < a_k$ and $B[k', j]$ dominates all of $T_{k'}^{q'}$.
    It remains to check the following cases:
    \begin{enumerate}
        \item $A[i, k'] \in T_{k'}^{a_{k'}}$.
        For every $B[k', j]$, there are most $\ceil{\frac{m_1}{n^b}}$ rows $i$ with $A[i, k'] \in T_{k'}^{a_{k'}}$.
        We compare $B[k', j]$ with elements $A[i, k'] \in T_{k'}^{a_{k'}}$ and increment $(A \dominance B)[i, j]$ if $A[i, k'] \leq B[k', j]$. %
        This takes $O(m_2 \cdot \frac{m_1}{n^b})$ time.
        \item $B[k', j]$ dominates some elements but not all in $T_{k'}^{q'}$.
        For every $B[k', j]$, this condition can only hold for one set $T_{k'}^{q'}$ as the parts are sorted in increasing order.
        We compare $B[k', j]$ with elements $A[i, k'] \in T_{k'}^{q'}$ and increment $(A \dominance B)[i, j]$ if $A[i, k'] \leq B[k', j]$.
        This takes $O(m_2 \cdot \frac{m_1}{n^b})$ time.
    \end{enumerate}
    
    The overall time complexity is $O(\frac{m_1 m_2}{n^b} + n^{\omega(a, b, c)})$.
\end{proof}

Given that we hope to construct a data structure for the $(\max, \min)$-product, we modify the computation of the sparse dominance product to compute a data structure separating the computation into preprocessing and query stages.

\begin{lemma}[Sparse Dominance Product Data Structure]
    \label{lemma:sparse-dominance-col-query}
    Suppose $A$ is an $n \times n^{b}$ matrix with $m_1$ entries less than $\infty$ and $B$ is an $n^{b} \times n$ matrix with $m_2$ entries greater than $- \infty$.
    There is a data structure with $O(n^{\omega(1, b, 1)})$ preprocessing time, answering column queries of the dominance product $A \dominance B$ in $\bigO{c_j \frac{m_2}{n}}$ time where the $j$-th column of $B$ contains $c_j$ finite elements.
\end{lemma}

\begin{proof}
    In the preprocessing phase, compute the row-balancing $(A', A'') = \colbal(A)$ as in Lemma \ref{lemma:sparse-dominance-product}.
    Similarly, construct matrices $\hat{A}, \hat{B}$, and compute the matrix product $\hat{A}\hat{B}$.
    This requires $n^{\omega(1, b, 1)}$ time.

    Now, suppose there is a query for the $j$-th column of $A \dominance B$.
    By assumption, the $j$-th colum of $B$ has $c_j$ finite elements $< \infty$.
    Our goal is to answer the column query i.e. output $(A \dominance B)[i, j]$ for all $i$.
    As in Lemma \ref{lemma:sparse-dominance-product}, $(\hat{A}\hat{B})[i, j]$ counts the number of $k'$ where $A[i, k'] \leq B[k', j]$ and $A[i, k'] \in T_{k'}^{q'}$ where $q' < a_k$ and $B[k', j]$ dominates all of $T_{k'}^{q'}$.
    It remains to check the following cases:
    \begin{enumerate}
        \item $A[i, k'] \in T_{k'}^{a_{k'}}$.
        For every finite $B[k', j]$, there are most $\ceil{\frac{m_1}{n^b}}$ rows $i$ with $A[i, k'] \in T_{k'}^{a_{k'}}$.
        We compare $B[k', j]$ with elements $A[i, k'] \in T_{k'}^{a_{k'}}$ and increment $(A \dominance B)[i, j]$ if $A[i, k'] \leq B[k', j]$.
        This requires $\bigO{c_j \cdot \frac{m_1}{n^b}}$ time.
        \item $B[k', j]$ dominates some elements but not all in $T_{k'}^{q'}$.
        For every finite $B[k', j]$, this condition can only hold for one set $T_{k'}^{q'}$ as the parts are sorted in increasing order.
        We compare $B[k', j]$ with elements $A[i, k'] \in T_{k'}^{q'}$ and increment $(A \dominance B)[i, j]$ if $A[i, k'] \leq B[k', j]$.
        This requires $\bigO{c_j \cdot \frac{m_1}{n^b}}$ time.
    \end{enumerate}

    The query time complexity is therefore $\bigO{c_j \cdot \frac{m_1}{n^b}}$.
\end{proof}

Using a similar argument, we can also construct a data structure answering row queries.

\begin{corollary}
    \label{cor:sparse-dominance--row-query}
    Suppose $A$ is an $n \times n^{b}$ matrix with $m_1$ entries less than $\infty$ and $B$ is an $n^{b} \times n$ matrix with $m_2$ entries greater than $- \infty$.
    There is a data structure with $O(n^{\omega(1, b, 1)})$ preprocessing answering row queries of the dominance product $A \dominance B$ in $\bigO{r_i \frac{m_2}{n}}$ time where the $i$-th row of $A$ contains $r_i$ finite elements.
\end{corollary}

Now, we prove Lemma \ref{lem:max-min-rect-datastructure}.

\begin{proof}[Proof of Lemma \ref{lem:max-min-rect-datastructure}]
    We modify the $\maxleq$-Product algorithm of \cite{DuanP09}, since the $\maxmin$-Product can be reduced to two $\maxleq$-Product instances.
    Fix a parameter $0 \leq g \leq b$.
    Let $L$ be a sorted list of all entries in $A, B$ and partition $L$ into $G$ groups, $L_1, \dotsc, L_{G}$ such that the first $G - 1$ parts have $\ceil{\frac{2 n^{1 + b}}{G}}$ elements and the final part has the remaining elements.
    For each $r \in [G]$, define the following matrices
    \begin{equation*}
        A_r[i, j] = \begin{cases}
            A[i, j] & A[i, j] \in L_r \\
            \infty & \otherwise
        \end{cases}, ~~ B_r[i, j] = \begin{cases}
            B[i, j] & B[i, j] \in L_r \\
            - \infty & \otherwise
        \end{cases}.
    \end{equation*}
    Compute the row-balancing $(A_r', A_r'') = \rowbal(A_r)$ and dominance products $A_r \dominance B, A_r' \dominance B, A_r'' \dominance B$ as follows.
    For the dominance product, we do not fully compute them, only preprocessing the input as in Lemma \ref{lemma:sparse-dominance-col-query}.
    \begin{enumerate}
        \item To compute $A_r \dominance B$, construct Boolean matrices $\hat{A}_r[i, k] = 1$ if $A[i, k] \in L_r$ and $\hat{B}_r[k, j] = 1$ if $B[k, j] \in \bigcup_{i = r + 1}^{g} L_i$.
        Compute $\hat{A}_r \hat{B}_r$ and the preprocess $A_r \dominance B_r$ according to Lemma \ref{lemma:sparse-dominance-col-query}.
        This takes $\bigO{n^{\omega(1, b, 1)}}$ time.

        \item To compute $A_r' \dominance B$, construct Boolean matrices $\hat{A'}_r[i, k] = 1$ if $A'[i, k] \in L_r$ and $\hat{B}_r$ identically.
        Then, preprocess $A_r' \dominance B_r$ and compute $\hat{A'}_r \hat{B}_r$ in $\bigO{n^{\omega(1, b, 1)}}$ time.

        \item To compute $A_r'' \dominance B$, construct Boolean matrices $\hat{A''}_r[i, k] = 1$ if $A''[i, k] \in L_r$ and $\hat{B}_r$ identically.
        Then, preprocess $A_r'' \dominance B_r$ and compute $\hat{A''}_r \hat{B}_r$ in $\bigO{n^{\omega(1, b, 1)}}$ time.
    \end{enumerate}

    We have computed $\hat{A}_r \hat{B}_r, \hat{A'}_r \hat{B}_r, \hat{A''}_r \hat{B}_r$ and preprocessed $A_r \dominance B_r$ and $A_r' \dominance B_r$ for all $r \in [G]$.
    This concludes the preprocessing phase which has overall time complexity
    \begin{equation*}
        \bigO{G n^{\omega(1, b, 1)}}.
    \end{equation*}

    Then, in the query phase, we hope to compute the $i$-th row of the $\maxleq$ product.
    We begin by computing the $i$-th row of the preprocessed dominance products.
    For each $r \in [G]$, let $c_r$ denote the number of finite elements in the $i$-th row of $A_r$.
    First, we observe 
    \begin{equation*}
        A_r \dominance B = A_r \dominance B_r + \hat{A}_r \hat{B}_r,
    \end{equation*} 
    since the first term counts dominance pairs where both entries are in $L_r$ while the second counts dominance pairs where $B[k, j] \in L_i$ for $i > r$.
    By Lemma \ref{cor:sparse-dominance--row-query}, we can compute the $i$-th row of $A_r \dominance B_r$ in time $\bigO{c_r \cdot \frac{n^{1 + b}}{G n}} = \bigO{\frac{c_r n^{b}}{G}}$ since $B_r$ has $\bigO{\frac{n^{1 + b}}{G}}$ finite elements.
    Adding this to $\hat{A}_r \hat{B}_r$ computes $A_r \dominance B$.
    Similarly, we use the following identities to compute $A_r' \dominance B, A_{r}'' \dominance B_r$ in $\bigO{\frac{c_r n^{b}}{G}}$ time:
    \begin{align*}
        A_r' \dominance B &= A_r' \dominance B_r + \hat{A'}_r \hat{B}_r, \\
        A_r'' \dominance B &= A_r'' \dominance B_r + \hat{A''}_r \hat{B}_r.
    \end{align*}
    We will in fact need to compute all rows of $A_{r}'' \dominance B_r$ indexed by $i' = \rho(i, q)$ for $q \in [a_i - 1]$, accounting for $\ceil{\frac{c_r G}{n^b}}$ rows in $(A_{r}'' \dominance B_r)$.
    However, each of these rows contains at most $\bigO{\frac{n^b}{G}}$ finite elements (and this is also true of the single $i$-th row in $A_r'$.
    Thus, querying all such rows takes time $\bigO{c_r \frac{n^{b}}{G}}$ time.
    Overall computing the rows of these dominance products takes time
    \begin{equation*}
        \bigO{\sum_{r = 1}^{G} \frac{c_r n^{b}}{G}} = \bigO{\frac{n^{1 + b}}{G}}.
    \end{equation*}
    
    For each entry $(i, j)$ we find the largest $r$ such that $(A_r \dominance B)[i, j] > 0$.
    Let $C$ denote the $\maxleq$ product of $A, B$.
    There are now two cases:
    \begin{enumerate}
        \item Suppose $(A_r' \dominance B)[i, j] > 0$.
        Then, since the $i$-th row of $A_r'$ consists of the largest entries of the $i$th row of $A_r$, we have $C[i, j] \in A_r'[i]$.
        By brute-force search over the elements of $A_r'$, we obtain $C[i, j]$ in $\bigO{\frac{n^b}{G}}$ time.

        \item Otherwise, find the largest $q$ such that $(A_r'' \dominance B)[\rho(i, q), j] > 0$ (this is in fact computed in the row-querying phase above) so that $C[i, j] \in T_{i}^{q}$.
        We compute $C[i, j]$ by brute force search in $\bigO{\frac{n^b}{G}}$ time.
    \end{enumerate}
    Since we compute this for all $j$, the row query requires $\bigO{n^{1 + b}{G}}$ time.
    To complete the proof, choose $G = \Theta(n^{b - g})$ so that $n^{g} = \bigTh{\frac{n^b}{G}}$. 
\end{proof}

We now present our incremental algorithm for \SSBP{}, which improves upon the simple framework used for our incremental \nwSSSP{} algorithms.
Recall that we processed updates in batches, recomputing \nwAPSP{} on the graph at the start of each batch.
However, this is wasteful: We do not need \emph{every} distance in the all-pairs solution, only the distances involving endpoints of the inserted edges.
For bottleneck paths, we showed that the corresponding $(\max, \min)$-product is amenable to a data structure with a preprocessing phase and a query phase.
Thus, at the end of each batch, we preprocess the $(\max, \min)$-product on the current graph.
Then, in future batches, whenever the capacity between a pair of vertices is required, we can simply query the relevant data structure.
Since we do not need to query every single pair of vertices, this should save time over a complete \APBP{} computation.

In the following, let $0 \le g \le t \le 1$ be parameters. 
Consider the whole sequence of updates in incremental \SSBP{}, and consider all intervals of the sequence of the form $(2^\ell \cdot (s - 1) \cdot n^t, 2^\ell \cdot s \cdot n^t]$, for $\ell \ge 0, s > 0$. 
These are essentially dyadic intervals with endpoints multiplied by $n^t$. 

\begin{lemma}
\label{lem:inc-SSBP-dyadic-intervals}
    For every interval $L$ of length $n^{T'} = 2^{\ell} n^{t}$ described above, we can create a data structure in $\OO(n^{1+T +g} + n^{T \cdot (3 + \omega) / 2} + n^{(1 + 2T + \omega(1, T, T)) / 2} + n^{\omega(1, T, 1) + T - g})$ preprocessing time where $T = \min\{1, T'\}$, so that given any vertex $v$, we can compute in $\OO(n^{1+g})$ time \SSBP{} to/from $v$ among paths whose latest updated edge is in $L$.
\end{lemma}
\begin{proof}
    We can prove the lemma by induction, and assume that the data structures have been created for all intervals strictly before $L$. 

    Let $U \subseteq V$ be the set of vertices that are among the endpoints of all updated edges in $L$. Clearly, $|U| = O(n^T)$. For each $u \in U$, we query the previous data structures to find out \SSBP{} to/from $u$  among paths whose latest updated edge is strictly before $L$. As the intervals are like dyadic intervals, we can find $O(\log n)$ intervals that cover all updates before $L$, so we only need to query $O(\log n)$ previous data structures. Therefore, the total time for this step is $\OO(n^{1+T +g})$. We use $\bottleneck_{<}(u, v)$ to denote bottleneck distances among paths whose latest updated edge is strictly before $L$.

    We create a $V \times U$ matrix $A$, a $U \times U$ matrix $S$, and a $U \times V$ matrix $B$. All entries of $A, S, B$ correspond to the $\bottleneck_{<}$ distances. Additionally, for entries of $S$, we also update it with the updated edge weights in the interval $L$. Then it is not difficult to see that $(A \varovee (S)^{\varovee n} \varovee B)_{i, j}$ is the bottleneck distance from $i$ to $j$ among paths whose latest updated edge is in $L$. Thus, we can do the following:
    \begin{enumerate}
        \item Compute $(S)^{\varovee n}$ in $\OO(n^{T \cdot (3 + \omega) / 2})$ time. 
        \item Compute $A \varovee (S)^{\varovee n}$ in $\OO(n^{(1 + 2T + \omega(1, T, T))/2})$ time using \cref{lem:max-min-rect-datastructure}. 
        \item Finally, use the data structure part of \ref{lem:max-min-rect-datastructure} to preprocess between $(A \varovee (S)^{\varovee n})$ and $B$ in $\OO(n^{\omega(1, T, 1) + T - g})$ time, so that we can later support row-queries of $(A \varovee (S)^{\varovee n}) \varovee B$ in $\OO(n^{1+g})$ time. 
    \end{enumerate}
    Summing over all terms obtains the desired time complexity.
\end{proof}

We are now ready to give the incremental \SSBP{} algorithm.

\begin{proposition}
Incremental SSBP can be solved in $\OO(n^{(1+\omega)/2})$ amortized time per update. 
\label{prop:inc-ssbp-alg}
\end{proposition}

\begin{proof}
    Besides the data structures in \cref{lem:inc-SSBP-dyadic-intervals}, we also perform the followings after each edge insertion. 
    \begin{enumerate}
        \item Let the new edge be $(u, v)$, and let $\tau$ be the previous timestamp that is a multiple of $n^t$. 
        \item Use the data structures in \cref{lem:inc-SSBP-dyadic-intervals} to compute the SSBP to/from $u, v$ and $s$ among paths whose latest updated edge is on or before $\tau$. This can be done in $\OO(n^{1+g})$ time. Now, it suffices to compute SSBP from $s$ among paths whose latest updated edge is after $\tau$.
        \item Create the following ``three-layer'' graph. 
        The first layer is $\{s\}$, the second layer consists of all vertices that are endpoints of some edge updated after $\tau$, and the third layer is $V$. 
        From the first layer to the second layer, the second layer to the third layer, and within the second layer, we add edges using distances computed in the previous bullet point (all these distances have been computed, either in this update or an earlier update). 
        Between vertices in the second layer, we also add edges updated after $\tau$ (if there was an edge before, we keep the edge with larger weight).
        Then we compute \SSBP{} of this graph in $\OO(n^{1+t})$ time (because the number of updates after $\tau$ is $O(n^t)$, so the number of vertices in the second layer is $O(n^t)$, and the total number of edges is $O(n^{1+t})$). 
    \end{enumerate}
    
    \paragraph{Runtime analysis.} 
    The amortized running time of the preprocessing of the data structures in \cref{lem:inc-SSBP-dyadic-intervals} is 
    \[
    \max_{T \in [t, 1]} \OO(n^{1 +g} + n^{T \cdot (1 + \omega) / 2} + n^{(1 + \omega(1, T, T)) / 2} + n^{\omega(1, T, 1) - g})
    = \OO(n^{1 +g} + n^{(1 + \omega) / 2} + n^{\omega - g})
    \]
    and the running time of the remaining parts described above is
    \[
    \OO(n^{1+g} + n^{1+t}). 
    \]
    Taking $t = g = \frac{\omega - 1}{2}$ gives the claimed $\OO(n^{(1+\omega)/2})$ running time. 
\end{proof}

\begin{remark}[Efficient Algorithm for Fully Dynamic Bottleneck Path in \textbf{Undirected} Graphs]
\rm
Given an undirected graph $G$, it was known how to compute the weights of the bottleneck paths using a maximum spanning tree $T$ of $G$ \cite{hu1961maximum}: The weight of the bottleneck path between two vertices $u$ and $v$ is exactly the weight of the smallest edge on the tree path between $u$ and $v$ in $T$. Therefore, in order to maintain the bottleneck paths for a dynamic graph, it suffices to maintain its maximum spanning tree, and to support minimum weight queries on tree paths. For instance, we can use \cite{holm2015faster} to maintain the maximum spanning tree, and use link-cut tree \cite{sleator1983data} to support minimum weight queries on tree paths, to get a deterministic algorithm with $O(\log^4 n / \log \log n)$ amortized update time, and $O(\log n)$ amortized query time. 
\label{remark:efficient-bp-undirected}
\end{remark}

\subsection{Partially Dynamic Algorithm for \SSBP{}}

Our lower bound in Theorems \ref{thm:s-t-bp-lb} and \ref{thm:s-t-bp-lb-omv-3} crucially uses $n^2$ distinct weights in the constructed graph.
The following proposition shows that this is necessary.
In particular, by an appropriate modification of ES-Trees \cite{EvenS81}, partially dynamic $\SSBP$ can be computed efficiently on graphs with fewer distinct weights.
Moreover, this algorithm is combinatorial.

\begin{proposition} 
    \label{prop:inc-dec-ssbp}
    There is an algorithm solving incremental/decremental \SSBP{} in $\bigtO{m W}$ total-time update time, where $W$ is the number of distinct weights.
    Furthermore, the algorithm answers queries $\bottleneck(s, v)$ in $O(\log W)$ time.
    This algorithm holds against oblivious, adaptive adversaries.
\end{proposition}

\begin{proof}
    We begin by describing the decremental setting.
    Let $G$ denote the initial graph.
    For every distinct weight $w$, consider the graph $G_{w} = \set{e \given \wt(e) \geq w}$.
    Given graph $G$, we initialize a decremental \SSR{} algorithm of \cite{BernsteinPW19} on $G_{w}$ for each $w$.
    Given an edge deletion, we delete $e$ from every instance that contains $w$.
    
    Given a query $\bottleneck(s, v)$, we proceed by binary search to find the maximum $w$ where $v$ is reachable from $s$ on $G_{w}$ and return the maximum such $w$ as the bottleneck capacity between $s, v$.
    
    Since the decremental \SSR{} instance of \cite{BernsteinPW19} answers queries in $O(1)$ time, we can answer queries in $O(\log W)$ time via binary search.
    Since there are at most $W$ instances, the total time required by the algorithm is $\tO{m W}$.

    Note that we can handle the incremental setting since incremental \SSR{} can also be solved in total time $\tO{m}$.
\end{proof}

\subsection{Fully Dynamic All-Pairs Bottleneck Paths}

We present our fully dynamic $\APBP$ algorithm.

\begin{proposition}
    \label{prop:fully-dynamic-apbp}
    Fully dynamic $\APBP$ can be solved in $\OO(n^2)$ amortized time per update deterministically.  
\end{proposition}
\begin{proof}

We first describe the intuition of our algorithm. Suppose all the edge weights are integers from $[1, 2n^2]$. Let $M$ be some sufficiently large integer, and we replace the weight $w$ of an edge by $M^{2n^2 - w}$. Consider a path from $u$ to $v$ where the edge weights on the path are $w_1, \ldots, w_\ell$, then the total length of the path under the new weight is $\sum_{i} M^{2n^2 - w_i}$, which is dominated by the edge with the smallest value of $w_i$. Thus, the shortest path from $u$ to $v$ under the new weight will maximize the minimum old edge weight, i.e., it will correspond to a bottleneck path. Intuitively, we could run Demetrescu and Italiano's dynamic APSP algorithm \cite{DemetrescuI04} under the new weight to achieve $\OO(n^2)$ amortized running time, assuming we can perform arithmetic operations between the new (huge) weights in constant time. Below, we describe algorithm in more details and the way to remove the assumptions. 

We first use online list labelling to remove the assumption that all edge weights are from $[1, 2n^2]$. 

\begin{theorem}[\cite{itai1981sparse}]
\label{thm:online-list-labelling}
    We can deterministically store a set of $t$ dynamically-changing items in an array of length $2t$, so that the sorted orders of the items are preserved, and we only move $\OO(1)$ items per operation in amortization. 
\end{theorem}

We use \cref{thm:online-list-labelling} to store the original edge weights of the graph as the items, and use the indices of the items stored in the arrays as their new weights. 
As the orders of indices of the items preserve the order of the items, this is without loss of generality for bottleneck paths. 
Now for every edge update in the graph, we might need to perform up to $\OO(1)$ weight updates in amortization, which is affordable. 
From now on, we assume all the edge weights are integers from $[1, n^2]$. 

Next, we aim to use Demetrescu and Italiano's dynamic APSP algorithm \cite{DemetrescuI04} in a white-box way. 
In particular, if we simply store the path weights as described in the intuition, it would be too costly. 
Instead, we use some data structure to maintain the path weights. 

The weight of any path in our graph (under the new weights described in the intuition) can be represented as a length $2n^2$ array, where the value on the $i$-th entry of the array equals the number of times the path use an edge of weight $i$. 
Therefore, say this array is $(a_i)_{i=1}^{2n^2}$, then the weight of the path is $\sum_{i=1}^{2n^2} a_i M^{2n^2 - i}$, for some large integer $M$. 
We use a persistent segment tree \cite{DBLP:journals/jcss/DriscollSST89} to maintain such arrays efficiently.

Next, we run Demetrescu and Italiano's dynamic APSP algorithm \cite{DemetrescuI04} that has $\OO(n^2)$ amortized running time per operation, but we use the data structure to support all operations their algorithm performs on path weights. 
Initially, we create a persistent segment tree maintaining the all-zero array, representing the empty path. Then, our data structure needs to support the followings efficiently (i.e., in $\OO(1)$ time per operation). 

\begin{enumerate}
    \item \textbf{Concatenate one edge to an existing path. } Say the new edge has weight $w$, and the array corresponding to the existing path is $a$, then the array corresponding to the new path is $a$, with $a_w$ incremented by $1$. Therefore, we can simply copy the segment tree for the existing path (as we maintain persistent segment trees, this takes $\OO(1)$ time), and then update the $w$-th entry of the new segment tree in $\OO(1)$ time. 
    \item \textbf{Compare the weight of two paths. } For two paths whose weight correspond to arrays $a$ and $b$ respectively, their weights are $\sum_{i=1}^{2n^2} a_i M^{2n^2 - i}$ and $\sum_{i=1}^{2n^2} b_i M^{2n^2 - i}$, for some large integer $M$. To compare these two weights, it suffices to find the longest common prefix of the arrays $a$ and $b$, and then compare the first differing position of $a$ and $b$.  

    Then similar to Aho, Hopcroft and Ullman's algorithm for Tree Isomorphism \cite{aho1974design}, we assign all subtrees of all persistent segment tree an identifying number, so that isomorphic trees will have the same identifying number while distinct trees will have a different number. We maintain a perfect hash function $H$ that maps integers or pairs of identifying numbers to identifying numbers, and we can maintain this hash function using a binary search tree. Leaf nodes in the segment trees represent an entry of an array, so it is associated with an integer. If this integer is $c$, then its identifying number would be $H(c)$. For internal nodes whose left subtree and right subtree have identifying numbers $L$ and $R$ respectively, its identifying number would be $H((L, R))$. 

    Then, we can perform binary search on two persistent segment trees augmented with identifying numbers to find the longest common prefix of the underlying arrays: To find the longest common prefix between two nodes $N_1$ and $N_2$, we first compare the identifying numbers of their left subtrees. If these two numbers are different, we know that there must be a different array entry corresponding to the left subtree, so we recursively compare the left subtrees; otherwise, we recursively compare the right subtrees. As the height of the segment trees is $\OO(1)$, this takes $\OO(1)$ time. 
\end{enumerate}
\end{proof}

\subsection{Conditional Lower Bounds for Bottleneck Paths}

We now present our conditional lower bound for partially dynamic bottleneck paths.
We observe that due to Remark \ref{remark:efficient-bp-undirected}, our lower bounds hold only for directed graphs.
In contrast, our lower bounds for shortest paths hold for undirected graphs as well.

\begin{theorem}
    \label{thm:s-t-bp-lb}
    Under the Combinatorial 4-Clique hypothesis, any combinatorial algorithm computing incremental/decremental $\stBP$ on directed graphs requires $n^{4 - o(1)}$ total update and query time.
\end{theorem}

\begin{proof}
    We describe the reduction for incremental $\stBP$, noting that in the decremental case we can simply run the reduction in reverse.
    
    Suppose for contradiction there is an algorithm $\innerAlg$ for incremental $\stBP$ with total time $O(n^{4 - c})$ for some $c > 0$.
    Consider a 4-Clique Detection instance with vertex sets $A, B, C, D$ of size $n$.
    Throughout the reduction, assume that the vertex sets $A, B, C, D$ are indexed from $0$ to $n - 1$.
    We design an algorithm with total time $O(n^{4 - c})$ for some positive $c > 0$, contradicting the Combinatorial 4-Clique hypothesis.

    We construct a graph with vertices
    \begin{equation*}
        \set{s} \cup (A_1 \cup A_2 \cup A_3) \cup (B_1 \cup B_2 \cup B_3) \cup (C_1 \cup C_2 \cup C_3) \cup (\hat{A}_1 \cup \hat{A}_2 \cup \hat{A}_3) \cup \set{t}.
    \end{equation*}
    Each vertex set $A_1, A_2, A_3, B_1, B_2, B_3, C_1, C_2, C_3, \hat{A}_1, \hat{A}_2$ and $\hat{A}_3$ has $n$ vertices.
    For each $a \in A$, we create a copy $a^{(i)} \in A_i$ and a copy $\hat{a}^{(i)} \in \hat{A}_i$ for $i \in \set{1, 2, 3}$.
    Similarly, for each $b \in B$ and $c \in C$, we create a copy $b^{(i)} \in B_i$ and $c^{(i)} \in C_i$ for $i \in \set{1, 2, 3}$.
    Initially, insert edges $(s, a^{(1)})$ and $(\hat{a}^{(3)}, t)$ with $\infty$ weight for all $a \in A$.
    Furthermore, insert edges with $\infty$ weight between $(a^{(3)}, b^{(1)})$ (resp. $(b^{(3)}, c^{(1)})$, $(c^{(3)}, \hat{a}^{(1)})$) if and only if $(a, b) \in E$ (resp. if $(b, c) \in E$,  $(c, a) \in E$). This creates a graph with $O(n)$ vertices

    We now proceed to more edge insertions of the reduction.
    In the outer loop we iterate over $d_i \in D$ in increasing order.
    \begin{enumerate}
        \item For each $b_j \in \neighborhood(d_i)$, insert edges $\left(b_j^{(1)}, b_{j + i \mod n}^{(2)}\right)$ and $\left(b_{j + i \mod n}^{(2)}, b_j^{(3)} \right)$ with weight $(i + 1) \cdot  n$.
        \item For each $c_j \in \neighborhood(d_i)$, insert edges $\left(c_j^{(1)}, c_{j + i \mod n}^{(2)}\right)$ and $\left(c_{j + i \mod n}^{(2)}, c_j^{(3)} \right)$ with weight $(i + 1) \cdot n$.
        \item In the inner loop, we iterate over $a_{k} \in A$ in increasing order.
        \begin{enumerate}
            \item If $a_k \in \neighborhood(d_i)$, insert the following edges of weight $i \cdot n + (k + 1)$:
            \begin{equation*}
                \left(a_{k}^{(1)}, a_{k + i \mod n}^{(2)}\right), \left(a_{k + i \mod n}^{(2)}, a_{k}^{(3)}\right), \left(\hat{a}_{k}^{(1)}, \hat{a}_{k + i \mod n}^{(2)}\right), \left(\hat{a}_{k + i \mod n}^{(2)}, \hat{a}_{k}^{(3)}\right).
            \end{equation*}
            \item Query $\innerAlg$ to get $\bottleneck(s, t)$. If $\bottleneck(s, t) \geq i \cdot n + (k + 1)$, return $\true$.
        \end{enumerate}
    \end{enumerate}
    Then, return $\false$ after all iterations.

    Clearly, the bottleneck of this algorithm is to run $\innerAlg$, so the running time is $O(n^{4-c})$. 

    We show the following lemma in order to show correctness:
    
\begin{lemma}
    \label{lemma:s-t-bottleneck-4-clique-equiv}
    Consider the query when iterating over $d_i$ and $a_k$.
    The result of the query is $\bottleneck(s, t) \geq i \cdot n + (k + 1)$ if and only if $(d_i, a_k)$ are in a 4-clique.
\end{lemma}

\begin{proof}
    Suppose $d_i, a_k$ are in a 4-clique with vertices $b_{j}, c_{\ell}$.
    Then, the following $(s, t)$-bottleneck path exists in the graph and has capacity $i \cdot n + (k + 1)$:
    \begin{equation*}
        \left(s, a_{k}^{(1)}, a_{k + i \mod n}^{(2)},  a_{k}^{(3)}, b_{j}^{(1)}, b_{j + i \mod n}^{(2)}, b_{j}^{(3)}, c_{\ell}^{(1)}, c_{\ell + i \mod n}^{(2)}, c_{\ell}^{(3)}, \hat{a}_{k}^{(1)}, \hat{a}_{k + i \mod n}^{(2)},  \hat{a}_{k}^{(3)}, t\right).
    \end{equation*}

    We verify that this path exists and has the required capacity.
    $(s, a_{k}^{(1)})$ has capacity $\infty$.
    Since $a_{k} \in N(d_i)$, each edge in the sub-path $\left(a_{k}^{(1)}, a_{k + i \mod n}^{(2)}, a_{k}^{(3)} \right)$ exists and has capacity $i \cdot n + (k + 1)$.
    Next, $(a_k, b_j) \in E$ implies that the edge $(a_{k}^{(3)}, b_{j}^{(1)})$ exists with $\infty$ capacity.
    Since $b_{j} \in N(d_i)$, each edge in the sub-path $\left(b_{j}^{(1)}, b_{j + i \mod n}^{(2)}, b_{j}^{(3)} \right)$ exists and has capacity $(i + 1) \cdot n$.
    Then, $(b_j, c_{\ell}) \in E$ implies that the edge $(b_{j}^{(3)}, c_{\ell}^{(1)})$ exists with $\infty$ capacity.
    Since $c_{\ell} \in N(d_i)$, each edge in the sub-path $\left(c_{\ell}^{(1)}, c_{\ell + i \mod n}^{(2)}, c_{\ell}^{(3)} \right)$ exists and has capacity $(i + 1) \cdot n$.
    Then, $(c_{\ell}, a) \in E$ implies that the edge $(c_{\ell}^{(3)}, \hat{a}_{k}^{(1)})$ exists with $\infty$ capacity.
    Finally, as $a_{k} \in N(d_i)$, we have the sub-path $\left(\hat{a}_{k}^{(1)}, \hat{a}_{k + i \mod n}^{(2)}, \hat{a}_{k}^{(3)} \right)$ with capacity $i \cdot n + (k + 1)$.
    We conclude by observing that the edge $(\hat{a}_{k}^{(3)}, t)$ has $\infty$ capacity.

    Conversely, suppose $\bottleneck(s, t) \geq i \cdot n + (k + 1)$.
    Then, since edges are directed left to right, there is a path of the form
    \begin{equation*}
        \left(s, a_{i_1}^{(1)}, a_{i_2}^{(2)},  a_{i_3}^{(3)}, b_{i_4}^{(1)}, b_{i_5}^{(2)}, b_{i_6}^{(3)}, c_{i_7}^{(1)}, c_{i_8}^{(2)}, c_{i_9}^{(3)}, \hat{a}_{i_{10}}^{(1)}, \hat{a}_{i_{11}}^{(2)},  \hat{a}_{i_{12}}^{(3)}, t\right)
    \end{equation*}
    
    achieving this capacity.
    At the time of the query in the $i$-th iteration over $D$ and $k$-th iteration over $A$ in the inner-loop, the maximum weight of any edge between $A_1 \times A_2$, $A_2 \times A_3$, $\hat{A}_1 \times \hat{A}_2$, and $\hat{A}_2 \times \hat{A}_3$ is $i \cdot n + (k + 1)$.
    Furthermore, the only edges attaining this weight are $\left(a_{k}^{(1)}, a_{k + i \mod n}^{(2)}\right)$, $\left(a_{k + i \mod n}^{(2)}, a_{k}^{(3)}\right)$, $\left(\hat{a}_{k}^{(1)}, \hat{a}_{k + i \mod n}^{(2)}\right)$, and $\left(\hat{a}_{k + i \mod n}^{(2)}, \hat{a}_{k}^{(3)}\right)$.
    If these edges exist, then $(a_k, d_i) \in E$ and $k = i_1 = i_3 = i_{10} = i_{12}$.
    The maximum weight of any edge between $B_1 \times B_2$ and $B_2 \times B_3$ is $(i + 1) \cdot n$.
    Furthermore, the only edges attaining this weight are $\left(b_{j}^{(1)}, b_{j + i \mod n}^{(2)}\right)$, $\left(b_{j + i \mod n}^{(2)}, b_{j}^{(3)}\right)$ for $b_j \in N(d_i)$.
    In particular, we have $i_4 = i_6 = j$ for some $b_j \in N(d_i)$.
    By a similar argument, $i_7 = i_9 = \ell$ for some $c_{\ell} \in N(d_i)$.
    
    Finally, it remains to show $a_k, b_j, c_{\ell}$ form a triangle.
    Since $\wt(a_{k}^{(3)}, b_{j}^{(1)}) = \infty$ we have $(a_{k}, b_{j}) \in E$.
    A similar argument shows that $(a_{k}, c_{\ell}), (b_j, c_{\ell}) \in E$.
\end{proof}

    From Lemma \ref{lemma:s-t-bottleneck-4-clique-equiv}, the above procedure solves the 4-Clique Detection instance, contradicting the Combinatorial 4-Clique hypothesis.
\end{proof}

Using an essentially identical reduction, we show that under the $\OMvThree$ hypothesis, any algorithm computing $\stBP$ with polynomial preprocessing requires $n^{\omega + 1 - o(1)}$ total update  time.

\begin{theorem}
    \label{thm:s-t-bp-lb-omv-3}
    Under the $\OMvThree$ hypothesis, any algorithm computing incremental/decremental $\stBP$ on directed graphs with polynomial preprocessing time requires $n^{\omega + 1 - o(1)}$ total update  time.
\end{theorem}

\begin{proof}
    Suppose for contradiction there is an incremental $\stBP$ algorithm $\innerAlg$ with polynomial preprocessing time and total update  time $O(n^{\omega + 1 - c})$ for some $c > 0$.
    The decremental case can be handled using similar modifications as \Cref{thm:s-t-sp-node-weight-lb-omv-3}.
    We design an efficient algorithm for $\OMvThree$.

    In the preprocessing phase we receive a Boolean matrix $A$.
    We construct a graph with vertices
    \begin{equation*}
        \set{s} \cup (A_1 \cup A_2 \cup A_3) \cup (B_1 \cup B_2 \cup B_3) \cup (C_1 \cup C_2 \cup C_3) \cup (\hat{A}_1 \cup \hat{A}_2 \cup \hat{A}_3) \cup \set{t}.
    \end{equation*}
    Each vertex set $A_1, A_2, A_3, B_1, B_2, B_3, C_1, C_2, C_3, \hat{A}_1, \hat{A}_2$, $\hat{A}_3$ has $n$ vertices, each indexed by $[n]$.
    Insert edges $(s, a^{(1)})$ and $(\hat{a}^{(3)}, t)$ with $\infty$ weight for all $a \in A$.
    Furthermore, insert edges with $\infty$ weight between $(a^{(3)}, b^{(1)})$ (resp. $(b^{(3)}, c^{(1)})$, $(c^{(3)}, \hat{a}^{(1)})$) if and only if $A[a, b] = \true$ (resp. if $A[b, c] = \true$, $A[a, c] = \true$).
    We then run the preprocessing step of $\innerAlg$ on this $O(n)$-vertex graph.

    We now proceed to the dynamic phase of the reduction.
    Throughout, we assume that both queries and coordinates are indexed starting at $0$ (just as vertices as indexed from $0$ in Theorem \ref{thm:s-t-bp-lb}).
    Suppose we have received the $i$-th query, $\vec{u}_i, \vec{v}_i, \vec{w}_i$.
    \begin{enumerate}
        \item For all $j$ with $\vec{v}_i[j] = \true$, insert edges $\left(b_j^{(1)}, b_{j + i \mod n}^{(2)}\right)$ and $\left(b_{j + i \mod n}^{(2)}, b_j^{(3)} \right)$ with weight $(i + 1) \cdot  n$.
        \item For all $j$ with $\vec{w}_i[j] = \true$, insert edges $\left(c_j^{(1)}, c_{j + i \mod n}^{(2)}\right)$ and $\left(c_{j + i \mod n}^{(2)}, c_j^{(3)} \right)$ with weight $(i + 1) \cdot n$.
        \item Now, we iterate over the coordinates $k$ of vector $\vec{v}_i$ in increasing order.
        \begin{enumerate}
            \item If $\vec{u}_i[k] = \true$, insert edges of weight $i \cdot n + (k + 1)$ between \begin{equation*}
                \left(a_{k}^{(1)}, a_{k + i \mod n}^{(2)}\right), \left(a_{k + i \mod n}^{(2)}, a_{k}^{(3)}\right), \left(\hat{a}_{k}^{(1)}, \hat{a}_{k + i \mod n}^{(2)}\right), \left(\hat{a}_{k + i \mod n}^{(2)}, \hat{a}_{k}^{(3)}\right).
            \end{equation*}
            \item Compute $b(s, t)$ by querying $\innerAlg$. If $\bottleneck(s, t) \geq i \cdot n + (k + 1)$, return $\true$.
        \end{enumerate}
        Otherwise, return $\false$ if none of the queries return $\true$.
    \end{enumerate}

    During the dynamic phase, the bottleneck of the algorithm is running $\innerAlg$, so the total update and query time is $O(n^{\omega + 1 - c})$.

    The following lemma shows that our algorithm correctly computes each $\OMvThree$ query.
    
    \begin{restatable}{lemma}{stBPOMvThreeEquiv}
        \label{lemma:st-bp-omv-3-equiv}
        Consider the query $(\vec{u}_i, \vec{v}_i, \vec{w}_i)$ and the query after updating the $k$-th coordinate of $\vec{u}$.
        The result of the query is $\bottleneck(s, t) \geq i \cdot n + (k + 1)$ if and only if
        \begin{equation*}
            \bigvee_{j, \ell} \left( \vec{u}_i[k] \wedge \vec{v}_i[j] \wedge \vec{w}_i[\ell] \wedge A[k, j] \wedge A[j, \ell] \wedge A[\ell, k] \right) = \true.
        \end{equation*}
    \end{restatable}
    
    Since this lemma is essentially identical to Lemma \ref{lemma:s-t-bottleneck-4-clique-equiv}, we defer the proof to Appendix \ref{app:omv-3-lb}.
    
    Note that at the query after the $i$-th query and the $k$-th iteration of the inner loop, we have constructed the same graph as in Theorem \ref{thm:s-t-bp-lb}.
    In particular, Lemma \ref{lemma:st-bp-omv-3-equiv} shows that this query returns true if and only if
    \begin{equation*}
        \bigvee_{j, \ell} \left( \vec{u}_i[k] \wedge \vec{v}_i[j] \wedge \vec{w}_i[\ell] \wedge A[k, j] \wedge A[j, \ell] \wedge A[\ell, k] \right) = \true.
    \end{equation*}
    Thus, if the above is satisfied for any $k$, we return $\true$.
    Otherwise, we return $\false$, answering the query correctly in either case.
    Therefore, using $\innerAlg$ we have obtained an algorithm for the $\OMvThree$ instance with polynomial preprocessing and total update and query time $O(n^{\omega + 1 - c})$, contradicting the $\OMvThree$ hypothesis.
\end{proof}

Furthermore, we show that any partially dynamic $\SSBP$ algorithm computes the min-witness matrix product with $n$ updates.
In particular, the following reduction shows that any algorithm with amortized update time $O(n^{1.5-c})$ for $c > 0$ gives an improved algorithm for computing the min-witness product when $\omega = 2$. 
We leave as an interesting open question how to extend the following lower bound to $n^2$ updates.

\begin{theorem}
    \label{thm:ssbp-lb-min-witness}
    Under the Minimum-Witness Product hypothesis, any algorithm computing incremental/decremental $\SSBP$ on directed graphs requires $n^{2.5 - o(1)}$ total update time over $n$ updates.
    Equivalently, any algorithm computing $\SSBP$ on directed graphs requires $n^{1.5 - o(1)}$ amortized update time over $n$ updates.
\end{theorem}

\begin{proof}
    Suppose for contradiction there is an incremental $\SSBP$ algorithm with pre-processing time $O(n^{2.5 - c})$ and amortized update time $O(n^{1.5 - c})$ for some $c > 0$, describing modifications for the decremental version as necessary.
    We design an algorithm computing a min-witness product of $n \times n$ matrices $A, B$ in $O(n^{2.5 - c})$ time.

    Construct a $4$ layered graph
    \begin{equation*}
        \set{s} \cup U \cup V \cup W,
    \end{equation*}
    where $|U| = |V| = |W| = n$ and index each vertex set from $1$ to $n$.
    We initialize the graph as follows.
    For every entry $A[i, k] = \true$, add an edge $(u_i, v_k)$ with weight $i \cdot n - (k - 1)$.
    For every entry $B[k, j] = \true$, add an edge $(v_k, w_j)$ with weight $\infty$.
    In the decremental reduction, we additionally add edges $(s, v_i)$ of weight $\infty$ for all $i$.
    This creates a graph with $n$ vertices, so we can preprocess in $O(n^{2.5 - c})$ time.

    We now proceed to the dynamic phase.
    We iterate over $i \in [n]$ in increasing order.
    For each $i$, we add an edge $(s, u_i)$ of weight $\infty$ and query $\SSBP$.
    Suppose $\bottleneck(s, w_j) > (i - 1) \cdot n$.
    Then, we note the minimum witness of $i, j$ as $i \cdot n + 1 - \bottleneck(s, w_j)$.
    Otherwise, we note that there is no witness for $i, j$.
    In the decremental setting, we iterate over $[n]$ in decreasing order and instead delete the edge $(s, u_i)$ before each query.

    We argue that this correctly computes the minimum-witness product.
    Consider the query after inserting edge $(s, u_i)$.
    We claim $\bottleneck(s, w_j) > (i - 1) \cdot n$ if and only if there is some $k$ for which $A[i, k] = B[k, j] = \true$, and furthermore this path has bottleneck capacity $i \cdot n - (k - 1)$.
    In particular, this shows that $i \cdot n + 1 - \bottleneck(s, w_j)$ is precisely the minimum such $k$ (if such a path exists) and otherwise there is no witness.
    
    Suppose there is some $k$ for which $A[i, k] = B[k, j] = \true$.
    Then, there is the path $(s, u_i, v_k, w_j)$ with bottleneck capacity $i \cdot n + k$.
    Conversely, if there is a path with bottleneck capacity at least $(i - 1) \cdot n$, then it must have the form $(s, u_i, v_k, w_j)$ for some $k$ since the graph is directed and there are no edges from $s_{i'}$ for $i' > i$. 
    Thus, for this $k$ it must be the case that $A[i, k] = B[k, j] = \true$, as desired.

    The overall reduction makes $n$ updates and computes the minimum-witness product in $O(n^{2.5 - c})$, contradicting the Minimum-Witness Product hypothesis.
\end{proof}

\section{Earliest Arrivals (or Non-decreasing Paths)}
\label{sec:earliest-arrivals}

For the earliest arrivals problem, we obtain a simple, efficient algorithm for the incremental single-source problem via a reduction to reachability (Theorem \ref{thm:partially-dynamic-SSEA}).
In contrast, under the weight dynamic model, we show that any partially dynamic algorithm can be reduced from $4$-Clique Detection.
The earliest arrivals problem (sometimes also called non-decreasing paths problem) is defined below.

\begin{definition}[Earliest Arrivals]
    \label{def:earliest-arrivals}
    Let $G = (V, E, \wt)$ be a directed, weighted graph. 
    For any path $P$ is a {\bf valid itinerary} if the edges on the path are non-decreasing in weight.
    For a valid itinerary, let $\arrival(P) = \max_{e \in P} \wt(e)$ be the weight of the final edge.
    For any pair of nodes $u, v$, let $\arrivalPath(u, v)$ denote the {\bf earliest arrival}, the valid itinerary between $u, v$ minimizing $\arrival(P)$ (breaking ties arbitrarily).
    Let $\arrival(u, v) = \arrival(\arrivalPath(u, v))$.

    The {\bf $(s, t)$-Earliest Arrival} problem ($\stEA$) asks to compute $\arrival(s, t)$ for fixed nodes $s, t$.
    The {\bf Single Source Earliest Arrivals} problem ($\SSEA$) asks to compute $\arrival(s, v)$ for for a single source $s$ and all $v \in V$.
    The {\bf All Pairs Earliest Arrivals} problem ($\APEA$) asks to compute $\arrival(u, v)$ for all $u, v \in V$.
\end{definition}

\subsection{Decremental Algorithm for \SSEA{}}

We give a simple near-linear total time algorithm for partially dynamic \SSEA{} using a reduction from \SSEA{} to Single-Source Reachability (SSR) in sparse graphs:
\begin{lemma}[Reduction from \SSEA{} to SSR]
\label{lem:reduction-from-SSEA-to-SSR}
We can reduce an instance of SSEA in $n$-vertex $m$-edge graph $G$ to an instance of SSR in an $O(m)$-vertex $O(m)$-edge graph. The reduction runs in near-linear time. 
\end{lemma}
\begin{proof}
    Let $G=(V, E, \wt)$ be the graph for SSEA. We create the following graph for SSR.
    \begin{enumerate}
        \item For every $e \in E$, we create two vertices $p_e$ and $q_e$. The vertex set of the SSR instance is $\{s\} \cup \{p_e, q_e\}_{e \in E}$. 
        \item For every $e \in E$, we add an edge from $p_e$ to $q_e$. 
        \item For every $v \in E$, let $i_1, i_2, \ldots, i_x$ be the set of incoming edges, and let $o_1, o_2, \ldots, o_y$ be the set of outgoing edges. We sort the vertices $q_{i_1}, q_{i_2}, \ldots, q_{i_x}, p_{o_1}, p_{o_2}, \ldots, p_{o_y}$ based on the weights of their corresponding edges (if there is a tie, we rank incoming edges before outgoing edges since we require the path to be non-decreasing), and add a chain from the smallest to the largest. This way, $q_{i_a}$ can reach $p_{o_b}$ if and only if $\wt(i_a) \leq \wt(o_b)$. 
        \item For every $(s, v) \in E$, we add an edge from $s$ to $p_{(s, v)}$. 
    \end{enumerate}
    
    We observe that the reduction satisfies the following property.
    
    \begin{claim}
        \label{claim:ssr-ssea-equiv}
        Let $e_1, e_2, \ldots, e_x$ be the set of incoming edges of $v$. 
        Then
        \begin{equation*}
            \arrival(s, v) = \min \set{\wt(e_i) \given p_{e_i} \textrm{ is reachable from $s$}}.
        \end{equation*}
    \end{claim}

    Then, to answer a query it suffices to keep track of the minimum $e_i$ such that $p_{e_i}$ is reachable from $s$.
    To bound the run-time, note that the set of reachable $p_{e_i}$ can only increase/decrease, so maintaining all the relevant heaps requires only $\tO{m}$ total time.
\end{proof}

\begin{proof}[Proof of \Cref{claim:ssr-ssea-equiv}]
    Suppose $p_{e_i}$ is reachable from $s$.
    Then, there is a valid itinerary (non-decreasing path) from $s$ to the origin of $e_i$ in the \SSEA{} instance, since every edge $(p_e, q_e)$ taken is an edge in the graph and they must be taken in non-decreasing order by the chain construction.
    In particular, $q_{e_i}$ is reachable from $s$ so that $v$ is reachable from $s$ via a non-decreasing path ending at $e_i$, or $\arrival(s, v) \leq \wt(e_i)$.

    Otherwise, let $P = \arrivalPath(s, v)$ to be the earliest arrival path with final edge $e_j$.
    We claim $p_{e_j}$ is reachable from $s$.
    This follows easily as the path is a valid itinerary (non-decreasing path) so that there is a path in the constructed graph that follows the appropriate edges $(p_e, q_e)$ as well as edges within the chain.
\end{proof}

By maintaining the reduction dynamically, we can obtain a faster algorithm for partially dynamic \SSEA{}. 
\begin{theorem}
\label{thm:partially-dynamic-SSEA}
There is an algorithm for incremental/decremental \SSEA{} in $\OO(m)$ total time. 
This algorithm holds against an oblivious, adaptive adversary.
\end{theorem}
\begin{proof}
    We can maintain the reduction in \cref{lem:reduction-from-SSEA-to-SSR} dynamically: For inserting $e$, we need to insert the edge from $p_e$ to $q_e$, and appropriately insert $p_e$ and $q_e$ to the chains (note that we do not need to delete edges that are already on the chain, i.e., if we have a chain $1 \rightarrow 2 \rightarrow 4 \rightarrow 5$, and need to insert $3$, we just add $(2, 3)$ and $(3, 4)$ and do not need to delete $(2, 4)$). For deleting $e$, we only need to delete the edge from $p_e$ to $q_e$, and do not need to update the chain. 

    Incremental SSR can be solved (easily) in $O(m)$ total time. Decremental SSR can be solved in $\OO(m)$ total time \cite{BernsteinPW19}. 
\end{proof}

\subsection{Conditional Lower Bound for Weight Dynamic Earliest Arrivals}

\begin{theorem}
    \label{thm:s-t-ea-lb}
    Under the Combinatorial 4-Clique hypothesis, any combinatorial algorithm computing $\stEA$ with incremental/decremental weights requires $n^{2 - o(1)}$ amortized update time. 
    Furthermore, each edge weight is modified only once.
\end{theorem}

\begin{proof}
    We describe the reduction for incremental $\stEA$ (weight non-decreasing), noting that the decremental reduction can be obtained by reversing the reduction.
    
    Suppose for contradiction there is an algorithm $\innerAlg$ with  $O(n^{2 - c})$ amortized update time for some $c > 0$.
    Consider a 4-Clique Detection instance with vertex sets $A, B, C, D$ of size $n$.
    Throughout the reduction, assume that the vertex sets $A, B, C, D$ are indexed from $1$ to $n$.
    We design an algorithm with total time $O(n^{4 - c})$ for some positive $c > 0$, contradicting the Combinatorial 4-Clique hypothesis.

    We construct a graph with vertices
    \begin{equation*}
        \set{s} \cup (A_1 \cup A_2 \cup A_3) \cup (B_1 \cup B_2 \cup B_3) \cup (C_1 \cup C_2 \cup C_3) \cup (\hat{A}_1 \cup \hat{A}_2 \cup \hat{A}_3) \cup \set{t}.
    \end{equation*}
    Each vertex set $A_1, A_2, A_3, B_1, B_2, B_3, C_1, C_2, C_3, \hat{A}_1, \hat{A}_2$ and $\hat{A}_3$ has $n$ vertices.
    For each $a \in A$, we create a copy $a^{(i)} \in A_i$ and a copy $\hat{a}^{(i)} \in \hat{A}_i$ for $i \in \set{1, 2, 3}$.
    Similarly, for each $b \in B$ and $c \in C$, we create a copy $b^{(i)} \in B_i$ and $c^{(i)} \in C_i$ for $i \in \set{1, 2, 3}$.
    Initially, insert edges $(s, a^{(1)})$ with weight $1$ and edges $(\hat{a}^{(3)}, t)$ with weight $9$ for all $a \in A$.
    Furthermore, insert edges with weight $3$  between $(a^{(3)}, b^{(1)})$ (resp. weight $5$ between $(b^{(3)}, c^{(1)})$, weight $7$ between $(c^{(3)}, \hat{a}^{(1)})$) if and only if $(a, b) \in E$ (resp. $(b, c) \in E$,  $(c, a) \in E$).
    We also insert all edges between $A_1 \times A_2, A_2 \times A_3, B_1 \times B_2, B_2 \times B_3, C_1 \times C_2, C_2 \times C_3, \hat{A}_1 \times \hat{A}_2, \hat{A}_2 \times \hat{A}_3$ with weight $0$. This creates a graph with $O(n)$ vertices.

    We now proceed to the more updates of the reduction.
    In the outer loop we iterate over $d_i \in D$ in increasing order.
    \begin{enumerate}
        \item For each $b_j \in \neighborhood(d_i)$, increase the weights of $\left(b_j^{(1)}, b_{j + i \mod n}^{(2)}\right)$ and $\left(b_{j + i \mod n}^{(2)}, b_j^{(3)} \right)$ to $4$.
        \item For each $c_j \in \neighborhood(d_i)$, increase the weights of $\left(c_j^{(1)}, c_{j + i \mod n}^{(2)}\right)$ and $\left(c_{j + i \mod n}^{(2)}, c_j^{(3)} \right)$ to $6$.
        \item In the inner loop, we iterate over $a_{k} \in A$ in increasing order.
        \begin{enumerate}
            \item If $a_k \in \neighborhood(d_i)$, increase the weights of 
            \begin{equation*}
                \left(a_{k}^{(1)}, a_{k + i \mod n}^{(2)}\right), \left(a_{k + i \mod n}^{(2)}, a_{k}^{(3)}\right)
            \end{equation*}
            to $2$ and increase the weights of
            \begin{equation*}
                \left(\hat{a}_{k}^{(1)}, \hat{a}_{k + i \mod n}^{(2)}\right), \left(\hat{a}_{k + i \mod n}^{(2)}, \hat{a}_{k}^{(3)}\right)
            \end{equation*}
            to $8$. 
            \item Computing $\arrival(s, t)$ by querying $\innerAlg$. If $\arrival(s, t) = 9$, return $\true$.
            \item After the query, increase the weights of 
            \begin{equation*}
                \left(a_{k}^{(1)}, a_{k + i \mod n}^{(2)}\right), \left(a_{k + i \mod n}^{(2)}, a_{k}^{(3)}\right), \left(\hat{a}_{k}^{(1)}, \hat{a}_{k + i \mod n}^{(2)}\right), \left(\hat{a}_{k + i \mod n}^{(2)}, \hat{a}_{k}^{(3)}\right)
            \end{equation*}
            to $10$.
        \end{enumerate}
        \item After all queries of this round, increase the weights of
        \begin{equation*}
            \left(b_j^{(1)}, b_{j + i \mod n}^{(2)}\right), \left(b_{j + i \mod n}^{(2)}, b_j^{(3)} \right), \left(c_j^{(1)}, c_{j + i \mod n}^{(2)}\right), \left(c_{j + i \mod n}^{(2)}, c_j^{(3)} \right)
        \end{equation*}
        for all $b_j, c_j \in N(d_i)$ to $10$.
    \end{enumerate}
    Then, return $\false$ after all iterations.

    For each $d$, we increase/decrease the weights of at most $16n$ edges, so the total number of updates is $O(n^2)$. The bottleneck of this algorithm is to run $\innerAlg$, so the running time is $O(n^{4-c})$.

\begin{lemma}
    \label{lemma:s-t-earliest-arrivals-4-clique-equiv}
    Consider the query when iterating over $d_i$ and $a_k$.
    The result of the query is $\arrival(s, t) = 9$ if and only if $(d_i, a_k)$ are in a 4-clique.
\end{lemma}

\begin{proof}
    Note that all edges into $t$ have weight $9$.
    Thus, the arrival time $\arrival(s, t) \in \set{9, \infty}$, and $\arrival(s, t) = 9$ if and only if there is a valid itinerary from $s$ to $t$.
    
    Suppose $d_i, a_k$ are in a 4-clique with vertices $b_{j}, c_{\ell}$.
    Then, the following valid itinerary exists in the graph
    \begin{equation*}
        \left(s, a_{k}^{(1)}, a_{k + i \mod n}^{(2)},  a_{k}^{(3)}, b_{j}^{(1)}, b_{j + i \mod n}^{(2)}, b_{j}^{(3)}, c_{\ell}^{(1)}, c_{\ell + i \mod n}^{(2)}, c_{\ell}^{(3)}, \hat{a}_{k}^{(1)}, \hat{a}_{k + i \mod n}^{(2)},  \hat{a}_{k}^{(3)}, t\right).
    \end{equation*}

    We verify that this path exists and is valid (i.e. has non-decreasing weights).
    $(s, a_{k}^{(1)})$ has weight $1$.
    Since $a_{k} \in N(d_i)$, each edge in the sub-path $\left(a_{k}^{(1)}, a_{k + i \mod n}^{(2)}, a_{k}^{(3)} \right)$ has weight $2$. 
    Next, $(a_k, b_j) \in E$ implies that the edge $(a_{k}^{(3)}, b_{j}^{(1)})$ exists with weight $3$.
    Since $b_{j} \in N(d_i)$, each edge in the sub-path $\left(b_{j}^{(1)}, b_{j + i \mod n}^{(2)}, b_{j}^{(3)} \right)$ has weight $4$.
    Then, $(b_j, c_{\ell}) \in E$ implies that the edge $(b_{j}^{(3)}, c_{\ell}^{(1)})$ exists with weight $5$.
    Since $c_{\ell} \in N(d_i)$, each edge in the sub-path $\left(c_{\ell}^{(1)}, c_{\ell + i \mod n}^{(2)}, c_{\ell}^{(3)} \right)$ has weight $6$.
    Then, $(c_{\ell}, a) \in E$ implies that the edge $(c_{\ell}^{(3)}, \hat{a}_{k}^{(1)})$ exists with weight $7$.
    Finally, as $a_{k} \in N(d_i)$, we have the sub-path $\left(\hat{a}_{k}^{(1)}, \hat{a}_{k + i \mod n}^{(2)}, \hat{a}_{k}^{(3)} \right)$ with weight $8$.
    We conclude by observing that edge $(\hat{a}_{k}^{(3)}, t)$ has weight $9$.

    Conversely, suppose $\arrival(s, t) = 9$.
    Then, since edges are directed left to right, there is a path of the form
    \begin{equation*}
        \left(s, a_{i_1}^{(1)}, a_{i_2}^{(2)},  a_{i_3}^{(3)}, b_{i_4}^{(1)}, b_{i_5}^{(2)}, b_{i_6}^{(3)}, c_{i_7}^{(1)}, c_{i_8}^{(2)}, c_{i_9}^{(3)}, \hat{a}_{i_{10}}^{(1)}, \hat{a}_{i_{11}}^{(2)},  \hat{a}_{i_{12}}^{(3)}, t\right)
    \end{equation*}
    and is a valid itinerary.
    
    At the time of the query in the $i$-th iteration over $D$ and $k$-th iteration over $A$ in the inner-loop, the only edges in $A_1 \times A_2, A_2 \times A_3, \hat{A}_1 \times \hat{A}_2, \hat{A}_2 \times \hat{A}_3$ with weight $2$ are $\left(a_{k}^{(1)}, a_{k + i \mod n}^{(2)}\right)$, $\left(a_{k + i \mod n}^{(2)}, a_{k}^{(3)}\right)$, $\left(\hat{a}_{k}^{(1)}, \hat{a}_{k + i \mod n}^{(2)}\right)$, and $\left(\hat{a}_{k + i \mod n}^{(2)}, \hat{a}_{k}^{(3)}\right)$.
    Note that all other weights have weight either $0$ or $10$, and thus cannot participate in any valid itinerary (since edges into $A_1$ have weight $1$ and edges out of $A_3$ have weight $3$).
    If these edges exist with weight $2$, then $(a_k, d_i) \in E$ and $k = i_1 = i_3 = i_{10} = i_{12}$.
    The only valid edges between $B_1 \times B_2$ and $B_2 \times B_3$ with weight $4$ are $\left(b_{j}^{(1)}, b_{j + i \mod n}^{(2)}\right)$, $\left(b_{j + i \mod n}^{(2)}, b_{j}^{(3)}\right)$ for $b_j \in N(d_i)$ (and no other edges can participate in a valid itinerary since edges into $B_1$ have weight $3$ and edges out of $B_3$ have weight $5$).
    Since these paths are vertex disjoint, we have $i_4 = i_6 = j$ for some $b_j \in N(d_i)$.
    By a similar argument, $i_7 = i_9 = \ell$ for some $c_{\ell} \in N(d_i)$.
    
    Finally, it remains to show $a_k, b_j, c_{\ell}$ form a triangle.
    Since $(a_{k}^{(3)}, b_{j}^{(1)})$ form an edge, we ahve $(a_{k}, b_{j}) \in E$.
    A similar argument shows that $(a_{k}, c_{\ell}), (b_j, c_{\ell}) \in E$.
\end{proof}

    From Lemma \ref{lemma:s-t-earliest-arrivals-4-clique-equiv}, the above procedure solves the 4-Clique Detection instance, contradicting the Combinatorial 4-Clique hypothesis.
\end{proof}

Again, using an essentially identical reduction we obtain a lower bound for arbitrary algorithms under the $\OMvThree$ hypothesis.

\begin{restatable}{theorem}{stEAlbOMvThree}
    \label{thm:s-t-ea-lb-omv-3}
    Under the $\OMvThree$ hypothesis, any algorithm computing incremental/decremental $\stEA$ on directed graphs with polynomial preprocessing time requires $n^{\omega - 1 - o(1)}$ amortized update and query time.
    Furthermore, each edge weight is incremented only once.
\end{restatable}

We defer the proof to Appendix \ref{app:omv-3-lb}.
In particular, for \stEA{}, we have shown that modifying edge weights and inserting/removing edges lead to different complexities in the partially dynamic model.
Specifically, while edge insertions/deletions can be handled efficiently, it is hard to maintain earliest arrivals under weight modifications.
Since the weight-dynamic is more general than the insertion/deletion model, our lower bounds for both shortest paths and bottleneck paths continue to hold in the weight-dynamic model.

\begin{remark}
    \label{remark:one-weight-inc-to-many-weight-inc}
    In both Theorems \ref{thm:s-t-ea-lb} and \ref{thm:s-t-ea-lb-omv-3}, we have restricted the lower bound to only modify each edge weight once.
    We can extend our lower bound to handle many increases per edge as follows.
    After modifying each edge weight once, increase all edge weights by some large enough constant (e.g. $10$) so that all edges that started at weight $0$ are now weight $10$, all weights that started at weight $1$ are weight $11$, and so on, and run the reduction again.
    Using this modification, we can compute Unbalanced $4$-Clique, or $\OMvThree$ with arbitrarily (polynomially) many queries.
    Note that this maintains the fact that all weight modifications are increase only.
\end{remark}

\bibliographystyle{alpha}
\bibliography{references}

\newpage
\appendix

\section{Omitted Proofs for Partially Dynamic Lower Bounds}
\label{app:partially-dynamic-lb}

In this appendix, we provide several omitted lower bounds for partially dynamic path problems.

\subsection{Lower Bounds from the \texorpdfstring{$\OMvThree$}{OMv3} Hypothesis}
\label{app:omv-3-lb}

First, we show that bottleneck queries have the desired equivalence with $\OMvThree$.

\stBPOMvThreeEquiv*

\begin{proof}[Proof of Lemma \ref{lemma:st-bp-omv-3-equiv}]
    First, consider the converse direction.
    Let $j, \ell$ denote the indices such that the statement is true.
    We claim the following $(s, t)$-bottleneck path exists in the graph and has capacity $i \cdot n + (k + 1)$,
    \begin{equation*}
        \left(s, a_{k}^{(1)}, a_{k + i \mod n}^{(2)},  a_{k}^{(3)}, b_{j}^{(1)}, b_{j + i \mod n}^{(2)}, b_{j}^{(3)}, c_{\ell}^{(1)}, c_{\ell + i \mod n}^{(2)}, c_{\ell}^{(3)}, \hat{a}_{k}^{(1)}, \hat{a}_{k + i \mod n}^{(2)},  \hat{a}_{k}^{(3)}, t\right)
    \end{equation*}

    We verify that this path exists and has the required capacity.
    $(s, a_{k}^{(1)})$ has capacity $\infty$.
    Since $\vec{u}_{i}[k] = \true$, each edge in the sub-path $\left(a_{k}^{(1)}, a_{k + i \mod n}^{(2)}, a_{k}^{(3)} \right)$ exists and has capacity $i \cdot n + (k + 1)$.
    Next, $A[k, j] = \true$ implies that the edge $(a_{k}^{(3)}, b_{j}^{(1)})$ exists with $\infty$ capacity.
    Since $\vec{v}_{i}[j] = \true$, each edge in the sub-path $\left(b_{j}^{(1)}, b_{j + i \mod n}^{(2)}, b_{j}^{(3)} \right)$ exists and has capacity $(i + 1) \cdot n$.
    Then, $A[j, \ell] = \true$ implies that the edge $(b_{j}^{(3)}, c_{\ell}^{(1)})$ exists with $\infty$ capacity.
    Since $\vec{w}_{i}[\ell] = \true$, each edge in the sub-path $\left(c_{\ell}^{(1)}, c_{\ell + i \mod n}^{(2)}, c_{\ell}^{(3)} \right)$ exists and has capacity $(i + 1) \cdot n$.
    Then, $A[\ell, k] = \true$ implies that the edge $(c_{\ell}^{(3)}, \hat{a}_{k}^{(1)})$ exists with $\infty$ capacity.
    Finally, as $\vec{u}_{i}[k] = \true$, we have the sub-path $\left(\hat{a}_{k}^{(1)}, \hat{a}_{k + i \mod n}^{(2)}, \hat{a}_{k}^{(3)} \right)$ with capacity $i \cdot n + (k + 1)$.
    We conclude by observing that edge $(\hat{a}_{k}^{(3)}, t)$ has $\infty$ capacity.

    Conversely, suppose $\bottleneck(s, t) \geq i \cdot n + (k + 1)$.
    Then, since edges are directed left to right, there is a path of the form,
    \begin{equation*}
        \left(s, a_{i_1}^{(1)}, a_{i_2}^{(2)},  a_{i_3}^{(3)}, b_{i_4}^{(1)}, b_{i_5}^{(2)}, b_{i_6}^{(3)}, c_{i_7}^{(1)}, c_{i_8}^{(2)}, c_{i_9}^{(3)}, \hat{a}_{i_{10}}^{(1)}, \hat{a}_{i_{11}}^{(2)},  \hat{a}_{i_{12}}^{(3)}, t\right)
    \end{equation*}
    achieving this capacity.
    In both incremental and decremental cases, at the time of the query after updating $k$-th coordinate of $\vec{u}_{i}$ during the $\OMvThree$ query $\vec{u}_{i}, \vec{v}_i, \vec{w}_i$, the maximum weight of any edge between $A_1 \times A_2$, $A_2 \times A_3$, $\hat{A}_1 \times \hat{A}_2$, and $\hat{A}_2 \times \hat{A}_3$ is $i \cdot n + (k + 1)$.
    Furthermore, the only edges attaining this weight are $\left(a_{k}^{(1)}, a_{k + i \mod n}^{(2)}\right)$, $\left(a_{k + i \mod n}^{(2)}, a_{k}^{(3)}\right)$, $\left(\hat{a}_{k}^{(1)}, \hat{a}_{k + i \mod n}^{(2)}\right)$, and $\left(\hat{a}_{k + i \mod n}^{(2)}, \hat{a}_{k}^{(3)}\right)$.
    If these edges exist, then $\vec{u}_{i}[k] = \true$ and $k = i_1 = i_3 = i_{10} = i_{12}$.
    Note also that $i_{2} = i_{11} = k + i \mod n$.
    
    During this query, in both incremental and decremental cases, the maximum weight of any edge between $B_1 \times B_2$ and $B_2 \times B_3$ is $(i + 1) \cdot n$.
    Furthermore, the only edges attaining this weight are $\left(b_{j}^{(1)}, b_{j + i \mod n}^{(2)}\right)$, $\left(b_{j + i \mod n}^{(2)}, b_{j}^{(3)}\right)$ for $b_j \in N(d_i)$.
    In particular, we have $i_4 = i_6 = j$ and $i_5 = j + i \mod n$ for some $\vec{v}_{i}[j] = \true$.
    By a similar argument, $i_7 = i_9 = \ell$ and $i_8 = \ell + i \mod n$ for some $\vec{w}_{i}[\ell] = \true$.
    
    Finally, it remains to show $A[k, j] = A[j, \ell] = A[\ell, j] = \true$.
    Since $\wt(a_{k}^{(3)}, b_{j}^{(1)}) = \infty$ we have $A[k, j] = \true$.
    A similar argument shows that $A[j, \ell] = A[\ell, k] = \true$.
\end{proof}

Finally, we prove the lower bound for partially weight dynamic $\stEA$ under the $\OMvThree$ hypothesis.

\stEAlbOMvThree*

\begin{proof}[Proof of Theorem \ref{thm:s-t-ea-lb-omv-3}]
    We describe the reduction for incremental $\stEA$ (weight increasing), noting that the decremental reduction can be obtained by modifying the reduction as in \Cref{thm:s-t-sp-node-weight-lb-omv-3}.
    
    Suppose for contradiction there is an algorithm $\innerAlg$ with polynomial preprocessing time and total update and query time $O(n^{\omega + 1 - c})$ for some $c > 0$.
    Consider an $\OMvThree$ instance with matrix $A$ and queries $(\vec{u}_{i}, \vec{v}_i, \vec{w}_i)_{i = 1}^{n}$.
    Throughout the lower bound, assume that the matrices and vectors are indexed $1$ to $n$.
    We design an algorithm with polynomial preprocessing time and total update and query time $O(n^{\omega + 1 - c})$, contradicting the $\OMvThree$ hypothesis.

    We construct a graph with vertices,
    \begin{equation*}
        \set{s} \cup (A_1 \cup A_2 \cup A_3) \cup (B_1 \cup B_2 \cup B_3) \cup (C_1 \cup C_2 \cup C_3) \cup (\hat{A}_1 \cup \hat{A}_2 \cup \hat{A}_3) \cup \set{t}
    \end{equation*}
    Each vertex set $A_1, A_2, A_3, B_1, B_2, B_3, C_1, C_2, C_3, \hat{A}_1, \hat{A}_2$ and $\hat{A}_3$ has $n$ vertices.
    For each $a \in A$, we create a copy $a^{(i)} \in A_i$ and a copy $\hat{a}^{(i)} \in \hat{A}_i$ for $i \in \set{1, 2, 3}$.
    Similarly, for each $b \in B$ and $c \in C$, we create a copy $b^{(i)} \in B_i$ and $c^{(i)} \in C_i$ for $i \in \set{1, 2, 3}$.
    Initially, insert edges $(s, a^{(1)})$ with weight $1$ and edges $(\hat{a}^{(3)}, t)$ with weight $9$ for all $a \in A$.
    Furthermore, insert edges with $3$ weight between $(a^{(3)}, b^{(1)})$ (respectively weight $5$ between $(b^{(3)}, c^{(1)})$ and weight $7$ between $(c^{(3)}, \hat{a}^{(1)})$) if and only if $A[a, b] = \true$ (respectively if $A[b, c] = \true$ and $A[c, a] = \true$).
    We also insert all edges between $A_1 \times A_2, A_2 \times A_3, B_1 \times B_2, B_2 \times B_3, C_1 \times C_2, C_2 \times C_3, \hat{A}_1 \times \hat{A}_2, \hat{A}_2 \times \hat{A}_3$ with weight $0$.

    This creates a graph with $O(n)$ vertices, so that by running $\innerAlg$ we can preprocess the graph in polynomial time.
    We now proceed to the dynamic phase of the reduction.

    Consider an $\OMvThree$ query $\vec{u}_i, \vec{v}_i, \vec{w}_i$.
    \begin{enumerate}
        \item For each $\vec{v}_{i}[j] = \true$, increase the weights of $\left(b_j^{(1)}, b_{j + i \mod n}^{(2)}\right)$ and $\left(b_{j + i \mod n}^{(2)}, b_j^{(3)} \right)$ to $4$.
        \item For each $\vec{w}_{i}[j] = \true$, increase the weights of $\left(c_j^{(1)}, c_{j + i \mod n}^{(2)}\right)$ and $\left(c_{j + i \mod n}^{(2)}, c_j^{(3)} \right)$ to $6$.
        \item In the inner loop, we iterate over coordinates $k$ of $\vec{u}$ in increasing order.
        \begin{enumerate}
            \item If $\vec{u}_{i}[k] = \true$, increase the weights of 
            \begin{equation*}
                \left(a_{k}^{(1)}, a_{k + i \mod n}^{(2)}\right), \left(a_{k + i \mod n}^{(2)}, a_{k}^{(3)}\right)
            \end{equation*}
            to $2$ and increase the weights of,
            \begin{equation*}
                \left(\hat{a}_{k}^{(1)}, \hat{a}_{k + i \mod n}^{(2)}\right), \left(\hat{a}_{k + i \mod n}^{(2)}, \hat{a}_{k}^{(3)}\right)
            \end{equation*}
            to $8$. 
            \item Compute $\arrival(s, t)$ by querying $\innerAlg$. If $\arrival(s, t) = 9$, return $\true$.
            \item After the query, increase the weights of 
            \begin{equation*}
                \left(a_{k}^{(1)}, a_{k + i \mod n}^{(2)}\right), \left(a_{k + i \mod n}^{(2)}, a_{k}^{(3)}\right), \left(\hat{a}_{k}^{(1)}, \hat{a}_{k + i \mod n}^{(2)}\right), \left(\hat{a}_{k + i \mod n}^{(2)}, \hat{a}_{k}^{(3)}\right)
            \end{equation*}
            to $10$.
        \end{enumerate}
        \item After all queries of this round, increase all weights of
        \begin{equation*}
            \left(b_j^{(1)}, b_{j + i \mod n}^{(2)}\right), \left(b_{j + i \mod n}^{(2)}, b_j^{(3)} \right), \left(c_j^{(1)}, c_{j + i \mod n}^{(2)}\right), \left(c_{j + i \mod n}^{(2)}, c_j^{(3)} \right)
        \end{equation*}
        for all $\vec{u}_{i}[k] = \vec{w}_{i}[k] = \true$ to $10$.
        Then, return $\false$.
    \end{enumerate}

    The bottleneck of this algorithm is to run $\innerAlg$, so the total query time is $O(n^2 \cdot n^{\omega - 1 - c}) = O(n^{\omega + 1 - c})$.

\begin{lemma}
    \label{lemma:s-t-earliest-arrivals-omv-3-equiv}
    Consider the query when iterating over $d_i$ and $a_k$.
    The result of the query is $\arrival(s, t) = 9$ if and only if,
    \begin{equation*}
        \bigvee_{j, \ell} \left( \vec{u}_i[k] \wedge \vec{v}_i[j] \wedge \vec{w}_i[\ell] \wedge A[k, j] \wedge A[j, \ell] \wedge A[\ell, k] \right) = \true
    \end{equation*}
\end{lemma}

\begin{proof}
    Note that all edges into $t$ have weight $9$.
    Thus, the arrival time $\arrival(s, t) \in \set{9, \infty}$, and $\arrival(s, t) = 9$ if and only if there is a valid arrival path from $s$ to $t$.
    
    Suppose there is clause indexed by $j, \ell$ that is $\true$.
    Then, the following arrival path exists in the graph,
    \begin{equation*}
        \left(s, a_{k}^{(1)}, a_{k + i \mod n}^{(2)},  a_{k}^{(3)}, b_{j}^{(1)}, b_{j + i \mod n}^{(2)}, b_{j}^{(3)}, c_{\ell}^{(1)}, c_{\ell + i \mod n}^{(2)}, c_{\ell}^{(3)}, \hat{a}_{k}^{(1)}, \hat{a}_{k + i \mod n}^{(2)},  \hat{a}_{k}^{(3)}, t\right)
    \end{equation*}

    We verify that this path exists and is valid.
    $(s, a_{k}^{(1)})$ has weight $1$.
    Since $\vec{u}_{i}[k] = \true$, each edge in the sub-path $\left(a_{k}^{(1)}, a_{k + i \mod n}^{(2)}, a_{k}^{(3)} \right)$ has weight $2$.
    Next, $A[k, j] = \true$ implies that the edge $(a_{k}^{(3)}, b_{j}^{(1)})$ exists with weight $3$.
    Since $\vec{v}_{i}[j] = \true$, each edge in the sub-path $\left(b_{j}^{(1)}, b_{j + i \mod n}^{(2)}, b_{j}^{(3)} \right)$ has weight $4$.
    Then, $A[j, \ell] = \true$ implies that the edge $(b_{j}^{(3)}, c_{\ell}^{(1)})$ exists with weight $5$.
    Since $\vec{w}_{i}[\ell] = \true$, each edge in the sub-path $\left(c_{\ell}^{(1)}, c_{\ell + i \mod n}^{(2)}, c_{\ell}^{(3)} \right)$ has weight $6$.
    Then, $A[\ell, k] = \true$ implies that the edge $(c_{\ell}^{(3)}, \hat{a}_{k}^{(1)})$ exists with weight $7$.
    Finally, as $\vec{u}_{i}[k] = \true$, we have the sub-path $\left(\hat{a}_{k}^{(1)}, \hat{a}_{k + i \mod n}^{(2)}, \hat{a}_{k}^{(3)} \right)$ with weight $8$.
    We conclude by observing that edge $(\hat{a}_{k}^{(3)}, t)$ has weight $9$.

    Conversely, suppose $\arrival(s, t) = 9$.
    Then, since edges are directed left to right, there is a path of the form,
    \begin{equation*}
        \left(s, a_{i_1}^{(1)}, a_{i_2}^{(2)},  a_{i_3}^{(3)}, b_{i_4}^{(1)}, b_{i_5}^{(2)}, b_{i_6}^{(3)}, c_{i_7}^{(1)}, c_{i_8}^{(2)}, c_{i_9}^{(3)}, \hat{a}_{i_{10}}^{(1)}, \hat{a}_{i_{11}}^{(2)},  \hat{a}_{i_{12}}^{(3)}, t\right)
    \end{equation*}
    and is a valid arrival path.
    
    In both incremental and decremental cases, at the time of the query after updating the $k$-th coordinate of the $i$-th $\OMvThree$ query $\vec{u}_i, \vec{v}_i, \vec{w}_i$, the only edges in $A_1 \times A_2, A_2 \times A_3, \hat{A}_1 \times \hat{A}_2, \hat{A}_2 \times \hat{A}_3$ with weight $2$ are $\left(a_{k}^{(1)}, a_{k + i \mod n}^{(2)}\right)$, $\left(a_{k + i \mod n}^{(2)}, a_{k}^{(3)}\right)$, $\left(\hat{a}_{k}^{(1)}, \hat{a}_{k + i \mod n}^{(2)}\right)$, and $\left(\hat{a}_{k + i \mod n}^{(2)}, \hat{a}_{k}^{(3)}\right)$.
    Note that all other weights have weight either $0$ or $10$, and thus cannot participate in any valid arrival path (since edges into $A_1$ have weight $1$ and edges out of $A_3$ have weight $3$).
    If these edges exist with weight $2$, then $\vec{u}_{i}[k] = \true$ and $k = i_1 = i_3 = i_{10} = i_{12}$.
    The only valid edges between $B_1 \times B_2$ and $B_2 \times B_3$ with weight $4$ are $\left(b_{j}^{(1)}, b_{j + i \mod n}^{(2)}\right)$, $\left(b_{j + i \mod n}^{(2)}, b_{j}^{(3)}\right)$ for $\vec{v}_{i}[j] = \true$ (and no other edges can participate in an arrival path since edges into $B_1$ have weight $3$ and edges out of $B_3$ have weight $5$).
    Since these paths are vertex disjoint, we have $i_4 = i_6 = j$ for some $\vec{v}_{i}[j] = \true$.
    By a similar argument, $i_7 = i_9 = \ell$ for some $\vec{w}_{i}[\ell] = \true$.
    
    Finally, it remains to show $A[k, j] = A[j, \ell] = A[\ell, k] = \true$.
    Since $(a_{k}^{(3)}, b_{j}^{(1)})$ form an edge, we have $A[k, j] = \true$.
    A similar argument shows that $A[j, \ell] = A[\ell, k] = \true$.
\end{proof}

    From Lemma \ref{lemma:s-t-earliest-arrivals-omv-3-equiv}, the above procedure solves the \OMvThree{} instance, contradicting the $\OMvThree$ hypothesis.
\end{proof}

\subsection{Lower Bounds for Sparse Graphs}

We have given strong conditional lower bounds for a variety of partially dynamic problems on dense graphs, where $m = \Theta(n^2)$.
The following lemma shows that our lower bounds similarly apply to graphs with all sparsties.
In fact, we obtain a simple reduction for any graph problem where the solution does not change when adding isolated vertices.

\begin{proposition}[Graph Sparsification]
    \label{prop:graph-sparsification}
    For each of the following problems:
    \begin{enumerate}
        \item $\stSP, \SSSP$ on undirected, weighted graphs
        \item \nwstSP{}, \nwSSSP{} on undirected, node-weighted graphs
        \item $\stBP, \SSBP$ on directed, weighted graphs
    \end{enumerate}
    
    Suppose any incremental/decremental algorithm for the given problem requires $n^{t - o(1)}$ total update time  on $n$-vertex graphs.
    
    Then, for any $m \leq \binom{n}{2}$,  any incremental/decremental algorithm for $\problem$ requires $m^{t/2 - o(1)}$ total update time on $n$-vertex graphs with $m$ edges.
\end{proposition}

\begin{proof}
    Let $m \leq \binom{n}{2}$ and set $n_0 = \floor{\sqrt{m}} \leq n$.
    By assumption, any incremental/decremental algorithm requires $n_0^{t - o(1)}$ total time on $n_0$-vertex graphs.
    Given any instance of a partially dynamic $n_0$-vertex graph we can construct a partially dynamic $n$-vertex graph by adding $n - n_0$ isolated vertices, which we will never add edges to. 
    Suppose for contradiction there is an algorithm with $O(m^{t/2 - c})$ total time on the augmented graph for some $c > 0$.
    Then, we obtain an algorithm for the $n_0$-vertex graph simply by outputting the queries on the $n_0$-vertex sub-graph of the augmented $n$-vertex graph, obtaining an $O(m^{t/2 - c}) = O(n_0^{t - 2c})$ total time algorithm, a contradiction.
\end{proof}

\end{document}